\documentclass[12pt]{article}
\makeatletter \@addtoreset{equation}{section} \makeatother
\renewcommand{\theequation}{\thesection.\arabic{equation}}
\newcommand{\ba}{\begin{array}}
\newcommand{\ea}{\end{array}}
\newcommand{\beq}{\begin{equation}}
\newcommand{\eeq}{\end{equation}}
\newcommand{\bea}{\begin{eqnarray}}
\newcommand{\eea}{\end{eqnarray}}

\def\bce{\begin{center}}
\def\ece{\end{center}}

\def\nonu{\nonumber}

\def\pa{\partial}

\def\be{\beta}

\def\ep{\epsilon}

\def\la{\lambda}

\def\eps6{{\displaystyle \mathop{\epsilon}^{6}}{}}
\def\g6{{\displaystyle \mathop{g}^{6}}{}}

\def\nab6{{\displaystyle \mathop{\nabla}^{6}}{}}

\def\0{{\sst{(0)}}}
\def\1{{\sst{(1)}}}
\def\2{{\sst{(2)}}}
\def\3{{\sst{(3)}}}
\def\4{{\sst{(4)}}}
\def\5{{\sst{(5)}}}
\def\6{{\sst{(6)}}}
\def\7{{\sst{(7)}}}
\def\8{{\sst{(8)}}}

\def\ba{\begin{array}}
\def\ea{\end{array}}
\def\beq{\begin{equation}}
\def\eeq{\end{equation}}
\def\be{\begin{equation}}
\def\ee{\end{equation}}

\def\la{\lambda}
\def\eps{\epsilon}

\def\ba{\begin{array}}
\def\ea{\end{array}}
\def\beq{\begin{equation}}
\def\eeq{\end{equation}}
\def\be{\begin{equation}}
\def\ee{\end{equation}}

\def\la{\lambda}
\def\eps{\epsilon}

\def\eps6{{\displaystyle \mathop{\epsilon}^{6}}{}}

\def\nab6{{\displaystyle \mathop{\nabla}^{6}}{}}

\newcommand{\bean}{\begin{eqnarray*}}
\newcommand{\eean}{\end{eqnarray*}}

\begin{document}
\thispagestyle{empty} \addtocounter{page}{-1}
   \begin{flushright}
\end{flushright}

\vspace*{1.3cm}
  
\centerline{ \Large \bf   
Three Point Functions in the  ${\cal N}=4$ Orthogonal Coset Theory } 
\vspace*{1.5cm}
\centerline{{\bf Changhyun Ahn}, {\bf Hyunsu Kim} and 
{\bf Jinsub Paeng}
} 
\vspace*{1.0cm} 
\centerline{\it 
Department of Physics, Kyungpook National University, Taegu
41566, Korea} 
\vspace*{0.8cm} 
\centerline{\tt ahn, kimhyun, jdp2r@knu.ac.kr 
} 
\vskip2cm

\centerline{\bf Abstract}
\vspace*{0.5cm}

We construct the lowest higher spin-$2$ current  
in terms of the spin-$1$ and the spin-$\frac{1}{2}$ currents
living in the orthogonal $\frac{SO(N+4)}{SO(N) \times SO(4)}$ 
Wolf space coset theory for general $N$.  
The remaining fifteen higher spin currents are determined.
We obtain the three-point functions of bosonic 
(higher) spin currents with two scalars for finite $N$ and $k$ (the level
of the spin-$1$ current).
By multiplying $SU(2) \times U(1)$ into the above Wolf space coset theory,
the other fifteen higher spin currents together with 
the above lowest higher spin-$2$ current 
are realized in the extension of the 
large ${\cal N}=4$ linear superconformal algebra.
Similarly, the three-point functions  
of bosonic 
(higher) spin currents with two scalars for finite $N$ and $k$
are obtained.
Under the large $N$ 't Hooft limit, 
the two types of three-point functions in the nonlinear and linear 
versions coincide as in the unitary coset theory found previously.

\baselineskip=18pt
\newpage
\renewcommand{\theequation}
{\arabic{section}\mbox{.}\arabic{equation}}

\section{Introduction}

By analyzing the zero-mode eigenvalue equations for the bosonic (higher spin)
currents in the extension of the 
large ${\cal N}=4$ (non)linear superconformal algebra,
its three-point functions with two scalars \cite{AK1506} were obtained 
in the context of the large ${\cal N}=4$ holography \cite{GG1305}.    
Even though the corresponding three-point functions in the 
nonlinear  and linear versions are different from each 
other for finite $N$ and $k$, 
where these two parameters characterize the 
${\cal N}=4$ unitary coset theory (or they correspond to two levels of
the above large ${\cal N}=4$ (non)linear superconformal algebra), 
they coincide under the large $N$ 't 
Hooft limit.
For example,
the central charge in the large ${\cal N}=4$
linear superconformal algebra \cite{GG1305} 
is given by $c = 6(1-\la)(N+1)$, where 
the $\la$ is the 't Hooft coupling constant ($0 < \la < 1$).
For fixed $\la$, the large $N$ 't Hooft limit 
is equivalent to the large $c$ limit.
Note that the central charge in the nonlinear version is reduced by $3$.
As long as the three-point functions under the large 
$N$ 't Hooft limit are concerned, the higher-order effects 
(or subleading orders) 
of $\frac{1}{c}$ is not important, for example, in the study of marginal 
deformation in the Higgs phenomenon (in the context of 
other holographic  model)  
\cite{CH1506,HR1503} because the leading order of
$\frac{1}{c}$ is taken. 
However, 
we should observe the finite $N$-effect
in order to see the quantum behavior (or subleading orders of 
$\frac{1}{c}$) in this large ${\cal N}=4$ holography \cite{GG1305} 
(or above other 
holographic model). 

It is natural, as raised in \cite{AK1506}, 
to consider the other type of coset theory in order to observe the 
consistency check in the other type of large ${\cal N}=4$ holography.
In \cite{AP1410}, the $16$ lowest 
higher spin currents (one higher spin-$2$ current, 
four higher spin-$\frac{5}{2}$ currents, six higher spin-$3$ currents, four 
higher spin-$\frac{7}{2}$ currents and 
one higher spin-$4$ current) in the extension of 
large ${\cal N}=4$ nonlinear superconformal algebra 
were constructed in the orthogonal
coset theory for fixed $N=4$ (and for general $k$).
What is so special to the orthogonal coset theory compared to the unitary
coset theory?
One of the findings in \cite{AP1410} was that the lowest higher spin current 
in the ${\cal N}=4$ multiplet has spin $2$ and this implies that 
the highest higher spin current has spin $4$ as above. 
Then we expect that we will obtain the three-point functions for the 
higher spin-$4$ current. Note that for the unitary coset theory
the corresponding three-point functions were obtained for the (higher spin) 
currents of spins $s=2,3$.
We did not calculate the three-point functions of spins $s$ greater than $3$.
We can expect the spin-dependence for the three-point functions 
in the unitary coset theory under the large $N$ 't Hooft limit
from the results of the orthogonal coset theory
because we expect that they share the common spin-behavior.   
Furthermore, the six higher spin-$3$ currents 
in the orthogonal coset theory transform as 
the adjoint of $SO({\cal N}=4)$ (we are considering the $SO({\cal N}=4)$ 
singlet 
${\cal N}=4$ multiplet)
while the one higher spin-$3$ current 
in the unitary coset theory transforms as a singlet under the 
$SO({\cal N}=4)$. In other words, the former appear in the 
quadratic in the fermionic coordinates in the ${\cal N}=4$ multiplet
and the latter appears in the quartic in the fermionic coordinates 
in the ${\cal N}=4$ multiplet 
\footnote{
Similarly,
the six higher spin-$2$ currents 
in the unitary coset theory transform as the adjoint of $SO({\cal N}=4)$
(quadratic in the fermionic coordinates in the ${\cal N}=4$ multiplet)
while the one higher spin-$2$ current 
in the orthogonal coset theory transforms as a singlet under the 
$SO({\cal N}=4)$ (and appears in the fermionic independent term in the 
${\cal N}=4$ multiplet).}.

Therefore, we should obtain the $16$ lowest higher spin currents implicitly
(or explicitly)
for generic $N$ in the extension of the large ${\cal N}=4$ (non)linear 
superconformal algebra (in the realization of orthogonal coset theory)
by generalizing the previous work in \cite{AP1410} to 
the $N$-generalization.
As long as the three-point functions are concerned, 
the several $N$ cases are enough to determine them completely. 
This feature is different from the one in the bosonic coset theory 
\cite{AK1308} (in the context of 
\cite{GG1011,GG1205,GG1207}) 
where the explicit results for the higher spin currents (for
generic $N$) are necessary.
In this construction, the four spin-$\frac{3}{2}$ currents 
in the  large ${\cal N}=4$ (non)linear 
superconformal algebra play an important role.
We follow the procedure in \cite{AK1506}, construct the zero-mode eigenvalue 
equations and obtain the three-point functions for finite $N$ and $k$
(and also under the large $N$ 't Hooft limit).
For the unitary coset theory, 
the conformal dimension of a coset primary can be calculated 
from the quadratic Casimirs of $su(N+2)$ and $su(N)$, the quantum numbers 
of $u(1)$ algebras and an excitation number in \cite{GG1305}.
For the orthogonal coset theory, as far as we know, there is no explicit 
formula for the conformal dimension of a coset primary because it is rather 
nontrivial to obtain the correct factors in the above last
two quantities. 
This is one of the reasons why we are interested in this particular 
orthogonal coset theory.
See also the description of \cite{FG,Ahn1208,Ahn1206} 
in different orthogonal coset
theory.     

The ${\cal N}=4$ orthogonal coset theory we are interested in 
is described by the following 
`supersymmetric' coset \cite{ST}:
\bea
\mbox{Wolf} \times SU(2) \times U(1) = 
\frac{SO(N+4)}{SO(N) \times SU(2)} 
\times U(1).
\label{coset}
\eea
The fundamental currents are given by 
the bosonic spin-$1$ current $V^a(z)$ and 
the fermionic spin-$\frac{1}{2}$
current $Q^b(z)$. 
The indices run over 
$a, b, \cdots =1, 2, \cdots, \frac{(N+4)(N+3)}{2}$ where the number 
$\frac{(N+4)(N+3)}{2}$   is the dimension of the $g=so(N+4)$ algebra. 
For the extension of  
the ${\cal N}=4$ `nonlinear' superconformal algebra,
the relevant  coset is given by 
the Wolf space itself 
$\frac{SO(N+4)}{SO(N) \times SU(2) \times SU(2)}$. 
For the extension of  the ${\cal N}=4$ `linear' superconformal algebra,
the corresponding  coset is given by 
the Wolf space multiplied by $SU(2) \times U(1)$, 
which is equivalent to the above coset in the right hand side of
(\ref{coset}).

As  in \cite{AK1411}, 
we can construct 
the explicit $16$ lowest higher spin currents 
(which are multiple products of the above 
fundamental currents together with their derivatives)
which are expressed in terms of the Wolf space (or Wolf space multiplied
by $SU(2) \times U(1)$) coset fields.
These findings will allow us to calculate 
the zero modes for the higher spin currents in terms of 
the generators of the $g=so(N+4)$ algebra
because the zero modes of the spin-$1$
current $V_0^a$ satisfy the defining commutation  relations of the 
underlying finite dimensional Lie algebra $so(N+4)$.
Furthermore, all the operator product expansions 
between the higher spin currents and the 
spin-$\frac{1}{2}$ current $Q^a(z)$ 
are determined explicitly by construction.    

The minimal representations 
are given by two representations.
See also the previous works in \cite{Ahn1106,GV1106,Ahn1202,AP1310,AP1301}.
One minimal representation is
given by $(0;v)$, where  the nonnegative 
integer mode of the spin-$1$ current $V^a(z)$  in $\hat{so}(N+4)$
acting on the state $|(0;v)>$ vanishes.
Under the decomposition of 
$so(N+4)$ into $so(N) \oplus su(2) \oplus su(2)$, the   
adjoint representation of 
$so(N+4)$ can be broken into the following representations \cite{Slansky}:
${\bf \frac{1}{2}(N+4)(N+3)} \rightarrow ({ \bf \frac{1}{2} N(N-1)}, 
{\bf 1}, {\bf 1}) 
\oplus ({\bf 1},{\bf 3}, {\bf 1} ) \oplus ({\bf 1},{\bf 1}, {\bf 3}) 
\oplus ({\bf N},{\bf 2},{\bf 2})$.
Among these representations, 
the vector representation for $so(N)$
is given by $({\bf N}, {\bf 2}, {\bf 2})$ \footnote{For $N=4$, we 
have the breaking ${\bf 28} \rightarrow 
({\bf 1},{\bf 3}, {\bf 1}, {\bf 1} ) \oplus ({\bf 3},{\bf 1},{\bf 1},{\bf 1}) 
\oplus ({\bf 1},{\bf 1}, {\bf 3},{\bf 1}) \oplus ({\bf 1},{\bf 1},{\bf 1},
{\bf 3}) 
\oplus ({\bf 2},{\bf 2}, {\bf 2},{\bf 2})$ under $su(2) \oplus su(2) 
\oplus su(2) \oplus su(2)$ where the $so(4)$ is replaced with the first two
$su(2)$ factors.}.
Therefore, the representation $(0;v)$ 
corresponds to the representations 
  $({\bf N}, {\bf 2}, {\bf 2})$.
Note that the extra $su(2)$ factor in the above branching rule 
comes from the one in the left hand side of (\ref{coset}). 
The corresponding states for the representation 
$(0;v)$ are given by the 
$-\frac{1}{2}$ mode of the spin-$\frac{1}{2}$ current 
$Q^a(z)$ acting on the vacuum $|0>$, where the index $a$
is restricted to the $4N$ coset index
\footnote{We can further classify the four independent states
denoted by $|(0;v)>_{++,+-,-+,--}$ with $4N$ coset indices  
(See also \cite{npb1989})
where  four linear combinations among $(++,+-,-+,--)$ 
refer to the $({\bf 2},{\bf 2})$ 
of $su(2) \times su(2)$.}. 
The eigenvalue for the zero mode in the (higher spin) currents (multiple products
of the above spin-$1$ and spin-$\frac{1}{2}$ currents)
acting on this state can be obtained 
from the highest pole of the OPE between    
the (higher spin) current and the spin-$\frac{1}{2}$ current
as in unitary coset theory
\footnote{
Furthermore, the
nontrivial states exist  for the 
negative half-integer mode (as well as the $\frac{1}{2}$ mode) 
of the spin-$\frac{1}{2}$ current acting on the state 
$|(0;v)>$ because the action of the negative mode of the spin-$\frac{1}{2}$ 
current
on the vacuum $|0>$ is nonzero.
The positive 
half-integer modes of the spin-$\frac{1}{2}$ current ($\frac{3}{2}, 
\frac{5}{2}, \cdots$ modes)
acting on the state $|(0;v)>$ vanish.}.

The other minimal representation is given by
$(v;0)$, where 
the positive half-integer mode of the spin-$\frac{1}{2}$ current $Q^a(z)$  
in $\hat{so}(N+4)$
acting on the state $|(v;0)>$ vanishes.
They are 
singlets with respect to $so(N)$ in the $so(N+4)$ representation 
based on the vector representation.
That is,
the vector representation $({\bf N+4})$ of $so(N+4)$ transforms
as a singlet $ ({\bf 1},{\bf 4})_{\pm \frac{1}{2}}$ 
with respect to $so(N)$ 
under the branching 
$({\bf N+4})  \rightarrow 
({\bf N}, {\bf 1})_0 \oplus ({\bf 1},{\bf 4})_{\pm \frac{1}{2}}$
with respect to $so(N) \oplus so(4) \oplus u(1)$.
The indices $0$ and $\pm \frac{1}{2}$  
denote the $U(1)$ charge, which will be described later
in (\ref{eigenvalueu})
\footnote{In this case, the states are further classified as
$|(v;0)>_{++,+-,-+,--}$ with explicit $su(2) \times su(2)$ 
double indices. That is the vector representation ${\bf 4}$ breaks into 
$({\bf 1},{\bf 2}) \oplus ({\bf 2},{\bf 1})$ under the 
$su(2) \times su(2)$.}.  
On the other hand, $({\bf N}, {\bf 1})_0$
refers to the vector representation with respect to 
$so(N)$ and describes the light state $|(v;v)>$ as in unitary coset theory. 
For the state  $|(v;0)>$, the $so(N+4)$ generator
$T_{a^{\ast}}$ corresponds to the zero mode of 
the spin-$1$ current $V^a(z)$ because 
the zero mode of the spin-$1$ current satisfies the commutation relation
of the underlying finite-dimensional Lie algebra $so(N+4)$. 
Then, the nontrivial contributions to the zero-mode (of (higher spin) 
currents) eigenvalue equation
associated with the state  $|(v;0)>$
come from the multiple product of the spin-$1$ current $V^a(z)$ 
in the (higher spin) currents. 
After substituting the $so(N+4)$ generator $T_{a^{\ast}}$ into the 
zero mode of spin-$1$ current $V_0^a$ in the multiple product 
of the (higher spin) currents, we obtain the $(N+4) \times (N+4)$
matrix acting on the state  $|(v;0)>$.
Then, the last $4 \times 4$ subdiagonal matrix is associated with 
the above $so(4) \oplus u(1)$ algebra. The eigenvalue 
can be obtained from each diagonal matrix element
in this $4 \times 4$ matrix.   
Furthermore, the first  $N \times N$ subdiagonal matrix
provides the corresponding eigenvalues (for the higher spin currents) 
for the light state $|(v;v)>$, as mentioned before.

In section $2$,
we review the $\hat{so}(N+4)$ current algebra
generated by the spin-$1$ and the spin-$\frac{1}{2}$ currents.
The $11$ currents of large ${\cal N}=4$ nonlinear 
superconformal algebra using these fundamental currents are obtained.
The lowest higher spin-$2$
current for generic $N$ and $k$ is given. 
Furthermore, the remaining $15$ higher spin 
currents can be obtained implicitly. 

In section $3$,  the eigenvalue equations of the spin-$2$ 
stress-energy tensor are given for the above 
two minimal states. 
The eigenvalue equations of higher spin currents with 
spins-$2, 3$, and $4$ 
for the above 
two minimal states are presented.
The corresponding three-point functions are also described.

In section $4$,
the $16$ currents of large ${\cal N}=4$ linear 
superconformal algebra using the above fundamental currents are obtained. 
Furthermore, the $16$ higher spin 
currents can be obtained implicitly. 

In section $5$,
 the eigenvalue equations of spin-$2$ 
stress--energy tensor 
are given for the above 
two minimal states. 
Next, the eigenvalue equations of higher spin currents with 
spins-$2, 3$ and $4$ 
for the above 
two minimal states are given.
The corresponding three-point functions are described.

In section $6$,
the summary of this paper is described, and future directions 
are explained briefly.

In Appendices $A-E$,
some details in sections $2,3,4,5$ are presented.

We use the Thielemans package \cite{Thielemans} in this paper
\footnote{For the (higher spin) currents of the extension of the large 
${\cal N}=4$ linear superconformal algebra, the boldface notation is used.
For the $11$ currents of the large ${\cal N}=4$ nonlinear superconformal 
algebra, the hatted notation is used. }.

\section{The extension of 
the large $\mathcal N = 4$ nonlinear superconformal algebra}

In this section, we review the $\hat{so}(N+4)$ current algebra
generated by the spin-$1$ and the spin-$\frac{1}{2}$ currents.
We construct the $11$ currents of large ${\cal N}=4$ nonlinear 
superconformal algebra using these fundamental currents.
As far as we know, this observation is new even though the tensorial 
structures in the $11$ currents are the same as the ones in the unitary 
Wolf space coset theory. We explicitly obtain the lowest higher spin-$2$
current for generic $N$ and $k$ by generalizing the $N=4$ case in 
\cite{AP1410}. Furthermore, we show how the remaining $15$ higher spin 
currents can be obtained implicitly starting from the above higher spin-$2$
current. Finally, the general procedure to obtain the 
next $16$ higher spin currents is given.

\subsection{ The $\mathcal N =1$ Kac-Moody current algebra }

Let us consider the $\hat{so}(N+4)$ current algebra generated 
by the spin-$1$ and the spin-$\frac{1}{2}$ currents. 
The generators of the Lie algebra $g=so(N+4)$
satisfy the commutation relation
$\left[ T_a, T_b \right] = f_{a b}^{\;\;\;\; c} T_c $
and some of them are given in Appendix $A$.  
The adjoint 
indices run over $a, b, \cdots =1, 2, \cdots, \frac{(N+4)(N+3)}{2}$. 
The normalization for the generators is consistent with the metric 
$
g_{ab} =  \frac{1}{2} \mbox{Tr} (T_a T_b) =
\frac{1}{2 c_g} f_{ac}^{\,\,\,\,\,\, d} f_{bd}^{\,\,\,\,\,\, c}
$
where $c_g$ is the dual Coxeter number of the Lie algebra $g=so(N+4)$
and is given by $c_g =(N+2)$.
The operator product expansions (OPEs) 
between the spin-$1$ and the spin-$\frac{1}{2}$
currents  are summarized as  \cite{KT1985}
\bea
V^a(z) \, V^b(w) & = & \frac{1}{(z-w)^2} \, k \, g^{ab}
-\frac{1}{(z-w)} \, f^{ab}_{\,\,\,\,\,\,c} \, V^c(w) 
+\cdots,
\nonu \\
Q^a(z) \, Q^b(w) & = & -\frac{1}{(z-w)} \, (k+N+2) \, g^{ab} + \cdots,
\nonu \\
V^a(z) \, Q^b(w) & = & + \cdots.
\label{opevq}
\eea
Here $k$ is the level and a positive integer.
Note that there is no singular term in the OPE between 
the spin-$1$ current $V^a(z)$ and the spin-$\frac{1}{2}$ current 
$Q^b(w)$.
The ${\cal N}=1$ superspace description can be obtained from (\ref{opevq}).
The $k$-dependence appears in the above nontrivial OPEs while the 
$N$-dependence appears in the OPE between the spin-$\frac{1}{2}$ currents.
Furthermore, as we consider the multiple product of these 
fundamental currents, the $N$-dependence occurs from the combinations of the 
inverse metric 
$g^{ab}$ and the structure constant $f^{ab}_{\,\,\,\,\,\,c}$. 

\subsection{ The large $\mathcal N = 4$ nonlinear superconformal algebra}

The Wolf space coset we describe is given by \cite{Wolf,Alek,Salamon}
\bea
\mbox{Wolf}= \frac{G}{H} = 
\frac{SO(N+4)}{SO(N) \times SO(4)}.
\label{cosetWolf}
\eea
The group indices are divided into 
\bea 
G \quad \mbox{indices} &:& a, b, c, \cdots=1,2, \cdots, 
\frac{1}{4} (N+4)(N+3),
1^{\ast}, 2^{\ast}, \cdots, \left( \frac{1}{4} (N+4)(N+3)  \right)^{\ast},
\nonu \\
\frac{G}{H} \quad \mbox{indices} &:& \bar{a},\bar{b},\bar{c},\cdots=
1,2, \cdots, 2N, 1^{\ast}, 2^{\ast}, \cdots, 2N^{\ast}.
\label{abnotation}
\eea
The total $4N$ coset indices in (\ref{abnotation}) are divided into 
$2N$ without $\ast$ and $2N$ with $\ast$.
We only consider even dimensional $G=SO(N+4)$. 
That is, $N=4n$ or $N=4n+1$ for integer $n$.
For given $(N+4) \times (N+4)$ matrix, 
we can associate the above $4N$ coset indices
as follows:

\bea
\left(\begin{array}{rrrrr|rrrr}
&&&&& {\ast} & {\ast} & {\ast} & {\ast} \\
&&&&& {\ast} & {\ast} & {\ast} & {\ast} \\
&& &&& \vdots & \vdots & \vdots & \vdots \\
&&&&& {\ast} & {\ast} & {\ast} & {\ast} \\
&&&&& {\ast} & {\ast} & {\ast} & {\ast} \\ \hline 
{\ast} & {\ast} & \cdots & {\ast} & {\ast} &&&& \\
{\ast} & {\ast} & \cdots & {\ast} & {\ast} &&&& \\ 
{\ast} & {\ast} & \cdots & {\ast} & {\ast} &&&& \\ 
{\ast} & {\ast} & \cdots & {\ast} & {\ast} &&&& \\
\end{array}\right)_{(N+4) \times (N+4)}.
\label{Matrix}
\eea
As described in Appendix $A$, for example, 
the generators with $2N$ coset indices have two nonzero 
elements located at the above $N \times 4$ and $4 \times N$ 
off diagonal matrices in (\ref{Matrix})
(the other half generators with $2N$ coset indices 
denoted by $\ast$ can be obtained via the transpose of the first half 
generators).  

As done in the unitary case of \cite{AK1411},
we would like to 
construct the $11$ currents for generic $N$
from the data of $N=4$ case in \cite{AP1410}.
By writing the spin-$\frac{3}{2}$ currents 
with unknown rank-$2$ tensor with the coset indices
as well as $SO({\cal N}=4)$ index and using the defining OPE of
the large ${\cal N}=4$ nonlinear superconformal algebra
with the help of (\ref{opevq}), 
we analyze each pole term in order to extract the above 
$11$ currents explicitly.  

Then we can write down the 
$11$ currents of large $\mathcal N = 4$ nonlinear superconformal 
algebra in terms of 
 $\mathcal N =1$ Kac-Moody currents $V^a(z)$ and $Q^{\bar{b}}(z)$ 
together with the structure constant, the metric (which corresponds to
the one component of above unknown rank $2$ tensor with coset indices) 
and 
the three almost complex structures
 $h^i_{\bar{a} \bar{b}} (i=1,2,3)$ where the index $i$ stands for
$SO(3)$ index.  
The three almost complex structures $(h^1,h^2,h^3)$ are 
 antisymmetric rank-two tensors and 
satisfy the algebra of imaginary quaternions \cite{Saulina}
$
h^{i}_{\bar{a} \bar{c} } \, h^{j \bar{c} }_{\,\,\,\,\,\, \bar{b} }
= 
\ep^{ijk} \, h^{k}_{\bar{a} \bar{b} } -  \delta^{ ij } \, g_{\bar{a} \bar{b}}$. 
The three almost complex structures 
using $4N \times 4N $ matrices
are given by
\footnote{We consider two cases where  
$N=4n$ and $N=4n+1$ 
cases with some integer $n$. For convenience, we 
only represent the almost complex structures for $N=4n$ case. 
In principle, we can write 
down the complex structures for $N=4n+1$ case also.}
\bea
h^1_{\bar{a} \bar{b}} = 
\left(
\begin{array}{cccc}
0 & 1  & 0 & 0 \\
-1 & 0 & 0 & 0 \\
0 & 0 & 0 & 1 \\
0 & 0 & -1 & 0 \\
\end{array}
\right), 
h^2_{\bar{a} \bar{b}} = 
\left(
\begin{array}{cccc}
0 &i  & 0 & 0 \\
-i & 0 & 0 & 0 \\
0 & 0 & 0 & -i \\
0 & 0 & i & 0 \\
\end{array}
\right), 
h^{3}_{\bar{a} \bar{b}}
=
\left(
\begin{array}{cccc}
0 & 0  & i & 0 \\
0 & 0 & 0 & i \\
-i & 0 & 0 & 0 \\
0 & -i & 0 & 0 \\
\end{array}
\right)
\label{himatrix},
\eea
where each entry in (\ref{himatrix}) is  $N \times N$ matrix
and the third almost complex structure can be 
written in terms of  a product of other two: $h^{3}_{\bar{a} \bar{b}}
\equiv h^{1}_{\bar{a} \bar{c}}  \, h^{2 \bar{c} }_{\,\,\,\,\,\, \bar{b}}$.
Note that $h^{0}_{\bar{a} \bar{b}}= g_{\bar{a} \bar{b}}$.

Now we obtain the final results for the $11$ currents as follows:
\bea
\hat{G}^{0}(z) &  = &   \frac{i}{(k+N+2)}  \, Q_{\bar{a}} \, V^{\bar{a}}(z),
\qquad
\hat{G}^{i}(z)  =  \frac{i}{(k+N+2)} 
\, h^{i}_{\bar{a} \bar{b}} \, Q^{\bar{a}} \, V^{\bar{b}}(z),
\nonu \\
\hat{A}_{i}(z) &  = & 
(-1)^{i+1} \frac{1}{4N} \, f^{\bar{a} \bar{b}}_{\,\,\,\,\,\, c} \, h^i_{\bar{a} \bar{b}} \, V^c(z), 
\qquad
\hat{B}_{i}(z)  =  
-\frac{1}{4(k+N+2)} \, h^i_{\bar{a} \bar{b}} \, Q^{\bar{a}} \, Q^{\bar{b}}(z),
\nonu \\
\hat{T}(z)  & = & 
\frac{1}{2(k+N+2)^2} \left[ (k+N+2) \, V_{\bar{a}} \, V^{\bar{a}} 
+k \, Q_{\bar{a}} \, \pa \, Q^{\bar{a}} 
+f_{\bar{a} \bar{b} c} \, Q^{\bar{a}} \, Q^{\bar{b}} \, V^c  \right] (z)
\nonu \\
&- & 
\frac{1}{(k+N+2)} \sum_{i=1}^3 \left( (-1)^{i} \hat{A}_{i} + \hat{B}_{i}  \right)^2 (z),
\label{closedform}
\eea
where the index $i(=1,2,3)$ in the spin-$1$ currents 
stands for $su(2)$ adjoint index respectively. 
The spin-$\frac{3}{2}$ currents 
 $\hat{G}^{\mu}(z)$ with $SO(4)$ index $\mu$ are the four supersymmetry 
generators \footnote{We have the following relations between the 
spin-$\frac{3}{2}$ currents with double index notation where 
$SU(2) \times SU(2)$ symmetry is manifest
and those with a single index notation where $SO(4)$ symmetry is manifest
\label{doubleindex}
\bea
\hat{G}_{11}(z) & = & \frac{1}{\sqrt{2}} (\hat{G}^1- i \hat{G}^2)(z), \qquad
\hat{G}_{12}(z) = -\frac{1}{\sqrt{2}} (\hat{G}^3- i \hat{G}^0)(z),
\nonu \\
\hat{G}_{22}(z) &  = & \frac{1}{\sqrt{2}} (\hat{G}^1+ i \hat{G}^2)(z), \qquad
\hat{G}_{21}(z) = -\frac{1}{\sqrt{2}} (\hat{G}^3+ i \hat{G}^0)(z).
\nonu
\eea
},
the spin-$1$ currents  $\hat{A}_{i}(z)$ and $\hat{B}_{i}(z)$   are 
six spin-$1$ generators of $\hat{su}(2)_{k} \times \hat{su}(2)_{N}$ and 
the current $\hat{T}(z)$ is the spin-$2$ stress energy tensor. 
The extra factors $(-1)^{i+1}$ or $(-1)^i$ in (\ref{closedform})
come from the sign change of the spin-$1$ currents \cite{AK1411}.
Note that the index of the fundamental 
spin-$1$ current has either $a$ or $\bar{a}$
while the index of the fundamental 
spin-$\frac{1}{2}$ current has only the coset index
$\bar{a}$.   

Then the large ${\cal N}=4$ nonlinear superconformal algebra (realized in 
the coset theory (\ref{cosetWolf}))
can be realized by the above $11$ currents and characterized by
the OPE between the spin-$2$ current, the OPEs between the 
spin-$\frac{3}{2}$ currents, the OPEs between the spin-$1$ currents and 
the spin-$\frac{3}{2}$ currents, the OPEs between the spin-$1$ currents 
and the OPEs between the spin-$2$ current and other $10$ currents 
\cite{GS,cqg1989,npb1989,GK}. 

\subsection{ The $16$ lowest higher spin currents}

In \cite{AP1410},  
the explicit results for the following higher spin currents (one spin-$2$ current,
four spin-$\frac{5}{2}$ currents, six spin-$3$ currents, four 
spin-$\frac{7}{2}$ currents and one spin-$4$ current) for $N=4$
were written as the fundamental spin-$1$ and spin-$\frac{1}{2}$ currents 
\bea
\left(2, \frac{5}{2}, \frac{5}{2}, 3 \right)
& : & (T^{(2)}, T_{+}^{(\frac{5}{2})}, T_{-}^{(\frac{5}{2})}, T^{(3)}),
\nonu \\
 \left(\frac{5}{2}, 3, 3, \frac{7}{2} \right) & : &
(U^{(\frac{5}{2})}, U_{+}^{(3)}, U_{-}^{(3)}, U^{(\frac{7}{2})} ), \nonu \\
\left(\frac{5}{2}, 3, 3, \frac{7}{2} \right) & : &
(V^{(\frac{5}{2})}, V^{(3)}_{+}, V^{(3)}_{-}, V^{(\frac{7}{2})}),  \nonu \\
\left(3, \frac{7}{2}, \frac{7}{2}, 4 \right) & : &
 (W^{(3)}, W_{+}^{(\frac{7}{2})}, W_{-}^{(\frac{7}{2})}, W^{(4)}).
\label{lowmultiplet}
\eea
It is very important to obtain the lowest higher spin current
from the experience in \cite{AK1411}.
We would like to determine the above higher spin currents for generic $N$.
The lowest spin in the ${\cal N}=4$ multiplet of 
(\ref{lowmultiplet}) is given by spin-$2$ rather than spin-$1$ 
because there was no higher spin-$1$ 
current satisfying the primary condition and 
the regular conditions when $N=4$ \cite{AP1410}. 
Can we prove this for general $N$?

Let us first try to consider the possibility of the higher spin-$1$ current.
We can use the results in \cite{AK1411} in order to analyze the 
existence of higher 
spin-$1$ current for the orthogonal case. 
The ansatz for the higher spin-1 current for general $N$ is
given by
\bea
T^{(1)} (z) &=&   A_a V^a  (z)
+ B_{\bar{a} \bar{b} } \, Q^{\bar{a}} \, Q^{\bar{b}} (z),
\label{possspin1}
\eea
where  the two coefficients 
$A_a$ and  $B_{\bar{a} \bar{b} }$ are undetermined constants.
The most nontrivial constraint for the higher spin-$1$ current is 
the primary condition that the higher spin-$1$ current should be 
primary  field under the stress energy tensor  $\hat{T}(z)$
as follows:
\bea
T^{(1)} (z) \,  \hat{T} (w)= \frac{1}{(z-w)^2} \, T^{(1)} (w) +\cdots, 
\label{T1primary}
\eea
where we change the order of the operators in the left hand side
compared to the standard expression.
Then the 
primary condition in (\ref{T1primary}) requires the following 
two tensor equations as follows:
\bea
\frac{k}{2(k+N+2)(N+2)} \, A_a \, f_{\bar{b} \bar{c} }^{\,\,\,\,\,\, a} 
 & = &
B_{\bar{b} \bar{c}}, \nonu \\
A_a \, f^{a \bar{b} }_{\,\,\,\,\,\, c} \, f^{c}_{\,\,\,\, \bar{b} d}
-\frac{k}{(N+2)} \, A_a \, 
f^a_{\,\,\,\, \bar{b} \bar{c}} \, f^{\bar{b} \bar{c}}_{\,\,\,\,\,\, d}
& = &  2 (k+N+2) A_d.
\label{primaryT1}
\eea
The second equation of (\ref{primaryT1}) 
is determined by the structure constant 
of $g=so(N+4)$. 
It is not hard to find the structure constant of $so(N+4)$ when $N$ is fixed
and we can test the existence of the solution.
In general, there is no nontrivial $A_a$ satisfying the second condition in 
the orthogonal case. Thus we obtain the trivial solution $A_a=0$.
Then the coefficient $B_{\bar{a} \bar{b}}$ is 
also zero from the first condition in (\ref{primaryT1}).
Thus the above higher spin-$1$  current 
$T^{(1)} (z)$ is identically zero and 
there is no higher spin-1 current (\ref{possspin1}) in the orthogonal case.

\subsubsection{The higher spin currents of spins
$\left(2, \frac{5}{2}, \frac{5}{2}, 3 \right)$}

Let us determine the first ${\cal N}=2$ multiplet in (\ref{lowmultiplet}).

$\bullet$ Construction of the lowest higher spin-$2$ current

The ansatz for the 
higher spin-$2$ current based on $N=4$ case  \cite{AP1410} is 
given by
\bea
T^{(2)}(z)&=&c_1  \, V_{\bar{a}} V^{\bar{a}}(z)+ c_2 \,  \sum_{a':so(N)} 
V_{a'} V^{a'}(z)+ 
c_3 \,  \sum_{a'':so(4)} V_{a''} V^{a''}(z)+
c_4 \, \sum_{i=1}^3 \hat{A}_i \hat{A}_i(z) \nonu \\
& + & c_5 \,
\sum_{i=1}^3 \hat{B}_i \hat{B}_i(z)
+  c_6 \, Q_{\bar{a}} \pa Q^{\bar{a}}(z)
+ c_7 \, \sum_{\mu =0}^3 h^{\mu}_{\bar{a} \bar{b}} h^{\mu}_{\bar{c} \bar{d}}
f^{\bar{a} \bar{c}}_{\,\,\,\,\,\, e} Q^{\bar{b}} Q^{\bar{d}} V^e(z),
\label{T2ansatz}
\eea
where $c_i(N,k)$ are the undetermined coefficient functions. 
Note that for general $N$, we should have 
the different coefficients $c_2$ and $c_3$ in (\ref{T2ansatz})
even though they correspond to the subgroup in the Wolf space. 
Furthermore, it is nontrivial to check the tensorial structure 
in the $c_7$-term. In the construction of the OPE between 
the spin-$\frac{3}{2}$ currents in the large ${\cal N}=4$
linear superconformal algebra, this kind of term occurs in \cite{AK1506}.
The indices in the almost complex structures 
are contracted with the ones in the structure constant and the 
spin-$\frac{1}{2}$ currents.
The index for spin-$1$ current runs over the $so(N+4)$ algebra.  
The higher spin-$2$ current should satisfy the following OPEs
\bea
\hat{T}(z) \, T^{(2)}(w) &=& \frac{1}{(z-w)^2} \, 2T^{(2)}(w)+  
\frac{1}{(z-w)} \, \pa T^{(2)}(w)+ \cdots,
\nonu \\
\phi(z)  \, T^{(2)}(w) &=& + \cdots,
\label{nonliT2condition}
\eea
where $\phi(z)=\hat{A}_i(z), \hat{B}_i(z), {\bf F^a}(z)$ and ${\bf U}(z)$.
Note that by construction, 
the higher spin currents should commute with both 
the four spin-$\frac{1}{2}$ currents ${\bf F^a}(z)$ and the spin-$1$ current
${\bf U}(z)$ 
of the large ${\cal N}=4$ linear superconformal algebra
\footnote{
From the Goddard-Schwimmer formula \cite{GS}, 
the conditions (\ref{nonliT2condition}) are equivalent to 
the conditions for the higher spin-$2$ current in the linear version
\cite{BCG1404},
\bea
{\bf T}(z) \, T^{(2)}(w) &=& \frac{1}{(z-w)^2} \,
2T^{(2)}(w)+  \frac{1}{(z-w)} \, \pa T^{(2)}(w)+ \cdots,
\nonu \\
 \Phi(z)  \, T^{(2)}(w) &=& + \cdots,
\nonu
\eea
where the current $\Phi(z)$ stands for 
the spin-$1$  currents ${\bf A}_i(z)$ and ${\bf B}_i(z)$, 
the spin-$\frac{1}{2}$ currents ${\bf F^{a}}(z)$, 
the spin-$1$ current ${\bf U}(z)$
of the large ${\cal N}=4$ linear superconformal algebra (where 
the current ${\bf T}(z)$ is the stress energy tensor). 
}.  
The requirement  (\ref{nonliT2condition}) determines every coefficient functions $c_i$ except
the overall factor for $N=4,5,8,9$ cases.
From those solutions, we obtain the general solution for the coefficients 
$c_i(N,k)$ as follows:
\bea
c_1  & = & -\frac{(2 k^2 N+k^2+4 k N^2+6 k N+2 k+11
   N^2-2 N-24)}{2 (k-1) N (k+N+2)^2}, 
   \nonu \\
   c_2 & = & \frac{6 (2 k N+3 k+3 N+4)}{(k-1) N
   (k+N+2)^2},
\qquad
c_3  =  \frac{3 (k+N-2) (2 k N+3 k+3 N+4)}{2
   (k-1) (k+2) (k+N+2)^2}, 
   \nonu \\
c_4 & = & \frac{2 (N+2) (2 k+N)}{(k+2) (k+N+2)^2},
\qquad
c_5  =  \frac{2 k (2 k+N)}{N (k+N+2)^2},
\nonu \\
c_6 & = & \frac{k (N+2) (2 k+N)}{N (k+N+2)^3},
\qquad
c_7  =  \frac{(N+2) (2 k+N)}{4 N (k+N+2)^3}.
\label{t2coeff}
\eea
Note that the coefficients 
$c_2$ and $c_3$ are different in general but it is easy to see that 
they are the same for $N=4$.

The appropriate  choice for the overall factor of the higher spin-$2$
current comes from the following OPE   
\bea
T^{(2)}(z) \, T^{(2)}(w)&=& \frac{1}{(z-w)^4} \, e_1
\label{t2t2} 
\\
& + & 
\frac{1}{(z-w)^2}
\Bigg[ 
e_2 \,
T^{(2)} + e_3 \left( 
\hat{T} + \frac{1}{(k+2)}  \, (\hat{A}_{3} \hat{A}_{3} 
+ 
\hat{A}_{+} \hat{A}_{-} + i \, \pa \hat{A}_{3})
\right.
\nonu \\
& + & \left.
\frac{1}{(N+2)}  ( \hat{B}_{3} \hat{B}_{3} 
+    \hat{B}_{+} \hat{B}_{-} 
 +  i  \, 
\pa \hat{B}_{3}) \right) \Bigg](w) + 
\frac{1}{(z-w)} \, \frac{1}{2}\pa (\mbox{pole-2})(w) + \cdots,
\nonu
\eea
where the central term or structure constants in (\ref{t2t2})
$e_i$ are given by
\bea
e_1&=&\frac{3 k (2 k+N) (2 k N+3 k+3 N+4) \left(2 k^2 N+k^2+4 k N^2+6
   k N+2 k+11 N^2-2 N-24\right)}{(k-1) (k+2) N (k+N+2)^3},
   \nonu \\
   e_2&=&\frac{2 \left(2 k^2 N+7 k^2-2 k N^2-6 k N-10 k-13 N^2-2
   N+24\right)}{(k-1) N (k+N+2)},
   \label{ecoeff} \\
   e_3&=&\frac{4 (N+2) (2 k+N) \left(2 k^2 N+k^2+4 k N^2+6 k N+2 k+11
   N^2-2 N-24\right)}{(k-1) N^2 (k+N+2)^2}. 
\nonu
\eea 
Because the maximum power of $k$ in 
the polynomial appearing in the numerators of the coefficients in 
(\ref{ecoeff})
is given by $2$, we could determine all the coefficients completely
with the data of $N=4,5,8,9$ cases.
We will see the three-point function with this choice of overall factor
in the higher spin-$2$ current later.

$\bullet$ Construction of the other higher spin currents

Now let us determine the other three higher spin currents 
in the first ${\cal N}=2$ multiplet in (\ref{lowmultiplet}).
As done in $N=4$ case in \cite{AP1410}, 
we can calculate the OPE between 
$\hat{G}_{21}(z)$ and $T^{(2)}$(w)
where the explicit forms are given in (\ref{closedform}) with the 
footnote \ref{doubleindex} and (\ref{T2ansatz}) with (\ref{t2coeff}).
Again the fundamental OPEs in (\ref{opevq}) are used heavily.
Three almost complex structures are given in (\ref{himatrix})
and the metric is also related to the following relation 
$h^{0}_{\bar{a} \bar{b}}= g_{\bar{a} \bar{b}}$.
Then it turns out that the following nontrivial first-order pole
is given by
\bea
\left(
\begin{array}{c}
\hat{G}_{21} \\
\hat{G}_{12}  \\
\end{array}
\right)(z) \, T^{(2)}(w)  & = &
\frac{1}{(z-w)}
T_{\pm}^{(\frac{5}{2})}(w) +
\cdots.
\label{g2112t2}
\eea
For $N=4,5,8,9$ cases, 
we have the explicit forms for the first-order pole in terms of 
the fundamental spin-$1$ and spin-$\frac{1}{2}$ currents. 
Even for generic $N$, 
we can express the explicit results for the higher spin-$\frac{5}{2}$
currents $T_{\pm}^{(\frac{5}{2})}(w)$ 
but we do not present them in this paper. 

Because the higher spin-$\frac{5}{2}$ current 
 $T_{-}^{(\frac{5}{2})}(w)$ is determined from the OPE 
(\ref{g2112t2}), let us calculate the OPE between 
$\hat{G}_{21}(z)$ and 
this higher spin-$\frac{5}{2}$ current 
 $T_{-}^{(\frac{5}{2})}(w)$ explicitly.
Then we obtain the following result
\bea
\hat{G}_{21}(z) \, T_{-}^{(\frac{5}{2})}(w) & = &
\frac{1}{(z-w)^2} \, 4 T^{(2)} (w)
 +  \frac{1}{(z-w)}  \left[
 \frac{1}{4} 
\pa (\mbox{pole-2})
+T^{(3)} \right](w) +  \cdots.
\label{g21t5half-other}
\eea
There are no quasiprimary fields in the first-order pole 
in (\ref{g21t5half-other}).
The numerical factor $\frac{1}{4}$ in the first term of the 
first-order pole is fixed 
by the spins of the two  currents in the left hand side of the above OPE
and the spin of the higher spin-$2$ current 
living in the second-order pole.
Then the higher spin-$3$ current $T^{(3)}(w)$ can be obtained from the 
explicit first-order pole from the OPE 
$\hat{G}_{21}(z) \, T_{-}^{(\frac{5}{2})}(w)$ and subtract 
the derivative of the higher spin-$2$ current $\pa T^{(2)}(w)$,
along the line of \cite{BBSS1,BBSS2,BS}. 
As before, for several $N$ case, the explicit results are found. 

Therefore, the first ${\cal N}=2$ multiplet in (\ref{lowmultiplet})
is determined for generic $N$ completely (and implicitly).

\subsubsection{The higher spin currents of spins
$\left(\frac{5}{2}, 3, 3, \frac{7}{2} \right)$}

Let us determine the second ${\cal N}=2$ multiplet in (\ref{lowmultiplet}).
As done in (\ref{g2112t2}), we 
calculate the OPE 
$\hat{G}_{11}(z) \, T^{(2)}(w)$.
The spin-$\frac{3}{2}$ current 
$\hat{G}_{11}(z)$ is given by 
 (\ref{closedform}) with the 
footnote \ref{doubleindex}.
The similar OPE
$\hat{G}_{22}(z) \, T^{(2)}(w)$
can be used for other higher spin-$\frac{5}{2}$ current
later.
The lowest higher spin-$\frac{5}{2}$ current $U^{(\frac{5}{2})} (w)$ 
of this ${\cal N}=2$ multiplet can be obtained 
from the first-order pole of the following OPE
\bea
\hat{G}_{11}(z) \, T^{(2)}(w)  & = &
\frac{1}{(z-w)}U^{(\frac{5}{2})} (w) +
\cdots.
\label{g11t2other}
\eea

Furthermore, from the above higher spin-$\frac{5}{2}$ current 
appearing in (\ref{g11t2other}) found for generic $N$, we 
can calculate 
the OPE 
between the spin-$\frac{3}{2}$ currents and 
this higher spin-$\frac{5}{2}$ current explicitly.
\bea
\left(
\begin{array}{c}
\hat{G}_{21} \\
\hat{G}_{12}  \\
\end{array}
\right)
(z) \, U^{(\frac{5}{2})}(w)  & = &
\frac{1}{(z-w)}  U^{(3)}_{\pm} (w) +
\cdots.
\label{g2112u5half}
\eea
There are no derivative terms or quasiprimary fields
in the first-order pole of (\ref{g2112u5half}).



Because the higher spin-$3$ current 
 $U_{-}^{(3)}(w)$ is obtained for generic $N$ from the OPE 
(\ref{g2112u5half}), let us calculate the OPE between 
$\hat{G}_{21}(z)$ and 
this higher spin-$3$ current 
 $U_{-}^{(3)}(w)$ explicitly.
\bea
\hat{G}_{21}(z) \, U_{-}^{(3)}(w) & = &
\frac{1}{(z-w)^2} \Bigg[\frac{2(2N+3+3k)}{(N+2+k)} U^{(\frac{5}{2})}\Bigg](w)
\nonu \\
& + & \frac{1}{(z-w)}  \left[ \frac{1}{5} 
\pa (\mbox{pole-2})
+U^{(\frac{7}{2})}
 \right](w)+ \cdots.
\label{g21u3-other}
\eea
It is not difficult to 
obtain the $N$-dependence on the structure constant in the 
second-order pole of (\ref{g21u3-other}).
We confirm this for $N=4,5,8,9$ as before.
The numerical factor $\frac{1}{5}$
appearing in the first term of the first-order pole in (\ref{g21u3-other})
can be determined using the previous argument.
There are no quasiprimary fields in the first-order pole 
in (\ref{g21u3-other}).
Then the higher spin-$\frac{7}{2}$ current $U^{(\frac{7}{2})}(w)$ 
can be obtained from the 
explicit first-order pole from the OPE 
$\hat{G}_{21}(z) \, U_{-}^{(3)}(w)$ and subtract 
the derivative of the higher spin-$\frac{5}{2}$ 
current $\frac{2(2N+3+3k)}{5(N+2+k)} \, \pa U^{(\frac{5}{2})}(w)$. 

Therefore, the second ${\cal N}=2$ multiplet in (\ref{lowmultiplet})
is found for generic $N$ implicitly.

\subsubsection{The higher spin currents of spins
$\left(\frac{5}{2}, 3, 3, \frac{7}{2} \right)$}

Let us determine the third ${\cal N}=2$ multiplet in (\ref{lowmultiplet}).
As done in previous subsection, we 
calculate the OPE 
$\hat{G}_{22}(z) \, T^{(2)}(w)$.
The spin-$\frac{3}{2}$ current 
$\hat{G}_{22}(z)$ is given by 
 (\ref{closedform}) with the 
footnote \ref{doubleindex}.
The lowest higher spin-$\frac{5}{2}$ current $V^{(\frac{5}{2})} (w)$ 
of this ${\cal N}=2$ multiplet can be obtained 
from the first-order pole of the following OPE
\bea
\hat{G}_{22}(z) \, T^{(2)}(w)  & = &
\frac{1}{(z-w)}  V^{(\frac{5}{2})} (w) +
\cdots.
\label{g22t2other}
\eea
We can combine the two OPEs (\ref{g11t2other}) and (\ref{g22t2other}). 

Furthermore, with the help of above higher spin-$\frac{5}{2}$ current 
appearing in (\ref{g22t2other}) found for generic $N$, we 
can calculate the following OPE
\bea
\left(
\begin{array}{c}
\hat{G}_{21} \\
\hat{G}_{12}  \\
\end{array}
\right)
(z) \, V^{(\frac{5}{2})}(w)  & = &
\frac{1}{(z-w)} V^{(3)}_{\pm}(w) +
\cdots.
\label{g2112v5half}
\eea
In this case also, we can combine 
the two OPEs (\ref{g2112u5half}) and (\ref{g2112v5half}).


From  the higher spin-$3$ current 
 $V_{-}^{(3)}(w)$ obtained for generic $N$ from the OPE 
(\ref{g2112v5half}), the OPE between 
$\hat{G}_{21}(z)$ and 
this higher spin-$3$ current 
 $V_{-}^{(3)}(w)$  can be obtained explicitly as follows:
\bea
\hat{G}_{21}(z) \, V_{-}^{(3)}(w) & = &
\frac{1}{(z-w)^2} \Bigg[\frac{2(3N+3+2k)}{(N+2+k)} V^{(\frac{5}{2})}\Bigg](w)
\nonu \\
& + & \frac{1}{(z-w)}  \left[ \frac{1}{5} 
\pa (\mbox{pole-2})
+V^{(\frac{7}{2})}
 \right](w)+ \cdots.
\label{g21v3-other}
\eea
The $N$-dependence on the structure constant in the 
second-order pole of (\ref{g21v3-other})
can be  confirmed for $N=4,5,8,9$ as before.
This structure constant and the corresponding one in (\ref{g21u3-other}) 
have the $N \leftrightarrow k$ symmetry. 
Note the numerical factor $\frac{1}{5}$
appearing in the first term of the first-order pole.
There are no quasiprimary fields in the first-order pole.
Then the higher spin-$\frac{7}{2}$ current $V^{(\frac{7}{2})}(w)$ 
can be obtained from the  
explicit  first-order pole from the OPE 
$\hat{G}_{21}(z) \, V_{-}^{(3)}(w)$ and subtract 
the derivative of the higher spin-$\frac{5}{2}$ 
current $\frac{2(3N+3+2k)}{5(N+2+k)} \, \pa V^{(\frac{5}{2})}(w)$. 

Therefore, the third ${\cal N}=2$ multiplet in (\ref{lowmultiplet})
is found from (\ref{g22t2other}), (\ref{g2112v5half}) and 
(\ref{g21v3-other}) for generic $N$ implicitly.

\subsubsection{The higher spin currents of spins
$\left(3, \frac{7}{2}, \frac{7}{2}, 4 \right)$}

Let us determine the fourth ${\cal N}=2$ multiplet in (\ref{lowmultiplet}).
We 
calculate the OPE 
$\hat{G}_{22}(z) \, U^{(\frac{5}{2})}(w)$.
The spin-$\frac{3}{2}$ current 
$\hat{G}_{22}(z)$ is given by 
 (\ref{closedform}) with the 
footnote \ref{doubleindex} and 
the higher spin-$\frac{5}{2}$ current 
$U^{(\frac{5}{2})}(w)$ is given by 
 (\ref{g11t2other}).
The lowest higher spin-$3$ current $W^{(3)} (w)$ 
of this ${\cal N}=2$ multiplet can be obtained 
from the first-order pole of the following OPE
\bea
\hat{G}_{22}(z) \, U^{(\frac{5}{2})}(w) & = &
\frac{1}{(z-w)^2}\, 4 T^{(2)}(w)
+  \frac{1}{(z-w)}  \left[ \frac{1}{4} 
\pa (\mbox{pole-2})
+W^{(3)}
 \right](w)+ \cdots.
\label{g22u5halfother}
\eea
There are no quasiprimary fields in the first-order pole 
in (\ref{g22u5halfother}).
The numerical factor $\frac{1}{4}$ in the first term of the 
first-order pole is fixed 
by the previous description.
Then the higher spin-$3$ current $W^{(3)}(w)$ can be obtained from the 
explicit first-order pole from the OPE 
$\hat{G}_{22}(z) \, U^{(\frac{5}{2})}(w)$ and subtract 
the derivative of the higher spin-$2$ current $\pa T^{(2)}(w)$. 
As before, for several $N$ case, the explicit results are found. 

From  the higher spin-$3$ current 
 $W^{(3)}(w)$ obtained for generic $N$ from the OPE 
(\ref{g22u5halfother}), the OPE between 
$\hat{G}_{21}(z)$ ($\hat{G}_{12}(z)$) and 
this higher spin-$3$ current 
 $W^{(3)}(w)$  can be obtained explicitly as follows:
\bea
\left(
\begin{array}{c}
\hat{G}_{21} \\
\hat{G}_{12} \nonu \\
\end{array}
\right)
(z) \,W^{(3)}(w) & = &
\pm \frac{1}{(z-w)^2} \Bigg[
\frac{(N-k)}{(N+2+k)} T_{\pm}^{(\frac{5}{2})} \Bigg](w)
\nonu \\
& + & \frac{1}{(z-w)}  \left[ \frac{1}{5}  
\pa (\mbox{pole-2})
+W_{\pm}^{(\frac{7}{2})}
 \right](w)+ \cdots.
\label{g2112w3}
\eea

From  the higher spin-$\frac{7}{2}$ current 
 $W_{-}^{(\frac{7}{2})}(w)$ obtained for generic $N$ from the OPE 
(\ref{g2112w3}), the OPE between 
$\hat{G}_{21}(z)$ and 
this higher spin-$\frac{7}{2}$ current 
 $W_{-}^{(\frac{7}{2})}(w)$  can be obtained explicitly as follows:
\bea
\hat{G}_{21}(z) \,W_{-}^{(\frac{7}{2})}(w) & = &
\frac{1}{(z-w)^3} \left[ -\frac{48(-N+k)}{5(N+2+k)}T^{(2)}
 \right](w)
\nonu \\
& + & \frac{1}{(z-w)^2}  \left[ -\frac{6(-N+k)}{5(N+2+k)}T^{(3)}+\frac{2(3N+4+3k)}{(N+2+k)}W^{(3)}
 \right. \nonu \\
&+ & \left.
\frac{16i}{(N+2+k)}(\hat{A_3}
-\hat{B_3})T^{(2)}
 \right](w)
 \nonu \\
 &+&\frac{1}{(z-w)} \left[ \frac{1}{6}
\pa (\mbox{pole-2})
-\frac{144(-N+k)}{((59N+88)+(30N+59)k)}
\left(\hat{T}T^{(2)}-\frac{3}{10} \pa^2 T^{(2)} \right)
 \right. \nonu \\
& + & \left.
 W^{(4)}
 \right](w)
 + \cdots.
\label{g21w7half}
\eea
The various $N$-dependent structure constants appearing in 
(\ref{g21w7half}) can be confirmed for $N=4,5,8,9$ as before.
In particular, the nonlinear terms appear in the second- and first-order
poles. 
In the first-order pole, the quasiprimary field of spin $4$ appears.
Then the higher spin-$4$ current $W^{(4)}(w)$ 
can be obtained from the  
explicit first-order pole from the OPE 
$\hat{G}_{21}(z) \, W_{-}^{(\frac{7}{2})}(w)$ and subtract 
both the derivative of the second-order pole with $\frac{1}{6}$ 
and the above quasiprimary field-term. 

Therefore, the fourth ${\cal N}=2$ multiplet in (\ref{lowmultiplet})
is found from (\ref{g22u5halfother}), (\ref{g2112w3}) and 
(\ref{g21w7half}) for generic $N$ implicitly.

\subsection{
The $16$ second lowest higher spin currents  }

Let us denote the next 16 higher spin currents by its spin contents 
as follows:
\bea
\left(3, \frac{7}{2}, \frac{7}{2}, 4 \right) &:& (P^{(3)},
P_{+}^{(\frac{7}{2})}, P_{-}^{(\frac{7}{2})}, P^{(4)}), \qquad
  \left(\frac{7}{2}, 4, 4, \frac{9}{2} \right):
 (Q^{(\frac{7}{2})},
Q_{+}^{(4)}, Q_{-}^{(4)}, Q^{(\frac{9}{2})}), \nonu \\
\left(\frac{7}{2}, 4, 4, \frac{9}{2} \right) & : &
 (R^{(\frac{7}{2})},
R_{+}^{(4)}, R_{-}^{(4)}, R^{(\frac{9}{2})}),  \qquad
 \left(4, \frac{9}{2}, \frac{9}{2}, 5 \right):
 (S^{(4)},
S_{+}^{(\frac{9}{2})}, S_{-}^{(\frac{9}{2})}, S^{(5)}).
\label{next16}
\eea
We expect that these higher spin currents in (\ref{next16}) will appear 
when we calculate the various OPEs between the lowest $16$ higher 
spin currents in (\ref{lowmultiplet}).
In this subsection we would like to construct only four 
higher spin-$\frac{7}{2}$ currents only.
The remaining ones will appear in Appendices $B$ and $C$.

\subsubsection{The four higher spin-$\frac{7}{2}$ currents}

From the experience of the unitary case \cite{Ahn1408},
we have the explicit OPE $T^{(2)}(z) \, U^{(\frac{5}{2})}(w)$
(and  $T^{(2)}(z) \, V^{(\frac{5}{2})}(w)$) where the higher spin currents 
belong to the lowest ${\cal N}=4$ multiplet in the unitary coset theory.  
The new higher spin-$\frac{7}{2}$ currents occur in the first-order pole.
This implies that 
we expect that we try to calculate the same OPE in the orthogonal case.
It turns out that
\bea
T^{(2)}(z)\,
\left(
\begin{array}{c}
U^{(\frac{5}{2})} \\
V^{(\frac{5}{2})}
\end{array} \right)(w)
& = & \frac{1}{(z-w)^3}
 c_1 \,
 \left(
 \begin{array}{c}
  \hat{G}_{11}\\
  \hat{G}_{22}
  \end{array} \right)(w)
 \nonu \\
&+&\frac{1}{(z-w)^2} \Bigg[\frac{1}{3}\pa (\mbox{pole-3})
+ c_2 \, \left(
 \begin{array}{c}
\hat{G}_{11}\\
-\hat{G}_{22}
  \end{array} \right)\hat{A}_{3}
+ c_3\, \left(
 \begin{array}{c}
-\hat{G}_{21} \\
\hat{G}_{12}
  \end{array} \right) \hat{A}_{\pm}
\nonu \\
&+& c_4 \, \left(
 \begin{array}{c}
\hat{G}_{11}\\
-\hat{G}_{22}
  \end{array} \right)\hat{B}_{3}
+ c_5 \, \left(
 \begin{array}{c}
\hat{G}_{12}\\
-\hat{G}_{21}
  \end{array} \right) \hat{B}_{\mp}
+ c_6 \, \pa \left(
 \begin{array}{c}
\hat{G}_{11}\\
\hat{G}_{22}
  \end{array} \right)
\nonu \\
& + & c_7 \, \left(
 \begin{array}{c}
U^{(\frac{5}{2})}\\
V^{(\frac{5}{2})}
  \end{array} \right) \Bigg](w)
\nonu \\
&+&\frac{1}{(z-w)}\Bigg[
\frac{2}{5} \pa (\mbox{pole-2})-\frac{1}{20} \pa^2 (\mbox{pole-3})
+ c_8 \,
\left(
\begin{array}{c}
\hat{T}\hat{G}_{11} - \frac{3}{8}\pa^2 \hat{G}_{11}\\
\hat{T}\hat{G}_{22} - \frac{3}{8}\pa^2 \hat{G}_{22}
\end{array} \right)
\nonu \\
& + & \left(
\begin{array}{c}
Q^{(\frac{7}{2})}\\
R^{(\frac{7}{2})}
\end{array} \right)
\Bigg](w)
+ \cdots.
\label{t2u5half}
\eea
We have the explicit structure constants $c_1$-$c_8$ for $N=4$ case 
appearing in (\ref{t2u5half}) 
but we do not present them here.
Note that the higher spin-$\frac{5}{2}$ currents (which appear in the
left hand side of this OPE) arise at the 
second-order pole. The quasiprimary fields of spin-$\frac{7}{2}$
appear in the first-order pole.
We can rearrange the two derivative terms in the first-order pole 
in order to express them in standard way where the first derivative term
is written usually without the descendant term from the third-order 
pole \cite{Ahn1211,Ahn1305}. 

Similarly, we can calculate the following OPE
\bea
T^{(2)}(z)\,
T^{(\frac{5}{2})}_{\pm}(w)
& = & \frac{1}{(z-w)^3}
 c_1 \, 
 \left(
 \begin{array}{c}
  \hat{G}_{21}\\
  \hat{G}_{12}
  \end{array} \right)(w)
 \nonu \\
&+&\frac{1}{(z-w)^2} \Bigg[\frac{1}{3}\pa (\mbox{pole-3})
+ c_2 \, \left(
 \begin{array}{c}
-\hat{G}_{21}\\
\hat{G}_{12}
  \end{array} \right)\hat{A}_{3}
+ c_3 \, \left(
 \begin{array}{c}
\hat{G}_{11}\\
-\hat{G}_{22}
  \end{array} \right) \hat{A}_{\mp}
\nonu \\
&+& c_4 \, \left(
 \begin{array}{c}
-\hat{G}_{21}\\
\hat{G}_{12}
  \end{array} \right)\hat{B}_{3}
+ c_5 \, \left(
 \begin{array}{c}
-\hat{G}_{22}\\
\hat{G}_{11}
  \end{array} \right) \hat{B}_{\mp}
+ c_6 \, \pa \left(
 \begin{array}{c}
\hat{G}_{21}\\
\hat{G}_{12}
  \end{array} \right)
+  c_7 \,
T^{(\frac{5}{2})}_{\pm}
\Bigg](w)
\nonu \\
&+&\frac{1}{(z-w)}\Bigg[
\frac{2}{5} \pa (\mbox{pole-2})-\frac{1}{20} \pa^2 (\mbox{pole-3})
+ c_8 \, 
\left(
\begin{array}{c}
\hat{T}\hat{G}_{21} - \frac{3}{8}\pa^2 \hat{G}_{21} \\
\hat{T}\hat{G}_{12} - \frac{3}{8}\pa^2 \hat{G}_{12}
\end{array} \right)
\nonu \\
& + & 
P^{(\frac{7}{2})}_{\pm}
\Bigg](w)
+ \cdots.
\label{t2tp5half}
\eea
In (\ref{t2tp5half}), the structure constants for $N=4$ are known
and the composite fields appearing in the right hand side look similar to
the ones in (\ref{t2u5half}).

Therefore, the four higher spin-$\frac{7}{2}$ currents in (\ref{next16})
are determined implicitly. Once the structure constants are written in terms of 
$N$ and $k$, then we can obtain them from the first-order poles explicitly.

\subsubsection{The remaining higher spin currents}

If we would like to construct the remaining $12$ higher spin currents
in (\ref{next16}), then we should calculate them with the help of 
the spin-$\frac{3}{2}$ currents and the known higher spin currents.
In Appendix $B$, we present the defining OPE equations for these higher spin currents and in Appendix $C$, we present how they appear  in the explicit OPEs 
between the $16$ lowest higher spin currents.

\section{ Three-point functions in the extension of 
the large ${\cal N}=4$ nonlinear superconformal algebra}

This section describes the three-point functions 
with scalars for the current of  
spin $s=2$ and the higher spin currents of spins $s=2,3,4$ 
explained in previous section.  
The large $N$ 't Hooft limit is defined by \cite{GG1305}
\bea
N,k \rightarrow \infty, \quad \lambda 
\equiv \frac{(N+1)}{(N+k+2)} \quad \mbox{fixed}.
\label{limit}
\eea
As described in the introduction, 
there are two simplest states  $|(v;0)>$ and $|(0;v)>$
we describe. The two levels of the $\hat{su}(2) \times \hat{su}(2)$
are given by $k$ and $N$ respectively.

\subsection{ Eigenvalue equations for the spin-$2$ current}

Let us focus on the eigenvalue equations for the stress energy tensor 
(\ref{closedform}) acting on the above two states.
We will see that the eigenvalues  lead to the ones in the unitary case 
\cite{AK1506}.

\subsubsection{ Eigenvalue equation for the spin-$2$ current acting on the 
state $|(v;0)>$}

The terms containing the fermionic spin-$\frac{1}{2}$ currents
$Q^{a }(z)$ do not contribute to the eigenvalue equation 
when we calculate the zero mode eigenvalues for the 
bosonic spin-$s$ current 
$J^{(s)}(z)$  acting on the state  $|(v;0)>$.
The zero mode of the spin-$1$ current $V_0^a$ 
satisfies the commutation relation 
of the underlying finite dimensional Lie algebra $g=so(N+4)$.
For the first state $|(v;0)>$, the generator $T_{a^{\ast}}$ corresponds to 
the zero mode $V_0^a$ as follows (See also \cite{GH1101}):
\bea
 V^a_0 |(v;0)>  =  T_{a^{\ast}} |(v;0)>. 
\label{generator}
\eea
Then the eigenvalues are encoded in the  last 
$4 \times 4$ diagonal matrix.

For example, 
we can calculate the conformal dimension of $|(v;0)>$ when $N=4$.
The explicit form for the stress energy tensor is given by
(\ref{closedform}). 
The only $Q^a(z)$-independent terms are given by 
the first term and the $\hat{A}_{i} \hat{A}_{i} (z)$-dependent term.
Then the eigenvalue equation for the zero mode of
the spin-$2$ current acting on the state $|(v;0)>$ leads to
\bea
\hat{T}_0 |(v;0)>  &\sim& 
\left[ \frac{1}{2(k+6)}   \, V_{\bar{a}} \, V^{\bar{a}}  
- \frac{1}{(k+6)} \sum_{i=1}^3  \hat{A}_{i} \hat{A}_{i} \right]_0   |(v;0)>
\nonu \\
&=&
\left[ 
\frac{1}{2(k+6)} \left(  \sum_{a=1}^{8 } T_{a^\ast} T_{a} 
+\sum_{a=1}^{8 } T_{a} T_{a^\ast} \right) \right] |(v;0)>
+ \frac{1}{(k+6)} l^{+} (l^{+}+1)  |(v;0)>
\nonu \\
&=&
\frac{4}{2(k+6)}  |(v;0)> + \frac{1}{(k+6)} \frac{3}{4}  |(v;0)>
= \Bigg[\frac{11}{4 (k+6)} \Bigg] |(v;0)>,
\label{hvzero}
\eea
where $\sim$ in the first line of (\ref{hvzero}) 
means that we ignore the terms including $Q^a(z)$.
In the second line, the summation over the coset indices 
$\bar{a} = 1, 2, \cdots, 8, 1^{\ast}, 2^{\ast}, \cdots, 8^{\ast}$
is taken explicitly and we used the condition (\ref{generator}).
Moreover the eigenvalue equation for the zero mode of the 
quadratic 
spin-$1$ currents is used where $l^{+}$ is the spin of the affine $\hat{su} (2)$ 
algebra.   
In the third line, we take 4 from the last  $4 \times 4$ diagonal matrix 
\footnote{
\label{footfoot}
The highest weight states of  the large $\mathcal N = 4$  (non)linear 
superconformal 
algebra can be characterized by
the conformal dimension $h$ and  two (iso)spins $l^{\pm}$ of 
$\hat{su}(2) \oplus \hat{su}(2)$ \cite{npb1989}
\bea
\left[ -\sum_{i=1}^3  \hat{A}_{i} \hat{A}_{i} \right]_0   |\mbox{hws}> 
=l^{+}(l^{+}+1) \, |\mbox{hws}>, 
\qquad
\left[ -\sum_{i=1}^3  \hat{B}_{i} \hat{B}_{i} \right]_0   |\mbox{hws}> 
=l^{-}(l^{-}+1) \, |\mbox{hws}>. 
\label{quadcasimir}
\eea
For example, in $g=so(8)$, the expressions (\ref{closedform})
imply that 
  \bea
 \left[ -\sum_{i=1}^3  \hat{A}_{i} \hat{A}_{i} \right]_0 |(v;\star)> =
 \left( \begin{array}{cccc}
 0 & 0 \\
 0 & \frac{3}{4} \end{array}
 \right)  |(v;\star)>, \quad
 \left[ -\sum_{i=1}^3  \hat{B}_{i} \hat{B}_{i} \right](z) \, 
Q^{\bar{A}^{\ast}}(w)|_{\frac{1}{(z-w)^2}}
 =\frac{3}{4} Q^{\bar{A}^{\ast}}(w),
 \nonu
\eea
where each element in matrix is $4 \times 4$ block matrix 
and the representation 
$\star=0$ (trivial representation) or $v$ (vector representation) of
$so(4)$.
We can see $l^{+}(v;0)=\frac{1}{2}$ (from the eigenvalues $\frac{3}{4}$ in matrix),  
$l^{+}(v;v)=0$ (from the first $0$ in matrix) and $l^{-}(0;v)=\frac{1}{2}$ (from 
the coefficient of the second order pole $\frac{3}{4}$).
Then the state $|(v;0)>$  has $l^{+}=\frac{1}{2}$, $l^{-}=0$, the state
$|(0;v)>$ has $l^{+}=0$, $l^{-}=\frac{1}{2}$
and the state $|(v;v)>$  has $l^{\pm}=0$. 
The eigenvalues for $l^{-}$ will be explained in next subsection.
Note the $(-1)$ sign in the left hand side of 
(\ref{quadcasimir}) comes from the anti-hermitian property
 \cite{npb1989,STVplb}.}.   

From the similar calculations for $N=5, 8, 9$, 
we can obtain the $N$-dependence of the eigenvalue in (\ref{hvzero})
as follows
\footnote{We can obtain the conformal dimension of light state from 
similar calculation
\bea
\hat{T}_0 |(v;v)> =\left[ \frac{2}{(k+N+2)}\right] |(v;v)>
\longrightarrow  \frac{2 \lambda}{(N+1)} |(v;v)>.
\nonu
\eea
As we expected, 
the conformal dimension of light state $|(v;v)>$ vanishes in the 
large $N$ 't Hooft limit.}:
\bea
\hat{T}_0 |(v;0)>  &=& \left[ \frac{(2N+3)}{4(k+N+2)}  \right] |(v;0)>,
\label{hvzero2}
\eea
where  the eigenvalue is the same value as 
the eigenvalue $h(f;0)$ given in unitary case \cite{GG1305}.
We can also check that 
this leads to the following 
reduced eigenvalue 
equation  $ T_0 |(v;0)>  =  \frac{\la}{2}  |(v;0)> $
under the large $N$ 't Hooft limit (\ref{limit}). 

\subsubsection{ Eigenvalue equation for the spin-$2$ current acting on the 
state $|(0;v)>$}

%

When we calculate the eigenvalue equations for the second state  $|(0;v)>$,  
we use the 
field representation which is similar to  \cite{GG1305,npb1989}
\bea
 |(0;v)>=
\frac{1}{\sqrt{k+N+2}} \, Q_{-\frac{1}{2}}^{\bar{a}}|0>,
\qquad \bar{a} = 1, 2, \cdots, 2N, 1^{\ast}, 2^{\ast}, \cdots, (2N)^{\ast}.
\label{state0v}
\eea
We need only the coefficient of highest-order pole $\frac{1}{(z-w)^s}$
in the OPE between the higher spin current $J^{(s)}(z)$ and 
the spin-$\frac{1}{2}$ current $Q^{\bar{a}}(w)$.
The lower singular terms do not contribute to the zero mode eigenvalue
equations.
Let us denote the highest-order pole as follows \cite{AKP,MZ1211}:
\bea 
J^{(s)}(z) \,
Q^{\bar{a}}(w) \Bigg|_\frac{1}{(z-w)^s}=  j(s) \, Q^{\bar{a}}(w),
\label{opejq}
\eea
where $j(s)$ stands for the corresponding coefficient of the 
highest order pole.
Then we obtain the following eigenvalue equation for the zero mode 
of the spin-$s$ current together with (\ref{state0v}) and (\ref{opejq})
\bea 
J_0^{(s)} |(0;v)> = j(s) |(0;v)>,
\label{j0f}
\eea
where the explicit relation between the 
current and its mode is given by
$J^{(s)}(z)=\sum_{n=-\infty}^\infty \frac{J_n^{(s)}}{z^{n+s}}$.
Therefore, in order to determine the above 
eigenvalue $j(s)$, one should calculate the explicit OPEs between 
the corresponding (higher spin) currents and the spin-$\frac{1}{2}$
current and read off the highest-order pole.

Let us consider the eigenvalue equation for the spin-$2$
current acting on the above state.
Since the OPE between 
the spin-$1$ current $V^a (z)$ and the spin-$\frac{1}{2}$ current 
$Q^{\bar{b}}(w)$ is regular, 
the terms containing $V^a (z)$  
do not contribute to the highest-order pole.
Therefore, the relevant terms in the spin-$2$ current $\hat{T}(z)$
are given by purely the spin-$\frac{1}{2}$ current-dependent terms.
Then the conformal dimension of the state $|(0;v)>$ is
\bea
\hat{T}_0 |(0;v)>  &\sim&
\left[ \frac{k}{2(k+N+2)^2} Q_{\bar{a}} \pa Q^{\bar{a}} 
-\frac{1}{(k+N+2)} \sum_{i=1}^3 \hat{B}_{i} \hat{B}_{i} \right]_0  |(0;v)> 
\nonu \\
&=&  \frac{k}{2(k+N+2)}   |(0;v)> 
+\frac{1}{(k+N+2)}  l^{-}( l^{-}+1)  |(0;v)> 
\nonu \\ 
&=&  \left[ \frac{(2k+3)}{4(N+k+2)} \right] |(0;v)>.
\label{hzerov}
\eea
In the first line of (\ref{hzerov}), 
the spin-$1$ current-dependent terms are 
ignored.
In the second line, 
we have used the fact that 
the eigenvalue equation 
$\left[ Q_{\bar{a}} \pa Q^{\bar{a}} \right]_0 |(0;v)>=(k+N+2)|(0;v)>$
(see (\ref{j0f})) can be obtained 
because  the highest-order pole gives the corresponding eigenvalue
$ Q_{\bar{a}} \pa Q^{\bar{a}}(z) \, Q^{\bar{b}}(w) |_{\frac{1}{(z-w)^2}}
=(k+N+2) Q^{\bar{b}}(w)$ (see (\ref{opejq})) which can be checked from the 
defining relation in (\ref{opevq}).
Furthermore, the characteristic eigenvalue equation 
for the affine $\hat{su}(2)$ algebra described in the footnote \ref{footfoot} 
is used.
The above eigenvalue  is exactly the same as the  eigenvalue 
$h(0;f)$ described 
in \cite{GG1305}.
Under the large $N$ 't Hooft limit (\ref{limit}), the eigenvalue 
equation implies that we have 
$\hat{T}_0 |(0;v)> = \frac{1}{2} (1-\la) |(0;v)>$.
There exists $N \leftrightarrow k$  
symmetry 
between the  eigenvalues in (\ref{hvzero2}) and 
(\ref{hzerov}). In the large 
$N$ 't Hooft limit, this is equivalent to  $ \la \leftrightarrow (1-\la)$ symmetry. 

\subsection{ Eigenvalue equations for the higher spin currents of spins 
$2,3$ and $4$ }

Now let us 
consider the eigenvalue equations for the higher spin currents
by following the descriptions in previous subsection.

\subsubsection{ Eigenvalue equations for the higher 
spin-$2$ current}

From the explicit expression for the higher spin-$2$  current 
$T^{(2)}$ (\ref{T2ansatz}) for several $N=4,5,8,9$,
we obtain  the eigenvalue equation for general $N$.
It turns out that
\bea
T^{(2)}_0 |(v;0)> &=& -\left[ \frac{(2 k N+k+4 N^2-4 N-12)}{2 (k+N+2)^2} 
\right] |(v;0)>,
\nonu \\
T^{(2)}_0 |(0;v)> &=& 
\left[ \frac{k (2 N+1) (2 k+N)}{2 N (k+N+2)^2}\right] |(0;v)>.
\label{T2eigenvalue}
\eea
Although there is no $ N \leftrightarrow k$ symmetry between the eigenvalues 
of the states 
$|(v;0)>$ and $|(0;v)>$ in (\ref{T2eigenvalue}), 
there exists the $ \lambda \leftrightarrow (1-\lambda)$ symmetry (up to sign) 
in 
the large $N$ 't Hooft limit. In other words, 
the eigenvalue equations reduce to 
\bea
T^{(2)}_0 |(v;0)> &=& - \lambda (1+\lambda) |(v;0)>,
\nonu \\
T^{(2)}_0 |(0;v)> &=&  (1-\lambda) (2-\lambda) |(0;v)>.
\label{t2eigen}
\eea
Compared to the corresponding eigenvalue equations for the 
higher spin-$2$ current for the 
unitary case \cite{AK1506}, the new last factor 
$(1+\la)$ and $(2-\la)$
in each eigenvalue 
occurs in (\ref{t2eigen}) respectively.
They have different $SO(4)$ representations as described in the 
introduction.

\subsubsection{ Eigenvalue equations for the higher 
spin-$3$ currents}

In order to represent the eigenvalue equations for the 
higher spin-$3$ currents, 
we should classify the $|(v;0)>$ states into the following four types
of column vectors
\bea
|(v;0)>_{++} & = & (0, \cdots, 0, 1, 0, 0, 0)^T, \qquad
|(v;0)>_{+-}= (0, \cdots, 0, 0, 1, 0, 0)^T, \nonu \\
|(v;0)>_{-+} & = & (0, \cdots, 0, 0, 0, 1, 0)^T, \qquad
|(v;0)>_{--}= (0, \cdots, 0, 0, 0, 0, 1)^T.
\label{eigenvector}
\eea
They have nontrivial $U(1)$ charges which will be described in section $5$.
 On the other hand, 
the $|(0;v)>$ states are  expressed by the following forms
\bea
|(0;v)>_{++} & : & \frac{1}{\sqrt{k+N+2}} Q_{-\frac{1}{2}}^{\bar{a}}|0>,
\qquad 
\bar{a}=1, 2, \cdots, N,
\nonu \\
|(0;v)>_{+-} & : & \frac{1}{\sqrt{k+N+2}} Q_{-\frac{1}{2}}^{\bar{a}}|0>,
\qquad 
\bar{a}=N+1, N+2, \cdots, 2N,
\nonu \\
|(0;v)>_{-+} & : & \frac{1}{\sqrt{k+N+2}} Q_{-\frac{1}{2}}^{\bar{a}}|0>,
\qquad 
\bar{a}=1^{\ast}, 2^{\ast}, \cdots, N^{\ast},
\nonu \\
|(0;v)>_{--} & : & \frac{1}{\sqrt{k+N+2}} Q_{-\frac{1}{2}}^{\bar{a}}|0>,
\qquad 
\bar{a}=(N+1)^{\ast}, (N+2)^{\ast}, \cdots, (2N)^{\ast}.
\label{plusminus}
\eea

Now we apply the eigenvalue equations for the zero mode of the 
higher spin-$3$
currents to these states.
It turns out that the 
eigenvalue equations for the higher spin-$3$
current $T^{(3)}(z)$ acting on (\ref{plusminus}) and (\ref{eigenvector}) 
are summarized by
\bea
T^{(3)}_0|(v;0)>_{\alpha \pm} &=& \pm \left[ \frac{(2 k N+k+4 N^2-4 N-12)}{
(k+N+2)^2} 
\right] |(v;0)>_{ \alpha \pm}, 
\nonu \\
T^{(3)}_0|(0;v)>_{\pm \alpha  } &=&\pm \left[ \frac{k (2 N+1) (2 k+N)}{N (k+N+2)^2}
\right] |(0;v)>_{\pm \alpha }, 
\label{T3eigen}
\eea
where the index $\alpha$ stands for $\alpha=+,-$.
The eigenvalues in (\ref{T3eigen}) are similar to the $T^{(2)}_0$ eigenvalues
in (\ref{T2eigenvalue}). The only overall factors are different from
each other.

For the higher spin-$3$ current $W^{(3)}_0$, we have the following 
relations
\bea
W^{(3)}_0|(v;0)>_{\alpha \pm} &=& \pm \left[ \frac{(2 k N+k+4 N^2-4 N-12)}{
(k+N+2)^2} 
\right] |(v;0)>_{ \alpha \pm}, 
\nonu \\
W^{(3)}_0|(0;v)>_{\pm \alpha } &=&\mp \left[ \frac{k (2 N+1) (2 k+N)}{N (k+N+2)^2}
\right] |(0;v)>_{\pm \alpha }.
\label{W3eigen}
\eea
We can easily see that the relations 
(\ref{W3eigen}) are the same as the ones in (\ref{T3eigen}) except the 
overall sign.

Furthermore, under the large $N$ 't Hooft limit (\ref{limit}), 
the above eigenvalue equations (\ref{T3eigen}) and 
(\ref{W3eigen}) become
\bea
T^{(3)}_0|(v;0)>_{\alpha \pm} &=& \pm 2 \lambda (1+\lambda)  |(v;0)>_{ \alpha \pm}, 
\nonu \\
T^{(3)}_0|(0;v)>_{\pm \alpha  } & = & \pm 2 (1-\lambda)(2-\lambda)  
|(0;v)>_{\pm \alpha }, 
\nonu \\
W^{(3)}_0|(v;0)>_{\alpha \pm} &=& \pm 2 \lambda (1+\lambda)  |(v;0)>_{ \alpha \pm}, 
\nonu \\
W^{(3)}_0|(0;v)>_{\pm \alpha  } &=&\mp 2 (1-\lambda)(2-\lambda)  
|(0;v)>_{\pm \alpha }. 
\label{spin3eigen}
\eea
Compared to the unitary case, 
the behavior of $\la(1+\la)$
and $(1-\la)(2-\la)$ in the eigenvalues (\ref{spin3eigen}) 
is the same the ones in \cite{AK1506}.

For the other remaining four higher spin-$3$ currents, 
we obtain the following nonzero results
\bea
\left[ U^{(3)}_{+} \right]_0 |(0;v)>_{- \pm}
& = & \mp 2i \left[ \frac{k (2 N+1) (2 k+N)}{N (k+N+2)^2}
\right]  |(0;v)>_{+ \mp} \rightarrow 
\mp 4i  (1-\lambda)(2-\lambda)  |(0;v)>_{+ \mp},
\nonu \\
\left[ U^{(3)}_{-} \right]_0 |(v;0)>_{ \pm + }
& = & \pm 2i \left[ \frac{(2 k N+k+4 N^2-4 N-12)}{(k+N+2)^2} 
\right]  |(v;0)>_{ \mp -} \rightarrow 
\pm 4i \lambda (1+\lambda)  |(v;0)>_{ \mp -},
\nonu \\
\left[ V^{(3)}_{+} \right]_0 |(v;0)>_{ \mp - }
& = & \pm 2i \left[ \frac{(2 k N+k+4 N^2-4 N-12)}{(k+N+2)^2} 
\right]  |(v;0)>_{ \pm +} \rightarrow 
\pm 4i \lambda (1+\lambda)  |(v;0)>_{ \pm +},
\nonu \\
\left[ V^{(3)}_{-} \right]_0 |(0;v)>_{+ \pm}
& = & \pm 2i \left[ \frac{k (2 N+1) (2 k+N)}{N (k+N+2)^2}
\right]  |(0;v)>_{- \mp} \nonu \\
& \rightarrow & 
\pm 4i  (1-\lambda)(2-\lambda)  |(0;v)>_{- \mp},
\label{otherthree}
\eea
where the large $N$ 't Hooft limit is taken.
Obviously,  they are not eigenvalue equations
and other relevant quantities (for example, the sum of quadratic
of the triplet) can be obtained from these relations (\ref{otherthree})
\footnote{More precisely, 
we have the following relations 
\bea
\left[ U^{(3)}_{+} \right]_0 \frac{1}{\sqrt{N+k+2}} Q^{a^*}_{-\frac{1}{2}} |0>
& = & -2i \left[ \frac{k (2 N+1) (2 k+N)}{N (k+N+2)^2}
\right]  \frac{1}{\sqrt{N+k+2}} Q^{a+N}_{-\frac{1}{2}} |0>, 
\nonu \\
\left[ U^{(3)}_{+} \right]_0 \frac{1}{\sqrt{N+k+2}} Q^{(a+N)^*}_{-\frac{1}{2}} |0>
& = &  2i \left[ \frac{k (2 N+1) (2 k+N)}{N (k+N+2)^2}
\right] \frac{1}{\sqrt{N+k+2}} Q^{a}_{-\frac{1}{2}} |0>,
\nonu \\
\left[ V^{(3)}_{-} \right]_0 \frac{1}{\sqrt{N+k+2}} Q^{a}_{-\frac{1}{2}} |0>
& = &  2i \left[ \frac{k (2 N+1) (2 k+N)}{N (k+N+2)^2}
\right]  \frac{1}{\sqrt{N+k+2}} Q^{(a+N)^*}_{-\frac{1}{2}} |0>, 
\nonu \\
\left[ V^{(3)}_{-} \right]_0 \frac{1}{\sqrt{N+k+2}} Q^{a+N}_{-\frac{1}{2}} |0>
& = &  -2i \left[ \frac{k (2 N+1) (2 k+N)}{N (k+N+2)^2}
\right]  \frac{1}{\sqrt{N+k+2}} Q^{a^*}_{-\frac{1}{2}} |0>, 
\nonu
\eea
where the index $a$ runs over $a=1, 2, \cdots, N$.}.

\subsubsection{ Eigenvalue equations for the higher 
spin-$4$ current}
 
It turns out that the eigenvalue equations  of the zero mode 
of the higher spin-$4$ current  $W^{(4)}(z)$
are described as
\bea
W^{(4)}_0 |(v;0)> &=& \left[   -\frac{2 \left(2 k N+k+4 N^2-4 N-12\right) }{(k+N+2)^3 (30 k N+59 k+59 N+88)} 
\times d_1   \right] |(v;0)>,
\nonu \\
W^{(4)}_0 |(0;v)> &=& \left[   -\frac{2 k (2 N+1) (2 k+N) }{N
   (k+N+2)^3 (30 k N+59 k+59 N+88)}  \times d_2  \right] |(0;v)>,
   \label{w4eigen1}
\eea
where we introduce two factors which show the $N \leftrightarrow k$ symmetry
\bea
d_1(N,k) &\equiv& \left(54 k N^2+81 N^2+36
   k^2 N+225 k N+176 N+78 k^2+206 k+88\right),
   \nonu \\
 d_2 (N,k) &\equiv&   \left(54 k^2 N+81 k^2+36
   k N^2+225 k N+176 k+78 N^2+206 N+88\right).
\nonu
\eea
There is no $N \leftrightarrow k$ symmetry between the two eigenvalues in 
(\ref{w4eigen1}).
But if we divide out the $T^{(2)}_{0}$ eigenvalues (denoted by
$t^{(2)}(v;0)$ and $t^{(2)}(0;v)$ respectively) from  the 
$W^{(4)}_{0}$ eigenvalues (denoted by $w^{(4)}(v;0)$ and $w^{(4)}(0;v)$ 
respectively), 
we can see the $N \leftrightarrow k$ symmetry and the following relation 
satisfies $
\left[ \frac{w^{(4)}(v;0)}{t^{(2)}(v;0)} \right]_{N \leftrightarrow k}
= -\frac{w^{(4)}(0;v)}{t^{(2)}(0;v)}$.
We will see that the eigenvalues become very simple
in different basis later.   

Under the large $N$ 't Hooft limit (\ref{limit}), we have 
\bea
W^{(4)}_0 |(v;0)> & = & -\frac{12}{5} \lambda (1+\lambda)(2+\lambda)|(v;0)>,
\nonu \\
W^{(4)}_0 |(0;v)> & = & -\frac{12}{5} (1-\lambda) (2-\lambda)(3-\lambda)|(0;v)>.
\label{limitw4eigen}
\eea
There exists the $\lambda \leftrightarrow (1- \lambda)$ symmetry.
We observe that the extra factors $(2+\la)$ and $(3-\la)$ 
in (\ref{limitw4eigen})
are present respectively 
compared to the corresponding eigenvalue equations in 
(\ref{spin3eigen}).

Let us describe the three point functions.
From the diagonal modular invariant with 
pairing up identical representations on the left (holomorphic)
and the right (antiholomorphic) 
sectors \cite{CY1106}, 
one of the primaries is given by 
$(v;0) \otimes (v;0)$
which is denoted by 
${\cal O}_{+}$ and the other 
is given by $(0;v) \otimes (0;v)$
which is denoted by ${\cal O}_{-}$.
Then the three point functions with these two scalars  
are obtained and their ratios can be written as 
\bea
\frac{<{\cal O}_{+ } 
{\cal O}_{+ }T^{(2)}>}{< {\cal O}_{- } 
{\cal O}_{- } T^{(2)}>}
 &= & 
 -\left[\frac{\lambda(1+\lambda)}{(1-\lambda)(2-\lambda)} \right],
\qquad
\frac{<{\cal O}_{+ } 
{\cal O}_{+ }T^{(3)}>}{< {\cal O}_{- } 
{\cal O}_{- } T^{(3)}>}
 =  \pm 
 \left[\frac{\lambda(1+\lambda)}{(1-\lambda)(2-\lambda)} \right],
 \label{three} 
\\
\frac{<{\cal O}_{+ } 
{\cal O}_{+ }W^{(3)}>}{< {\cal O}_{- } 
{\cal O}_{- } W^{(3)}>}
 &= & \pm
 \left[\frac{\lambda(1+\lambda)}{(1-\lambda)(2-\lambda)} \right],
 \qquad
\frac{<{\cal O}_{+ } 
{\cal O}_{+ }W^{(4)}>}{< {\cal O}_{- } 
{\cal O}_{- } W^{(4)}>}
 =  
 \left[\frac{\lambda(1+\lambda)(2+\lambda)}{(1-\lambda)(2-\lambda)(3-\lambda)} \right],
\nonu
\eea
where the states in the three-point functions for the 
higher spin-$3$ currents are assumed from (\ref{spin3eigen}). 
Depending on the states, the ratios can be plus sign or minus sign.
The behavior for  the ratios for the three-point functions 
is the same as the one in the unitary case up to the overall sign.
Furthermore, 
we see that  the  ratio for the three-point function
for the higher spin-$4$ current (\ref{three})
contains the factor $\left[ \frac{(2+\la)}{(3-\la)} \right]$
and the remaining factor appears in the corresponding three-point function 
for the higher spin-$3$ current.
We expect that 
 the  ratio for the three-point function
for the higher spin-$5$ current
contains the factor $ \left[\frac{\la(1+\la)(2+\la)(3+\la)}
{(1-\la)(2-\la)(3-\la)(4-\la)}\right]$ 
only after the analysis in the subsection 
$2.4$ has been done.
Recall that in the bosonic unitary (or orthogonal) coset theory
studied in \cite{GH1101,Ahn1111,Ahn1202,AK1308},
the ratios of three-point functions behave as 
$\frac{(1+\la)}{(1-\la)}$ for the spin-$2$ current
corresponding to the stress energy tensor, 
$-\frac{(1+\la)(2+\la)}{(1-\la)(2-\la)}$ for the higher spin-$3$ current,
$\frac{(1+\la)(2+\la)(3+\la)}{(1-\la)(2-\la)(3-\la)}$ for the higher spin-$4$
current, and 
$-\frac{(1+\la)(2+\la)(3+\la)(4+\la)}{(1-\la)(2-\la)(3-\la)(4-\la)}$ for the 
higher spin-$5$ current. Then by shifting the $\la $ appearing in the 
numerator as $\la \rightarrow -(1-\la)$, we can see the behavior of the above 
results in (\ref{three}) up to sign.  

Therefore, the ratios of the three-point functions can be summarized 
by (\ref{three}). In order to obtain these results, 
the equations (\ref{t2eigen}), (\ref{spin3eigen}), (\ref{limitw4eigen}) 
were crucial. Not that the ratio for the three-point function for the higher 
spin-$2$ current in (\ref{three})
has the factor $\left[\frac{(1+\la)}{(2-\la)} \right]$ 
which does not appear in the unitary 
case \cite{AK1506}.

\section{ The extension of 
the large $\mathcal N = 4$ linear superconformal algebra}

We construct the $16$ currents of large ${\cal N}=4$ linear 
superconformal algebra using the fundamental currents as in section $2$.
With the lowest higher spin-$2$
current found in section $2$, we show how the remaining $15$ higher spin 
currents can be obtained implicitly. 

\subsection{The large ${\cal N}=4$ linear superconformal algebra}

From the $N=4,5,8,9$ cases, we can obtain the following 
four spin-$\frac{1}{2}$ currents and the spin-$1$ current as 
follows:
\bea
{\bf F}_{11}(z)&=& \frac{i }{\sqrt{2}} Q^{(2N+3)}(z),
\qquad
{\bf F}_{22}(z)=-\frac{i }{\sqrt{2}} Q^{(2N+3)^\ast}(z),
\nonu \\
{\bf F}_{12}(z)&=& \frac{(1-i) }{2} Q^{(2N+2)^\ast}(z),
\qquad
{\bf F}_{21}(z)= \frac{(1+i) }{2} Q^{(2N+2)}(z),
\nonu \\
{\bf U}(z)&=& 
\frac{(1+i)}{2 \sqrt{2}} V^{(2N+2)} (z)+ \frac{(-1+i)}{2 \sqrt{2}} V^{(2N+2)^*}(z)
+\frac{i}{(N+k+2)} Q^{(2N+1)} Q^{(2N+1)^\ast}(z)
\nonu \\
&-& \frac{i}{2(N+k+2)} \left(  \sum_{a=1}^N Q^{a} Q^{a^\ast}
- \sum_{a=N+1}^{2N} Q^{a} Q^{a^\ast} \right)(z).
\label{1andhalfexp}
\eea
The corresponding $so(4)$ generators 
with indices, $(2N+1), (2N+2)$ and $(2N+3)$ (and their conjugates), 
are given in Appendix $A$. 
Note that the $N$-dependence in (\ref{1andhalfexp})
appears in the quadratic term in the spin-$\frac{1}{2}$ current.
Furthermore, the presence of the third term in ${\bf U}(z)$
is rather new feature in the orthogonal coset theory 
because we do not see the quadratic term with the index 
living in the lower $2 \times 2$ matrix for the unitary case. 

Then from the Goddard-Schwimmer formula \cite{GS},
we have 
\bea
{\bf T}(z) & = & \hat{T}(z) -\frac{1}{(N+k+2)} \left( {\bf U} {\bf U} + \pa 
{\bf F^a} {\bf F_a} \right)(z),
\nonu \\
{\bf G_a}(z) & = & \hat{G}_a(z) - \frac{2}{(N+k+2)}
\left(  {\bf U} {\bf F_a} -\frac{1}{3(N+k+2)} \ep_{abcd} {\bf F^b} {\bf F^c}
{\bf F^d} + 2 {\bf F^b} ( \alpha_{ba}^{+i} \hat{A}_i - 
\alpha_{ba}^{-i} \hat{B}_i) \right)(z), 
\nonu \\
{\bf A_i}(z) & = & \hat{A}_i(z) +\frac{1}{(N+k+2)} \alpha_{ab}^{+i} {\bf F^a}{
\bf F^b}(z), 
\nonu \\
{\bf B_i}(z) & = & \hat{B}_i(z) +\frac{1}{(N+k+2)} \alpha_{ab}^{-i} {\bf F^a}{
\bf F^b}(z), \qquad  a, b =11, 12, 21, 22.
\label{16expression}
\eea 
Here the $11$ currents, $\hat{T}(z)$, $\hat{G}_a(z)$, $\hat{A}_i(z)$
and $\hat{B}_i(z)$, in the nonlinear version 
are given in (\ref{closedform}) with the 
footnote \ref{doubleindex}.
Then the $16$ currents of the large ${\cal N}=4$ linear superconformal 
algebra \cite{STVplb,npb1988,Schoutensnpb}
are written in terms of the fundamental spin-$1$ and spin-$\frac{1}{2}$
currents living in the orthogonal coset theory via (\ref{16expression}),
(\ref{1andhalfexp}) and (\ref{closedform}) together with the footnote 
\ref{doubleindex}.

\subsection{The $16$ lowest higher spin currents}

As in (\ref{lowmultiplet}), we 
present the higher spin currents with boldface notations 
as follows: 
\bea
\left(2, \frac{5}{2}, \frac{5}{2}, 3 \right)
& : & ({\bf T^{(2)}}, {\bf T_{+}^{(\frac{5}{2})}}, {\bf T_{-}^{(\frac{5}{2})}}, 
{\bf T^{(3)}}),
\nonu \\
 \left(\frac{5}{2}, 3, 3, \frac{7}{2} \right) & : &
({\bf U^{(\frac{5}{2})}}, {\bf U_{+}^{(3)}}, {\bf U_{-}^{(3)}}, 
{\bf U^{(\frac{7}{2})}} ), \nonu \\
\left(\frac{5}{2}, 3, 3, \frac{7}{2} \right) & : &
({\bf V^{(\frac{5}{2})}}, {\bf V^{(3)}_{+}}, {\bf V^{(3)}_{-}}, 
{\bf V^{(\frac{7}{2})}}),  \nonu \\
\left(3, \frac{7}{2}, \frac{7}{2}, 4 \right) & : &
 ({\bf W^{(3)}}, {\bf W_{+}^{(\frac{7}{2})}}, {\bf W_{-}^{(\frac{7}{2})}}, 
{\bf W^{(4)}}).
\label{lowmultipletlin}
\eea
We take the lowest higher spin-$2$ current ${\bf T^{(2)}}(z)$ 
as the one $T^{(2)}(z)$ in the 
nonlinear version.
From the explicit results on the $16$ currents of 
the large ${\cal N}=4$ linear superconformal algebra  
in the previous subsection,
we would like to construct the higher spin currents in the linear 
version as in section $2$.

\subsubsection{The higher spin currents of spins
$\left(2, \frac{5}{2}, \frac{5}{2}, 3 \right)$}

Let us consider the first ${\cal N}=2$ multiplet
(\ref{lowmultipletlin}).
Because the nonlinear version for the appearance of 
the higher spin-$\frac{5}{2}$ currents was obtained in (\ref{g2112t2}),
we calculate the similar OPEs. 
The following OPEs satisfy
\bea
\left(
\begin{array}{c}
G_{21} \\
G_{12} 
\end{array}
\right)(z) \, \mathbf{T^{(2)}}(w)  & = &
\frac{1}{(z-w)}
\mathbf{T_{\pm}^{(\frac{5}{2})}}(w) +
\cdots.
\label{ling2112t2}
\eea
We expect that 
we have the extra terms for the higher spin-$\frac{5}{2}$ currents,  
coming from the 
OPEs between the difference of the spin-$\frac{3}{2}$ currents
in the nonlinear and linear versions  
and the higher spin-$2$ current, when we compare with the ones in 
(\ref{g2112t2}).
However, these OPEs do not have any singular terms according to 
(\ref{16expression}), the footnote \ref{doubleindex} and 
(\ref{nonliT2condition}).
Therefore, we have ${\bf T^{(\frac{5}{2})}_{\pm}}(w) = T^{(\frac{5}{2})}_{\pm}(w)$.  
Now we can calculate the last component higher spin-$3$ current in this
${\cal N}=2$ multiplet.
By taking the similar OPE in (\ref{g21t5half-other}), we obtain the 
following OPE where the first-order pole in (\ref{ling2112t2}) 
is used
\bea
G_{21}(z) \, \mathbf{T_{-}^{(\frac{5}{2})}}(w) & = &
\frac{1}{(z-w)^2} 4 \mathbf{T^{(2)}} (w)
 +  \frac{1}{(z-w)}  \left[
 \frac{1}{4} \pa (\mbox{pole-2}) +\mathbf{T^{(3)}} \right](w) +  \cdots.
\label{ling21t5half-}
\eea
In the second-order pole of (\ref{ling21t5half-}), we can see 
the same expression as in (\ref{g21t5half-other}) even though the left 
hand sides of these OPEs 
are different from each other.
However, the first-order pole provides 
the new higher spin-$3$ current which is different from the one appearing 
in (\ref{g21t5half-other})  in the 
nonlinear version. 

\subsubsection{The higher spin currents of spins
$\left(\frac{5}{2}, 3, 3, \frac{7}{2} \right)$}

Let us describe the next ${\cal N}=2$ multiplet in (\ref{lowmultipletlin}).
Again the previous OPE (\ref{g11t2other}) allows us to calculate the 
following OPE
\bea
G_{11}(z) \, \mathbf{T^{(2)}}(w)  & = &
\frac{1}{(z-w)}\mathbf{U^{(\frac{5}{2})}} (w) +
\cdots.
\label{ling11t2}
\eea
In general, we 
can extract the extra terms in the first-order pole in 
(\ref{ling11t2}) compared to the corresponding quantity in (\ref{g11t2other})   
by noting the 
difference of the spin-$\frac{3}{2}$ currents
in the nonlinear and linear versions in the left hand side of the OPE.  
However, according to the previous analysis (there are no singular terms), 
we have that the corresponding higher spin-$\frac{5}{2}$ current in 
linear version is the same as the one in the 
nonlinear version ${\bf U^{(\frac{5}{2})}}(w) = U^{(\frac{5}{2})}(w)$. 
The next two higher spin-$3$ currents 
can be obtained from the above higher spin-$\frac{5}{2}$ current
appearing in (\ref{ling11t2}). It turns out that 
\bea
\left(
\begin{array}{c}
G_{21} \\
G_{12} 
\end{array}
\right)(z) \, \mathbf{U^{(\frac{5}{2})}}(w)  & = &
\frac{1}{(z-w)}  \mathbf{U^{(3)}_{\pm}} (w) +
\cdots.
\label{ling21u5half}
\eea
We can see the similar nonlinear version in (\ref{g2112u5half}). 
Finally the last component higher spin-$\frac{7}{2}$ current 
can be obtained with the help of the first-order pole in 
(\ref{ling21u5half}) 
as follows:
\bea
G_{21}(z) \, \mathbf{U_{-}^{(3)}}(w) & = &
\frac{1}{(z-w)^2} \frac{2(2N+5+3k)}{(N+2+k)} \mathbf{U^{(\frac{5}{2})}}(w)
\nonu \\
& + & \frac{1}{(z-w)}  \left[ \frac{1}{5}
\pa (\mbox{pole-2})
+\mathbf{U^{(\frac{7}{2})}}
 \right](w)+ \cdots.
\label{ling21u3-}
\eea
Note that the structure constant appearing in the second-order pole 
in (\ref{ling21u3-}) is different from the one in (\ref{g21u3-other}).
Therefore, the second ${\cal N}=2$ multiplet is found for generic 
$N$ implicitly. 

\subsubsection{The higher spin currents of spins
$\left(\frac{5}{2}, 3, 3, \frac{7}{2} \right)$}

For the third ${\cal N}=2$ multiplet, we can start with the following OPE
\bea
G_{22}(z) \, \mathbf{T^{(2)}}(w)  & = &
\frac{1}{(z-w)}  \mathbf{V^{(\frac{5}{2})}} (w) +
\cdots.
\label{ling22t2}
\eea
The corresponding nonlinear version is given by (\ref{g22t2other}).
We can easily see that 
the extra terms in the first-order pole in (\ref{ling22t2}) 
compared to the one in (\ref{g22t2other})
can be read off from the difference in the spin-$\frac{3}{2}$
currents in the nonlinear and linear versions.
Similarly we have that 
the corresponding higher spin-$\frac{5}{2}$ current in 
the linear version is the same as the one in the 
nonlinear version ${\bf V^{(\frac{5}{2})}}(w) = V^{(\frac{5}{2})}(w)$. 
Similarly we can calculate the following OPEs
\bea
\left(
\begin{array}{c}
G_{21} \\
G_{12} 
\end{array}
\right)(z) \, \mathbf{V^{(\frac{5}{2})}}(w)  & = &
\frac{1}{(z-w)} \mathbf{V^{(3)}_{\pm}}(w) +
\cdots.
\label{ling21v5half}
\eea
Then
the final higher spin-$\frac{7}{2}$ current can be determined from 
the first-order pole in (\ref{ling21v5half})
as follows
\bea
G_{21}(z) \, \mathbf{V_{-}^{(3)}}(w) & = &
\frac{1}{(z-w)^2}\frac{2(3N+5+2k)}{(N+2+k)} \mathbf{V^{(\frac{5}{2})}}(w)
\nonu \\
& + & \frac{1}{(z-w)}  \left[ \frac{1}{5}  
\pa (\mbox{pole-2})
+\mathbf{V^{(\frac{7}{2})}}
 \right](w)+ \cdots.
\label{ling21v3-}
\eea
Again, 
 the structure constant appearing in the second-order pole 
in (\ref{ling21v3-}) is different from the one in (\ref{g21v3-other})
and is the same as the one in (\ref{ling21u3-}) 
by $N \leftrightarrow k$ symmetry.

\subsubsection{The higher spin currents of spins
$\left(3, \frac{7}{2}, \frac{7}{2}, 4 \right)$}

Let us consider the final ${\cal N}=2$ multiplet.
As in (\ref{g22u5halfother}), we calculate the following OPE
with (\ref{ling11t2})
\bea
G_{22}(z) \, \mathbf{U^{(\frac{5}{2})}}(w) & = &
\frac{1}{(z-w)^2}4\mathbf{T^{(2)}}(w)
+  \frac{1}{(z-w)}  \left[ \frac{1}{4} 
\pa (\mbox{pole-2})
+\mathbf{W^{(3)}}
 \right](w)+ \cdots.
\label{g22u5half}
\eea
From the first-order pole, we obtain the higher spin-$3$ current. 
Based on this result in (\ref{g22u5half}), 
we can calculate the following OPEs
\bea
\left(
\begin{array}{c}
G_{21} \\
G_{12} 
\end{array}
\right)(z)
 \,\mathbf{W^{(3)}}(w) & = & \pm
\frac{1}{(z-w)^2} \frac{(N-k)}{(N+2+k)} \mathbf{T_{\pm}^{(\frac{5}{2})}}(w)
\nonu \\
& + & \frac{1}{(z-w)}  \left[ \frac{1}{5}  
\pa (\mbox{pole-2})
+\mathbf{W_{\pm}^{(\frac{7}{2})}}
 \right](w)+ \cdots,
\label{ling21w3}
\eea
which is the same form as the one in (\ref{g2112w3}).
Now the final higher spin-$4$ current can be obtained by considering the 
following OPE
with (\ref{ling21w3})
\bea
G_{21}(z) \,\mathbf{W_{-}^{(\frac{7}{2})}}(w) & = &
\frac{1}{(z-w)^3} \left[ -\frac{48(-N+k)}{5(N+2+k)}\mathbf{T^{(2)}}
 \right](w)
\nonu \\
& + & \frac{1}{(z-w)^2}  \left[ -\frac{6(-N+k)}{5(N+2+k)}\mathbf{T^{(3)}}+6\mathbf{W^{(3)}}
 \right](w)
 \nonu \\
 &+&\frac{1}{(z-w)} \left[ \frac{1}{6}\pa 
(\mbox{pole-2})
 \right. \nonu \\
& - & \left. \frac{72(-N+k)}{((37N+59)+(15N+37)k)}
\left(T\mathbf{T^{(2)}}-\frac{3}{10} \pa^2 \mathbf{T^{(2)}} \right)
 +\mathbf{W^{(4)}}
 \right](w)
\nonu \\
& + & \cdots.
\label{ling21w7half-}
\eea
Compared to the one in (\ref{g21w7half}), 
the second-order pole in (\ref{ling21w7half-})
does not contain the nonlinear terms.
We can see that the combination of the 
quasiprimary field of spin $4$ and 
the primary higher spin-$4$ current (appearing in the second line of the 
first-order pole) can be identified with 
the quasiprimary field of spin $4$ in \cite{BCG1404}.
 
\subsection{The next $16$ lowest higher spin currents}

We can describe the next $16$ higher spin currents
by following the method in the subsection $2.4$ in the nonlinear version.

\subsection{The higher spin currents in different basis}

As in the unitary case \cite{AK1506}, 
we can obtain the following explicit relations
where we can have the higher spin currents in the basis of \cite{BCG1404}
\bea
V_0^{(2)}(z) &= & {\bf T^{(2)}},
\nonu \\
V_{\frac{1}{2}}^{(2), 0}(z) & = & -\frac{i}{\sqrt{2}} \left( - 
{\bf T_{+}^{(\frac{5}{2})}} + {\bf T_{-}^{(\frac{5}{2})}} \right),
\qquad
V_{\frac{1}{2}}^{(2), 1}(z)  =  \frac{1}{\sqrt{2}} \left(  
{\bf U^{(\frac{5}{2})}} + {\bf V^{(\frac{5}{2})}} \right),
\nonu \\
V_{\frac{1}{2}}^{(2), 2}(z) & = & \frac{i}{\sqrt{2}} \left(  
{\bf U^{(\frac{5}{2})}} - {\bf V^{(\frac{5}{2})}} \right),
\qquad
V_{\frac{1}{2}}^{(2), 3}(z)  =  -\frac{1}{\sqrt{2}} \left(  
{\bf T_{+}^{(\frac{5}{2})}} + {\bf T_{-}^{(\frac{5}{2})}} \right),
\nonu \\
V_{1}^{(2), \pm 1}(z) & = & i \left(  
{\bf U_{\mp}^{(3)}} - {\bf V_{\pm}^{(3)}} \right),
\qquad
V_{1}^{(2), \pm 2}(z)  =  - \left(  
{\bf U_{\mp}^{(3)}} + {\bf V_{\pm}^{(3)}} \right),
\nonu \\
V_{1}^{(2), \pm 3}(z) & = & \pm i \left(  
{\bf T^{(3)}} \pm  {\bf W^{(3)}} \right),
\nonu \\
V_{\frac{3}{2}}^{(2), 0}(z) & = & i \sqrt{2} \left(  
{\bf W_{+}^{(\frac{7}{2})}} + {\bf W_{-}^{(\frac{7}{2})}} \right),
\qquad
V_{\frac{3}{2}}^{(2), 1}(z)  =  - \sqrt{2} \left(  
{\bf U^{(\frac{7}{2})}} - {\bf V^{(\frac{7}{2})}} \right),
\nonu \\
V_{\frac{3}{2}}^{(2), 2}(z) & = & -i \sqrt{2} \left(  
{\bf U^{(\frac{7}{2})}} + {\bf V^{(\frac{7}{2})}} \right),
\qquad
V_{\frac{3}{2}}^{(2), 3}(z)  =  - \sqrt{2} \left(  
{\bf W_{+}^{(\frac{7}{2})}} - {\bf W_{-}^{(\frac{7}{2})}} \right),
\nonu \\
V_{2}^{(2)}(z) & = & -2 \left[  
{\bf W^{(4)}} - \frac{72(-N+k)}{((37N+59)+(15N+37)k)} 
\left( {\bf T} {\bf T^{(2)}} -\frac{3}{10} \pa^2 {\bf T^{(2)}}
  \right) \right].
\label{16basis}
\eea
In doing this, Appendices $D$ and $E$ are necessary to check these 
relations explicitly.
Of course, we can further reexpress the above $16$ higher spin 
currents (\ref{16basis}) in the manifest $SO(4)$ symmetry by introducing 
the derivative terms as done in \cite{AK1509}. 

\section{  
 Three-point functions in the extension of 
the large ${\cal N}=4$ linear superconformal algebra
}

As in section $3$, we calculate the three-point functions 
for the higher spin currents (obtained in previous section) 
in the extension of 
large ${\cal N}=4$ linear superconformal algebra. 

\subsection{  Eigenvalue equations for the spin-$2$ current  }

Let us  define the ${\bf U}$-charge as in  \cite{npb1989} as follows:
\bea
i {\bf U}_0 |(v;0)> & = & {\bf u}(v;0) |(v;0)>, \qquad
i {\bf U}_0 |(0;v)>   =  {\bf u}(0;v) |(0;v)>.
\label{eigenu}
\eea
We obtain the eigenvalues  $ {\bf u}(v;0)  $ and $ {\bf u}(0;v)  $
in (\ref{eigenu}) 
as follows
\footnote{The ${\bf U}$-charge 
of light state $(v;v)$ is zero from the explicit matrix acting on the 
states as in the unitary case \cite{AK1506}. 
The conformal dimension for the light state 
in the linear and the nonlinear version 
is the same.  
That is, 
$h'(v;v)=h(v;v)$ for finite $N$ and $k$. 
Furthermore, the coset components of spin-$1$ and 
spin-$\frac{1}{2}$ currents in the nonlinear version 
satisfy
 following OPEs
\bea
i {\bf U}(z)   \left(
\begin{array}{c}
Q^{\bar{a}}  \\
V^{\bar{a}}
 \end{array}
\right)(w)&=&\frac{1}{(z-w)} \left[ - \frac{1}{2} \left(
\begin{array}{c}
Q^{\bar{a}}  \\
V^{\bar{a}}
 \end{array}
\right) \right](w) + \cdots, 
\quad \bar{a}=1,2, \cdots, N, (N+1)^{\ast},
 (N+2)^{\ast}, \cdots,  (2N)^{\ast},
\nonu \\
i {\bf U}(z)   \left(
\begin{array}{c}
Q^{\bar{b}}  \\
V^{\bar{b}}
 \end{array}
\right)(w)&=&\frac{1}{(z-w)} \left[ \frac{1}{2} \left(
\begin{array}{c}
Q^{\bar{b}}  \\
V^{\bar{b}}
 \end{array}
\right) \right](w) + \cdots, 
\quad \bar{b}=1^{\ast},2^{\ast}, \cdots, N^{\ast},
 N+1,
 N+2, \cdots, 2N.
\nonu
\eea
We can obtain the ${\bf U}$-charges of $|(0;v)>_{\pm \pm}$ states
from the above OPEs between ${\bf U}(z)$ and $Q^{\bar{a}}(w)$.
}
: 
\bea
{\bf u}(v;0)_{a}  & = & {\bf u}(0;v)_{a} =   -\frac{1}{2} , 
\qquad
{\bf u}(v;0)_{b}   =    {\bf u}(0;v)_{b} =  \frac{1}{2},
\label{eigenvalueu}
\eea
where $a=++,- -$ and $b=+-,-+$. 
We need to know the value of ${\bf u}^2$ in this section
and 
we have 
${\bf u}^2 (v;0)={\bf u}^2 (0;v)=\frac{1}{4}$ 
for all $(v;0)$ and $(0;v)$ states.

From the Goddard-Schwimmer formula \cite{GS},
the following relation satisfies
\bea
{\bf T}_0 |(v;0)>
&\sim& \left[ \hat{T}-\frac{1}{(k+N+2)} 
 {\bf UU} \right]_0 |(v;0)>
 \nonu \\
&=& \left[ h(v;0)+\frac{1}{(k+N+2)} 
 {\bf u}^2 (v;0) \right]  |(v;0)>
 \nonu \\
 &=&
 \left[ \frac{(N+2)}{2(k+N+2)} \right] |(v;0)>.
\label{tv0linear}
\eea
In the first line of (\ref{tv0linear}), 
the spin-$\frac{1}{2}$ current-dependent 
terms are ignored as in (\ref{hvzero}).  
In the second line, the result $ h(v;0) = \frac{(2N+3)}{4(k+N+2)}$
appearing in (\ref{hvzero2}) 
is substituted  
and the fact  that  $ {\bf u}^2 (v;0)=\frac{1}{4} $ is used.

Because the OPE between 
$  {\bf F^{a}} (z)$ and $ Q^{\bar{a}}(w)$ is regular, 
$ \pa {\bf F^{a} F_{a}}(z)$ term  in the precise 
relation between the stress energy tensors in the nonlinear and linear 
versions (\ref{16expression}) does not contribute 
to the eigenvalue equation.
Then  we obtain  
the zero mode eigenvalue  equation of ${\bf T}(z) $ for the state $|(0;v)>$
as follows:
\bea
{\bf T}_0 |(0;v)>
&\sim& \left[ \hat{T}-\frac{1}{(k+N+2)} 
 {\bf UU} \right]_0 |(0;v)>
 \nonu \\
&=& \left[ h(0;v)+\frac{1}{(k+N+2)} 
 {\bf u}^2 (0;v) \right]  |(0;v)>
 \nonu \\
 &=&
 \left[ \frac{(k+2)}{2(k+N+2)} \right] |(0;v)>.
\label{t0vlinear}
\eea
In the first line of (\ref{t0vlinear}), the trivial contribution described 
before is ignored.
In the second line, the result $ h(0;v) = \frac{(2k+3)}{4(k+N+2)}$
appearing in (\ref{hzerov}) 
is substituted  
and the fact  that  $ {\bf u}^2 (0;v)=\frac{1}{4} $ is used.
As we expect, there exists $N \leftrightarrow k$  
symmetry 
between the  eigenvalues in (\ref{tv0linear}) and 
(\ref{t0vlinear}) because the nonlinear version has this symmetry and 
the extra term coming from ${\bf u}^2$ preserves this symmetry as 
above.

The large $N$ limit (\ref{limit}) for (\ref{tv0linear}) 
and (\ref{t0vlinear}) leads to
\bea
{\bf T}_0 |(v;0)> & = & \frac{1}{2} \lambda  |(v;0)>, \qquad
{\bf T}_0 |(0;v)>  =  \frac{1}{2} (1-\lambda)  |(0;v)>,
\label{Expexp}
\eea
which are  exactly the same as the ones  in the nonlinear version.
Because there are no $N$-dependence in the ${\bf U}$-charge of
 $(v;0)$ and $(0;v)$, the second terms in ($\ref{tv0linear}$) and 
 ($\ref{t0vlinear}$) behave as $\frac{1}{N}$. Therefore the second terms
 vanish in the large 
$N$ 't Hooft limit.
We obtain the equations (\ref{Expexp}). 
  
\subsection{  Eigenvalue equations for the 
higher spin currents of spins $2, 3, 4$}

As in the nonlinear version, we can analyze the three-point functions 
for the higher spin currents. 

\subsubsection{  Eigenvalue equations for the higher spin-$2,3$ currents}

Because the higher spin-$2$ current ${\bf T}^{(2)}(z)$ in the 
linear version is the same as the   
higher spin-$2$ current $T^{(2)}(z)$ in the nonlinear version,
we have 
the equations (\ref{T2eigenvalue}) and (\ref{t2eigen}). 

Although the six higher spin-$3$ currents in the linear version 
are not the same as
the corresponding higher spin-$3$ currents in the nonlinear version, 
their eigenvalues for the states $|(v;0)>$ and $|(0;v)>$ are exactly the 
same. 
Then, we have 
(\ref{T3eigen}) with ${\bf T_0^{(3)}}$, (\ref{W3eigen}) 
with ${\bf W_0^{(3)}}$ and (\ref{spin3eigen}) with ${\bf T_0^{(3)}}$ and 
 ${\bf W_0^{(3)}}$.
We do not repeat them here.

\subsubsection{  Eigenvalue equations for the higher spin-$4$ current}

For the final higher spin-$4$ current,
the following eigenvalue equations
hold
\bea
{\bf W}^{(4)}_0 |(v;0)> &=&\left[  -\frac{6 \left(2 k N+k+4 N^2-4 N-12\right) }
{(k+N+2)^3 (15 k N+37 k+37 N+59)} \times d_3  \right] |(v;0)>,
\nonu \\
{\bf W}^{(4)}_0 |(0;v)> &=& \left[   -\frac{6 k (2 N+1) (2 k+N) }
{N (k+N+2)^3 (15 k   N+37 k+37 N+59)}  \times d_4   \right]
|(0;v)>,
\label{w4eigen}
\eea
where we introduce two factors showing $N \leftrightarrow k$ symmetry
\bea
d_3(N,k) & \equiv &  \left(6 k^2 N+16 k^2+9 k N^2+55 k N+69 k+18 N^2+64
   N+59\right),
   \nonu \\
   d_4(N,k) & \equiv & 
   \left(6 k N^2+16 N^2+9 k^2 N+55 k N+69 N+18 k^2+64 k+59\right).
\nonu
\eea
In the large $N$ 't Hooft limit, the above eigenvalue equations
(\ref{w4eigen}) lead to  
\bea
{\bf W}^{(4)}_0 |(v;0)> & = & -\frac{12}{5} \lambda (1+\lambda)(2+\lambda)
|(v;0)>,
\nonu \\
{\bf W}^{(4)}_0 |(0;v)> & = &-\frac{12}{5} (1-\lambda)(2-\lambda)(3-\lambda)
|(v;0)>.
\label{eigenspinfourlargen}
\eea

From the explicit relations (\ref{16basis}), we 
can write the above eigenvalue equations in the basis of \cite{BCG1404}.
For example, the higher spin-$4$ current which is a quasiprimary field
$V_2^{(2)}(z)$ is given by the last equation of (\ref{16basis}). Then we 
can calculate the following eigenvalue equation 
\bea
\left[ V_2^{(2)} \right]_0 |(v;0)>&=& 
-2 \left[ {\bf W}^{(4)}_0 + \frac{72(N-k)}{(37N+37k+15N k+59)}
\left( {\bf T}_0 {\bf  T}^{(2)}_0 + \frac{1}{5} {\bf T}^{(2)}_0 \right) \right]  |(v;0)>
\nonu \\
&=&\left[ \frac{12 (2 k+3 N+5) \left(2 k N+k+4 N^2-4 N-12\right)}{5 (k+N+2)^3} \right] |(v;0)>.
\label{bcgw4}
\eea
Note that 
the factor $\left[\frac{1}{(37N+37k+15N k+59)} \right]$ 
in the quasiprimary field
containing ${\bf T^{(2)}}$ 
also appears in (\ref{w4eigen}).
Compared to the previous eigenvalue equation (\ref{w4eigen}), 
the above expression (\ref{bcgw4}) is very simple because 
of the contribution from the extra zero mode in the quasiprimary field.
The factor $\left[\frac{(2 k N+k+4 N^2-4 N-12)}
{(k+N+2)^2} \right]$ in 
(\ref{bcgw4}) appears in (\ref{T3eigen}) and (\ref{W3eigen}).
Then the remaining factor $\left[\frac{(2 k+3 N+5)}{(k+N+2)}\right]$
occurs in (\ref{bcgw4}).
Similarly, for other state we have the following eigenvalue equation
\bea
\left[ V_2^{(2)} \right]_0 |(0;v)>&=& 
\left[ \frac{12(2 N+3 k+5) k (2 N+1) (2 k+N) }{5 N (k+N+2)^3} \right]
 |(0;v)>.
\label{bcgw4other}
\eea
The factor $\left[\frac{k (2 N+1) (2 k+N)}
{N (k+N+2)^2}\right]$ in (\ref{bcgw4other}) 
appears in (\ref{T3eigen}) and (\ref{W3eigen}).
Then the remaining factor $\left[\frac{(2 N+3 k+5)}{(k+N+2)}\right]$
occurs in (\ref{bcgw4other}).
In the large $N$ 't Hooft limit, the second and third terms in (\ref{bcgw4}) 
do not contribute the eigenvalue equation. Therefore,
we obtain
\bea
\left[ V_2^{(2)} \right]_0 
|(v;0)>&=& \frac{24}{5} \lambda (1+\lambda)(2+\lambda) |(v;0)>,
\nonu \\
\left[ V_2^{(2)} \right]_0 
|(0;v)>&=& \frac{24}{5}  (1-\lambda)(2-\lambda)(3-\lambda) |(0;v)>.
\label{Eigeneigen}
\eea
The eigenvalue equations (\ref{Eigeneigen}) 
have common $\la$ dependence of 
(\ref{eigenspinfourlargen}).
Therefore, the eigenvalue equations for the higher spin-$4$ currents
${\bf W^{(4)}}(z)$ and $V_2^{(2)}(z)$, under the large $N$ 't Hooft limit,
are equal  to each other up to the overall numerical factor.

As in the nonlinear version, the ratios of the three-point functions 
can be summarized by (\ref{three}) where all the higher spin currents
are replaced with the corresponding higher spin currents in the 
linear  version.

\section{Conclusions and outlook }

In this paper,
the lowest higher spin-$2$ current 
in the orthogonal $\frac{SO(N+4)}{SO(N) \times SO(4)}$ 
Wolf space coset theory for general $N$
was obtained in (\ref{T2ansatz}) and (\ref{t2coeff}).  
The remaining fifteen higher spin currents were determined implicitly
in the subsection $2.3$.
The three-point functions of bosonic 
(higher) spin currents with two scalars for finite $N$ and $k$ were
obtained.
The other type of fifteen higher spin currents together with 
the above lowest higher spin-$2$ current in the extension of the 
large ${\cal N}=4$ linear superconformal algebra
was determined implicitly in the subsection $4.2$.
The three-point functions  
of bosonic 
(higher) spin currents with two scalars for finite $N$ and $k$
were found.
Under the large $N$ 't Hooft limit, 
the two types of three-point functions in the nonlinear and linear 
versions coincided and their ratios were in (\ref{three}).

Further directions can be found as follows:

$\bullet$ Three-point function in the bulk 

It is an open problem to obtain 
the asymptotic symmetry algebra of the higher spin theory 
on the $AdS_3$ space. One of the motivations in this direction is 
to determine the three-point functions in the bulk theory 
and compare the results of this paper with them.

$\bullet$ The general spin $s$-dependence of the three-point function   

It is a good exercise to see whether we find 
the above three-point function with $s=5$ 
by considering the next higher spin currents and determine whether 
the behavior looks like that in \cite{AK1308}.
It would be interesting to obtain the three-point functions for the 
higher spin-$s$ current for general $N$ and $k$. 

$\bullet$ The operator product expansion of the $16$ higher spin currents
in ${\cal N}=4$ superspace 

It is known in \cite{AK1509} that the corresponding OPEs were
found for the unitary coset theory. It is an open problem to 
obtain the similar OPEs for the orthogonal coset theory.
We expect that the one single ${\cal N}=4$ OPE behaves differently 
compared to the unitary case because the lowest 
${\cal N}=4$ multiplet has a superspin $2$.
From the experience of \cite{AK1509}, 
it is enough to determine the basic $16$ OPEs between the 
lowest higher spin-$2$ current and the $16$ higher spin currents
living in the ${\cal N}=4$ multiplet. Moreover, the change of the higher spin
currents
is necessary to express them in $SO(4)$ symmetric way. See Appendix 
$D$ and $E$.  
Furthermore, it is an open problem to describe the ${\cal N}=4$ 
Kac-Moody algebra which generalizes the OPEs in the subsection 
$2.1$ and construct the ${\cal N}=4$ stress energy tensor (and the higher spin
${\cal N}=4$ multiplet) in terms of 
these ${\cal N}=4$ Kac-Moody currents. See also the 
${\cal N}=2$ description in 
\cite{Ahn1311}. 

$\bullet$ An extension of small ${\cal N}=4$ linear superconformal algebra

In this construction, the complete OPEs 
between the $16$ currents (of 
large ${\cal N}=4$ linear superconformal algebra) 
and the $16$ lowest higher spin currents for general $N$ and $k$
should be obtained.
In particular, the OPEs between 
the $16$ lowest higher spin currents should be determined.
After that we can take the appropriate limits.

$\bullet$  Oscillator formalism for the higher spin currents

It is an open problem to see whether we can see the oscillator formalism
in an extension of the large ${\cal N}=4$ linear superconformal algebra
in the context of the orthogonal coset theory along the line of 
\cite{GG1305}.

$\bullet$ The next $16$ higher spin currents 

We can consider the next $16$ higher spin currents, where 
the bosonic currents contain the higher spin currents with spins $3, 4, 5$.
We would like to analyze the behaviors of the three-point functions 
to determine whether they behave as what we expect. 
Furthermore, the basis in \cite{BCG1404} is more useful because 
the defining OPEs between the $16$ currents 
(in the large ${\cal N}=4$ linear superconformal algebra) and the 
next $16$ higher spin currents have already been presented. 
In the present paper, each eigenvalue equation for the 
six higher spin-$3$ currents in the nonlinear and linear versions 
has the same expression for general $N$ and $k$.
It would be interesting to observe 
this behavior for the six higher spin-$4$ currents. 

$\bullet$ Three-point functions involving the fermionic (higher spin)
currents

It would be interesting (and an open problem) 
to explicitly obtain the three-point functions 
with fermionic (higher spin) currents as raised in the unitary case. 

$\bullet$ Other approach in order to obtain the conformal dimensions of 
the orthogonal coset primaries

As described in the introduction, it is an open problem 
to obtain  the conformal dimensions of 
the orthogonal coset primaries.


\vspace{.7cm}

\centerline{\bf Acknowledgments}

This work was supported by the Mid-career Researcher Program through
the National Research Foundation of Korea (NRF) grant 
funded by the Korean government (MEST) 
(No. 2012-045385/2013-056327/2014-051185).
CA acknowledges warm hospitality from 
the School of  Liberal Arts (and Institute of Convergence Fundamental
Studies) at Seoul National University of Science and Technology.

\newpage

\appendix

\renewcommand{\thesection}{\large \bf \mbox{Appendix~}\Alph{section}}
\renewcommand{\theequation}{\Alph{section}\mbox{.}\arabic{equation}}

\section{ The coset generators with $so(N+4)$ algebra in complex basis }

In this Appendix, we present the coset generators. 
Let us focus on the $N=4n$ case. 
Based on the $N=4$ case \cite{AP1410}, we can 
rearrange them in order to describe the eigenvalue equations efficiently
in sections $3$ and $5$.
We describe them as follows:
\bea
&&T_1=
\left(\begin{array}{rrrrr|rrrr}
&&&&& 0 & -1 & 0 & 0 \\
&&&&& 0 & 0 & 0 & 0 \\
&&&&& \vdots & \vdots & \vdots & \vdots \\
&&&&& 0 & 0 & 0 & 0 \\
&&&&& 0 & 0 & 0 & 0 \\ \hline 
0 & 1 & \cdots & 0 & 0 &&&& \\
0 & 0 & \cdots & 0 & 0 &&&& \\ 
0 & 0 & \cdots & 0 & 0 &&&& \\ 
0 & 0 & \cdots & 0 & 0 &&&& \\
\end{array}\right),
\quad
T_2=
\left(\begin{array}{rrrrr|rrrr}
&&&&& 0 & 0 & 0 & 0 \\
&&&&& 0 & -1 & 0 & 0 \\
&&&&& \vdots & \vdots & \vdots & \vdots \\
&&&&& 0 & 0 & 0 & 0 \\
&&&&& 0 & 0 & 0 & 0 \\ \hline 
1 & 0 & \cdots & 0 & 0 &&&& \\
0 & 0 & \cdots & 0 & 0 &&&& \\ 
0 & 0 & \cdots & 0 & 0 &&&& \\ 
0 & 0 & \cdots & 0 & 0 &&&& \\
\end{array}\right), \cdots,
\nonu \\
&&T_N=
\left(\begin{array}{rrrrr|rrrr}
&&&&& 0 & 0 & 0 & 0 \\
&&&&& 0 & 0 & 0 & 0 \\
&&&&& \vdots & \vdots & \vdots & \vdots \\
&&&&& 0 & 0 & 0 & 0 \\
&&&&& 0 & -1 & 0 & 0 \\ \hline 
0 & 0 & \cdots & 1 & 0 &&&& \\
0 & 0 & \cdots & 0 & 0 &&&& \\ 
0 & 0 & \cdots & 0 & 0 &&&& \\ 
0 & 0 & \cdots & 0 & 0 &&&& \\
\end{array}\right), \quad
T_{N+1}=
\left(\begin{array}{rrrrr|rrrr}
&&&&& 0 & 0 & 0 & 0 \\
&&&&& 0 & 0 & 0 & -1 \\
&&&&& \vdots & \vdots & \vdots & \vdots \\
&&&&& 0 & 0 & 0 & 0 \\
&&&&& 0 & 0 & 0 & 0 \\ \hline 
0 & 0 & \cdots & 0 & 0 &&&& \\
0 & 0 & \cdots & 0 & 0 &&&& \\ 
1 & 0 & \cdots & 0 & 0 &&&& \\ 
0 & 0 & \cdots & 0 & 0 &&&& \\
\end{array}\right),
\nonu \\
&&T_{N+2}=
\left(\begin{array}{rrrrr|rrrr}
&&&&& 0 & 0 & 0 & -1 \\
&&&&& 0 & 0 & 0 &  0 \\
&&&&& \vdots & \vdots & \vdots & \vdots \\
&&&&& 0 & 0 & 0 & 0 \\
&&&&& 0 & 0 & 0 & 0 \\ \hline 
0 & 0 & \cdots & 0 & 0 &&&& \\
0 & 0 & \cdots & 0 & 0 &&&& \\ 
0 & 1 & \cdots & 0 & 0 &&&& \\ 
0 & 0 & \cdots & 0 & 0 &&&& \\
\end{array}\right), \cdots,
T_{2N}=
\left(\begin{array}{rrrrr|rrrr}
&&&&& 0 & 0 & 0 & 0 \\
&&&&& 0 & 0 & 0 &  0 \\
&&&&& \vdots & \vdots & \vdots & \vdots \\
&&&&& 0 & 0 & 0 & -1 \\
&&&&& 0 & 0 & 0 & 0 \\ \hline 
0 & 0 & \cdots & 0 & 0 &&&& \\
0 & 0 & \cdots & 0 & 0 &&&& \\ 
0 & 0 & \cdots & 0 & 1 &&&& \\ 
0 & 0 & \cdots & 0 & 0 &&&& \\
\end{array}\right).
\nonu
\eea
The nonzero component $-1$ for the first $N$ generators  
appears in the $(1,N+2)$-element, $(2,N+2)$-element, $\cdots$, 
and $(N,N+2)$-element, respectively. 
The corresponding nonzero component $1$
appears in the $(N+1, 2)$-element, 
$(N+1,1)$-element, $\cdots$, $(N+1,N)$, $(N+1,N-1)$-element,
respectively.
The nonzero component $1$ for the last $N$ generators  
appears in the $(N+3,1)$-element, $(N+3,2)$-element, $\cdots$, 
and $(N+3,N)$-element, respectively. 
The corresponding nonzero component $-1$
appears in the $(2,N+4)$-element, 
$(1,N+4)$-element, $\cdots$, $(N,N+4)$, $(N-1,N+4)$-element,
respectively.

The remaining $2N$ coset generators 
can be obtained from 
the above coset generators by transposing.
Therefore, we have the $4N$ coset generators as follows: 
\bea
T_1, \quad T_2, \quad \cdots, \quad  T_{2N}, \quad 
T^{\dagger}_{1}( \equiv T_{1^\ast}), 
\quad T^{\dagger}_{2}(\equiv T_{2^\ast}), \quad \cdots,  \quad 
T^{\dagger}_{2N}(\equiv T_{2N^\ast}).
\nonu
\eea
For the $N=4n+1$ case, we can do the similar rearrangement but we are 
not presenting them here.

In the coset theory of section $4$, the extra generators 
are located at the last $4 \times 4$ diagonal submatrix. 
We can generalize these for $N=4$ \cite{AP1410} to the general
$N$ as follows:
\bea
T_{2N+1} & = &
\left(\begin{array}{ccccc|cccc}
 0& 0 & \cdots &0 & 0 & 0 & 0 & 0 & 0 \\
 0& 0 &\cdots & 0 & 0 & 0 & 0 & 0 &  0 \\
 \vdots & \vdots & \vdots &\vdots & \vdots & 
\vdots & \vdots & \vdots & \vdots \\
 0 & 0 & \cdots &0 &0 & 0 & 0 & 0 & 0 \\
 0 & 0 & \cdots & 0 & 0 & 0 & 0 & 0 & 0 \\ \hline 
0 & 0 & \cdots & 0 & 0 & 0 & 0 & 0 & 0 \\
0 & 0 & \cdots & 0 & 0 & 0 & 0 & 0 & -1 \\ 
0 & 0 & \cdots & 0 & 0 & 1& 0& 0& 0 \\ 
0 & 0 & \cdots & 0 & 0 &0 &0 &0 & 0 \\
\end{array}\right),
\nonu \\
T_{2N+2} & = & 
\left(\begin{array}{ccccc|cccc}
 0& 0 & \cdots &0 & 0 & 0 & 0 & 0 & 0 \\
 0& 0 &\cdots & 0 & 0 & 0 & 0 & 0 &  0 \\
 \vdots & \vdots & \vdots &\vdots & \vdots & 
\vdots & \vdots & \vdots & \vdots \\
 0 & 0 & \cdots &0 &0 & 0 & 0 & 0 & 0 \\
 0 & 0 & \cdots & 0 & 0 & 0 & 0 & 0 & 0 \\ \hline 
0 & 0 & \cdots & 0 & 0 & \frac{1}{\sqrt{2}} & 0 & 0 & 0 \\
0 & 0 & \cdots & 0 & 0 & 0 & -\frac{1}{\sqrt{2}} & 0 & 0 \\ 
0 & 0 & \cdots & 0 & 0 & 0& 0& \frac{i}{\sqrt{2}}& 0 \\ 
0 & 0 & \cdots & 0 & 0 &0 &0 &0 & -\frac{i}{\sqrt{2}} \\
\end{array}\right),
\nonu \\
T_{2N+3} & = &
\left(\begin{array}{ccccc|cccc}
 0& 0 & \cdots &0 & 0 & 0 & 0 & 0 & 0 \\
 0& 0 &\cdots & 0 & 0 & 0 & 0 & 0 &  0 \\
 \vdots & \vdots & \vdots &\vdots & \vdots & 
\vdots & \vdots & \vdots & \vdots \\
 0 & 0 & \cdots &0 &0 & 0 & 0 & 0 & 0 \\
 0 & 0 & \cdots & 0 & 0 & 0 & 0 & 0 & 0 \\ \hline 
0 & 0 & \cdots & 0 & 0 & 0 & 0 & 0 & -1 \\
0 & 0 & \cdots & 0 & 0 & 0 & 0 & 0 & 0 \\ 
0 & 0 & \cdots & 0 & 0 & 0& 1& 0& 0 \\ 
0 & 0 & \cdots & 0 & 0 &0 &0 &0 & 0 \\
\end{array}\right).
\label{lineargen}
\eea
The nonzero components in (\ref{lineargen}) 
appear in the last $4\times 4$ matrices.
The remaining $N \times N$, $N \times 4$, and $4 \times N$ matrix elements 
in (\ref{lineargen})
are trivially zero. 
The remaining three generators are obtained from
the action of both transposing and complex conjugate 
$T_{(2N+1)^\ast}=T_{2N+1}^\dagger$,  
$T_{(2N+2)^\ast}=T_{2N+2}^\dagger$, and   
$T_{(2N+3)^\ast}=T_{2N+3}^\dagger$.
We can see that the generators 
$T_{2N+1}$, $T_{(2N+1)^\ast}$ and  $-\frac{(1+i)}{\sqrt{2}}
(T_{2N+2}- i T_{(2N+2)^\ast})$
consist of the $su(2)$ algebra.
Similarly, 
 the generators 
$T_{2N+3}$, $T_{(2N+3)^\ast}$ and  $\frac{(1-i)}{\sqrt{2}}
(T_{2N+2}+ i T_{(2N+2)^\ast})$
consist of the other $su(2)$ algebra.
The former is the coset generators while the latter is 
the subgroup generators of the coset theory.

\section{ The remaining next lowest higher spin currents }

In section $2$,
the four next higher spin-$\frac{7}{2}$ currents in (\ref{next16})
were obtained.
In this Appendix, the remaining $12$ next higher spin currents  in
(\ref{next16})
are obtained. 
This Appendix is kind of the defining OPEs for these higher spin currents.
Once these higher spin currents are determined explicitly, then 
we can easily describe the OPEs between the $16$ lowest higher spin currents
which will be studied in next Appendix $C$.
All the structure constants appearing in the OPEs for $N=4$ are known.
We do not present them (which are rather complicated fractional functions 
of $k$) in this paper.
For generic $N$, we expect that the structures appearing in the 
OPEs will be the same except the structure constants replaced by 
$N$-dependent expressions. 

\subsection{The 
six higher spin-$4$ currents and 
the higher spin-$3$ current}

Recall that in the unitary case \cite{Ahn1408}, the higher spin-$4$
current was obtained from the OPE between the particular 
spin-$\frac{3}{2}$ current 
and the higher spin-$\frac{7}{2}$ current which is the third component 
of the ${\cal N}=2$ multiplet (which contains the last component
as the above higher spin-$4$ current).
We can describe here similarly for the first ${\cal N}=2$ multiplet 
in (\ref{next16}).
 
Let us consider the OPE $\hat{G}_{21}(z)\,
P^{(\frac{7}{2})}_{-}(w)$ which gives the higher spin-$3$  current
$P^{(3)}(w)$ and the higher spin-$4$ current $P^{(4)}(w)$.
Recall that the higher spin-$\frac{7}{2}$ current was obtained 
from (\ref{t2tp5half}) in the section $2$.
The spin-$\frac{3}{2}$ current is given in (\ref{closedform}) with 
the footnote \ref{doubleindex}.
It turns out that 
\bea
\hat{G}_{21}(z)\,
P^{(\frac{7}{2})}_{-}(w)
 & = & \frac{1}{(z-w)^4}
 \Bigg[ c_1 \, \hat{A}_{3} + c_2 \, \hat{B}_3 \Bigg](w)
 \nonu \\
&+&\frac{1}{(z-w)^3} \Bigg[ -\frac{1}{2} \pa (\mbox{pole-4})
+  c_3 \, T^{(2)}
 +   c_4 \, \hat{T}
+  c_5 \, \hat{A}_{3}\hat{B}_{3}
\nonu \\
&+&  
c_6 \, \left(\hat{A}_{-}\hat{A}_{+}+\hat{A}_{3}\hat{A}_{3} -i\pa \hat{A}_3 
\right)
+ 
c_7 \,
\left(\hat{B}_{-}\hat{B}_{+}+\hat{B}_{3}\hat{B}_{3} -i\pa \hat{B}_3 \right) \Bigg](w)
\nonu \\
&+&\frac{1}{(z-w)^2}
 \Bigg[ 
 c_8 \, \left(\hat{T} \hat{A}_3-\frac{1}{2}\pa^2 \hat{A}_3 \right)
+ c_9 \, \left( \hat{T} \hat{B}_3-\frac{1}{2}\pa^2 \hat{B}_3 \right)
+P^{(3)} \Bigg](w)
\nonu \\
&+& \frac{1}{(z-w)} \Bigg[ \frac{1}{6} \pa  (\mbox{pole-2})
+ c_{10} \, 
\left(
\hat{T} \pa \hat{A}_3 -\frac{1}{2}\pa\hat{T} \hat{A}_3-\frac{1}{4}\pa^3 
\hat{A}_3 \right)
\nonu \\
&+& c_{11} \,
\left(
\hat{T} \pa \hat{B}_3 -\frac{1}{2}\pa\hat{T} \hat{B}_3-\frac{1}{4}\pa^3 
\hat{B}_3 \right)
+
c_{12} \, \left( \hat{T} T^{(2)} -\frac{3}{10} \pa^2 T^{(2)} \right)
\nonu \\
&+& c_{13} \, \left( 
\hat{T} \hat{T}-\frac{3}{10} \pa^2\hat{T} \right)
+ c_{14} \,
\left(
\hat{T}\hat{A}_3\hat{A}_3-\frac{3}{10}\pa^2 (\hat{A}_3\hat{A}_3) \right)
\nonu \\
&+&
c_{15} \, \left(
\hat{T}\hat{A}_3\hat{B}_3-\frac{3}{10}\pa^2 (\hat{A}_3\hat{B}_3) \right)
+ c_{16} \,
\left(
\hat{T}\hat{A}_{-}\hat{A}_{+} - \frac{3}{2} \pa\hat{A}_{-}\pa\hat{A}_{+} -
  \frac{1}{2}i \pa \hat{T}\hat{A}_3 \right)
  \nonu \\
&+& c_{17} \, 
\left(\hat{T}\hat{B}_3\hat{B}_3-\frac{3}{10}\pa^2 (\hat{B}_3\hat{B}_3) \right)
+ c_{18} \, 
 \left(
\hat{T}\hat{B}_{-}\hat{B}_{+} - \frac{3}{2} \pa\hat{B}_{-}\pa\hat{B}_{+} -
  \frac{1}{2}i \pa \hat{T}\hat{B}_3 \right)
\nonu \\
&+&P^{(4)} \Bigg](w)+\cdots.
\label{g21p7half}
\eea
We do not present all the $k$-dependent structure constants $c_1$-$c_{18}$.  
In the third-order pole, the coefficient $-\frac{1}{2}$
in the descendant field of spin-$1$ current located at the fourth-order pole
can be obtained from the standard procedure 
for given spins of the left hand side ($h_i=\frac{3}{2}$ and 
$h_j=\frac{7}{2}$) and the spin ($h_k=1$) of 
the spin-$1$ current appearing in the fourth-order pole. 
We realize that there are no new currents in the third-order pole.
There is no descendant field for the spin-$2$ field (appearing in 
the third-order pole) in the second-order pole \cite{Ahn1408} 
and we see the presence of 
higher spin-$3$ current $P^{(3)}(w)$ as well as two quasiprimary fields.
In the first-order pole, we can calculate the numerical 
coefficient $\frac{1}{6}$ ($h_k=3$) described before.
Furthermore, there exists the new higher spin-$4$ current $P^{(4)}(w)$.
In order to extract this higher spin-$4$ current, 
we should consider the correct nine quasiprimary fields. 
Most of these quasiprimary fields occurred in the unitary case 
\cite{Ahn1408} where the corresponding OPE is more complicated.

Let us calculate  OPE $\hat{G}_{21}(z) \, Q^{(\frac{7}{2})}(w)$ 
which gives the higher spin current $Q^{(4)}_{+}(w)$. 
Again this is what we expect because 
the second component of ${\cal N}=2$ stress energy tensor, 
the spin-$\frac{3}{2}$ current, 
provides the second component of the corresponding ${\cal N}=2$ multiplet
containing the first component as the above higher spin-$\frac{7}{2}$ 
current \cite{Ahn1408}. 
With the help of
(\ref{closedform}) with 
the footnote \ref{doubleindex} and (\ref{t2u5half}), we obtain the 
following OPE
\bea
\hat{G}_{21}(z)\,
Q^{(\frac{7}{2})}(w)
 & = & \frac{1}{(z-w)^4}
\Bigg[ c_1 \, \hat{B}_{-}
\Bigg](w)
 \nonu \\
&+&\frac{1}{(z-w)^3} \Bigg[ -\frac{1}{2} \pa (\mbox{pole-4})
+ c_2 \, \hat{B}_{-} \hat{A}_3 \Bigg](w)
\nonu \\
&+&\frac{1}{(z-w)^2}
 \Bigg[  
c_3 \, U^{(3)}_{+}
+ c_4 \, \hat{A}_{3}\pa\hat{B}_{-}
+ c_5 \,
\hat{B}_{3}
\pa\hat{B}_{-}
+ c_6 \, \hat{B}_{-}T^{(2)}
+ c_7 \, \hat{B}_{-}\hat{T}
\nonu \\
&+ &
 c_8 \, \hat{B}_{-}(\hat{A}_{3}\hat{A}_{3}+\hat{A}_{-}\hat{A}_{+})
 + c_9 \, 
\hat{B}_{-}(\hat{B}_{3}\hat{B}_{3}+\hat{B}_{-}\hat{B}_{+})
+
c_{10} \, \hat{B}_{-}\pa\hat{A}_{3}
\nonu \\
& + & c_{11} \,
\hat{B}_{-}\pa\hat{B}_{3}
+ c_{12} \,
\hat{G}_{21}\hat{G}_{11}
+ c_{13} \,  \pa^2 \hat{B}_{-}
\Bigg](w)
\nonu \\
&+& \frac{1}{(z-w)} \Bigg[
 \frac{1}{6} \pa (\mbox{pole-2})
+ c_{14}\, \hat{A}_3 \hat{A}_3 \pa \hat{B}_{-}
+ c_{15} \, \hat{B}_{-} \hat{A}_3 \hat{A}_3 \hat{B}_3 + 
c_{16} \, \hat{B}_{-} \hat{A}_{-}
\hat{A}_3 \hat{A}_{+} \nonu \\
& + &  c_{17} \,
\hat{B}_{-} \hat{A}_{-} \hat{B}_3 \hat{A}_{+}
+   c_{18} \, \hat{B}_{-} \hat{A}_{-} \pa \hat{A}_{+} + 
c_{19} \,  \hat{B}_{-} \hat{B}_{-}
\hat{A}_3 \hat{B}_{+} +
c_{20} \,  \hat{B}_{-} \hat{A}_{-} \hat{B}_3 \hat{A}_{+}
\nonu \\
&+& c_{21} \, \hat{B}_{-} \hat{A}_{-} \pa \hat{A}_{+} + 
c_{22} \,  \hat{B}_{-} \hat{B}_{-} \hat{A}_3 
\hat{B}_{+} + c_{23} \, \hat{B}_{-} \pa^2 \hat{A}_3
+ Q_{+}^{(4)}
\Bigg](w) + \cdots.
\label{g21q7half}
\eea
In the second-order pole of (\ref{g21q7half}), there is no new primary field.
In the first-order pole, we can see the higher spin-$4$ current
$Q^{(4)}_{+}(w)$.

Let us calculate  OPE $\hat{G}_{12}(z) \, R^{(\frac{7}{2})}(w)$  
which gives the higher spin current $R^{(4)}_{-}(w)$. 
The third component of ${\cal N}=2$ stress energy tensor 
provides the third component of the corresponding ${\cal N}=2$ multiplet
containing the first component as the above higher spin-$\frac{7}{2}$ 
current \cite{Ahn1408}. 
With the help of
(\ref{closedform}) with 
the footnote \ref{doubleindex} and (\ref{t2u5half}), we obtain the 
following OPE
\bea
\hat{G}_{12}(z)\,
R^{(\frac{7}{2})}(w)
 & = & \frac{1}{(z-w)^4}
\Bigg[ c_1 \, \hat{B}_{+}
\Bigg](w)
 \nonu \\
&+&\frac{1}{(z-w)^3} \Bigg[ -\frac{1}{2} \pa (\mbox{pole-4})
+ c_2 \,  \hat{A}_3 \hat{B}_{+} \Bigg](w)
\nonu \\
&+&\frac{1}{(z-w)^2}
 \Bigg[  
c_3 \, V^{(3)}_{-}
+ c_4 \, \hat{A}_{3}\pa\hat{B}_{+}
+ c_5 \,
\hat{B}_{3}\pa\hat{B}_{+}
+ c_6 \, T^{(2)}\hat{B}_{+}
+ c_7 \, \hat{T}\hat{B}_{+}
\nonu \\
&+ & c_8 \, 
(\hat{B}_{3}\hat{B}_{3}+\hat{B}_{-}\hat{B}_{+})\hat{B}_{+}
+ c_9 \, (\hat{A}_{3}\hat{A}_{3}+\hat{A}_{-}\hat{A}_{+})\hat{B}_{+}
+ c_{10} \, \hat{B}_{+}\pa\hat{A}_{3}
\nonu \\
& + &  c_{11} \, \hat{B}_{+}\pa\hat{B}_{3}
+ c_{12} \,  \hat{G}_{22}\hat{G}_{12}
+ c_{13} \, \pa^2 \hat{B}_{+}
\Bigg](w)
\nonu \\
&+& \frac{1}{(z-w)} \Bigg[
 \frac{1}{6} \pa (\mbox{pole-2})
  + c_{14} \, \hat{A}_{3} \hat{A}_{3}\pa \hat{B}_{+}
    +  c_{15} \, \hat{A}_{3} \hat{B}_{+}\pa \hat{A}_{3}
+  c_{16} \,  \hat{A}_{-} \hat{A}_{3}\hat{A}_{+} \hat{B}_{+} 
\nonu \\
& + & 
         c_{17} \, \hat{B}_{+} \hat{A}_{+}\pa\hat{A}_{-} +
       c_{18} \,  \hat{B}_{+} \pa^2 \hat{A}_{3}
+ R_{-}^{(4)}
\Bigg](w) + \cdots.
\label{g12r7half}
\eea
In this case, the first-order pole in (\ref{g12r7half})
gives the higher spin-$4$ current $R_{-}^{(4)}(w)$. 

Now we can consider the other spin-$\frac{3}{2}$ current in the left hand side 
of (\ref{g21q7half}). Then we obtain the following OPE
\bea
\hat{G}_{12}(z)\,
Q^{(\frac{7}{2})}(w)
 & = & \frac{1}{(z-w)^4}
 \Bigg[ c_1 \,  \hat{A}_{+}   \Bigg](w)
 \nonu \\
&+&\frac{1}{(z-w)^3} \Bigg[ -\frac{1}{2} \pa 
(\mbox{pole-4})+
c_2 \, \hat{B}_3 \hat{A}_{+}
 \Bigg](w)
\nonu
\\
&+&\frac{1}{(z-w)^2}
 \Bigg[
c_3 \, \left(
4\hat{A}_{3}\hat{A}_{3}\hat{A}_{+} + 2 i\hat{A}_{3}\pa\hat{A}_{+}  +
 \hat{A}_{-}\hat{A}_{+}\hat{A}_{+}\right)
 +\widetilde{Q}_{-}^{(3)}\Bigg](w)
\nonu \\
&+& \frac{1}{(z-w)}
\Bigg[ \frac{1}{6} \pa (\mbox{pole-2})
+  
c_4 \,  \hat{A}_{+}\hat{B}_{+}\pa \hat{B}_{-} +
    c_5 \, \hat{A}_{+}\pa^2\hat{B}_{3}
\nonu \\
& + & c_6 \, \hat{B}_3 \hat{A}_{+}\pa \hat{B}_{3}  +
    c_7 \, \hat{B}_{3}\hat{B}_{3}\pa \hat{A}_{+} + c_{8} \,
 \hat{B}_{-}\hat{B}_{3}\hat{A}_{+}\hat{B}_{+}
+Q_{-}^{(4)} \Bigg](w) +\cdots.
\label{g12q7half}
\eea
In the second-order pole of (\ref{g12q7half}), there exists 
a new higher spin-$3$ current $\widetilde{Q}_{-}^{(3)}(w)$.
It is not clear at the moment how this appears in the different 
${\cal N}=4$ multiplet. In the OPE $T^{(2)}(z) \, U_{-}^{(3)}(w)$
in next Appendix $C$, we also observe the appearance of this 
higher spin-$3$ current $\widetilde{Q}_{-}^{(3)}(w)$.
The first-order pole in (\ref{g12q7half})
gives the higher spin-$4$ current $Q_{-}^{(4)}(w)$.

Now we can consider the other spin-$\frac{3}{2}$ current in the left hand side 
of (\ref{g12r7half}). Then we obtain the following OPE
\bea
\hat{G}_{21}(z)\,
R^{(\frac{7}{2})}(w)
 & = & \frac{1}{(z-w)^4}
 \Bigg[ c_1 \,  \hat{A}_{-}   \Bigg](w)
 \nonu \\
&+&\frac{1}{(z-w)^3} \Bigg[ -\frac{1}{2} \pa 
(\mbox{pole-4})+
c_2 \, \hat{B}_3 \hat{A}_{-}
 \Bigg](w)
\nonu
\\
&+&\frac{1}{(z-w)^2}
 \Bigg[
c_3 \, \left(
 -6 \hat{A}_{3}\hat{A}_{3}\hat{A}_{-} + 4 i\hat{A}_{3}\pa\hat{A}_{-}  +
 \hat{A}_{-}\hat{A}_{-}\hat{A}_{+}
\right)
 +\widetilde{R}_{+}^{(3)}\Bigg](w)
\nonu \\
&+& \frac{1}{(z-w)}
\Bigg[ \frac{1}{6} \pa (\mbox{pole-2})
+  
c_4 \, \hat{A}_{-}\hat{B}_{+}\pa \hat{B}_{-} +
   c_5 \, \hat{A}_{-}\pa^2 \hat{B}_{3}
\nonu \\
& + &
   c_6\, \hat{B}_{3}\hat{A}_{-}\pa \hat{B}_{3} +
     c_7 \, \hat{B}_{3}\hat{B}_{3}\pa \hat{A}_{-} +
   c_8 \, \hat{B}_{-}\hat{B}_{3}\hat{A}_{-} \hat{B}_{+}
+R_{+}^{(4)} \Bigg](w) +\cdots.
\label{g21r7half}
\eea
In the second-order pole of (\ref{g21r7half}), there exists 
a new higher spin-$3$ current $\widetilde{R}_{+}^{(3)}(w)$.
It is not clear at the moment, as before, how this appears in the different 
${\cal N}=4$ multiplet. In the OPE $T^{(2)}(z) \, V_{+}^{(3)}(w)$
in next Appendix $C$, we also observe the appearance of this 
higher spin-$3$ current $\widetilde{R}_{+}^{(3)}(w)$.
The first-order pole in (\ref{g21r7half})
gives the higher spin-$4$ current $R_{+}^{(4)}(w)$.

Recall that the OPE between the spin-$\frac{3}{2}$ current
$\hat{G}_{22}(z)$
and the higher spin current living in the lowest component of 
${\cal N}=2$ multiplet gives the other higher spin current 
which belongs to the lowest component of other ${\cal N}=2$  
multiplet.
Let us consider 
the OPE $\hat{G}_{22}(z)\,
Q^{(\frac{7}{2})}(w)$ which gives  the higher spin-$4$ current $S^{(4)}(w)$
with the help of section $2$. 
We obtain the following OPE
\bea
\hat{G}_{22}(z)\,
Q^{(\frac{7}{2})}(w)
 & = & \frac{1}{(z-w)^4}
 \Bigg[ c_1 \,  \hat{A}_{3} + c_2 \, \hat{B}_3  \Bigg](w)
 \nonu \\
&+&\frac{1}{(z-w)^3} \Bigg[ -\frac{1}{2} \pa 
(\mbox{pole-4})+
c_3 \, T^{(2)}
 +   c_4 \, \hat{T}
+ c_5 \, \hat{A}_{3}\hat{B}_{3}
\nonu \\
&+& c_6 \, \left(\hat{A}_{-}\hat{A}_{+}+\hat{A}_{3}\hat{A}_{3} 
-i\pa \hat{A}_3 \right)
+ c_7 \, \left(
\hat{B}_{-}\hat{B}_{+}+\hat{B}_{3}\hat{B}_{3} -i\pa \hat{B}_3\right)
 \Bigg](w)
\nonu
\\
&+&\frac{1}{(z-w)^2}
 \Bigg[
 c_8 \, (T^{(3)}-W^{(3)})
+ c_9\, \hat{A}_3 \hat{T}
+c_{10} \, \hat{B}_3 T^{(2)}
+ c_{11} \, 
\hat{B}_3 \hat{T}
\nonu \\
&+& c_{12} \, \left(
\hat{A}_3 \hat{A}_3 \hat{B}_3+\hat{A}_{-} \hat{B}_3 \hat{A}_{+}-
   i\hat{B}_3 \pa \hat{A}_3 \right)
+ c_{13} \,
\hat{B}_3 \pa \hat{B}_3
\nonu \\
&+&
c_{14} \, \left(
\hat{B}_3 \hat{B}_3 \hat{B}_3+\hat{B}_{-} \hat{B}_3 \hat{B}_{+}\right)
+
c_{15} \, \hat{B}_{-} \pa \hat{B}_{+}
+
c_{16} \, \hat{B}_{+} \pa \hat{B}_{-}
\nonu \\
&+&c_{17} \, \left(4 \hat{A}_3 \pa \hat{A}_3 +2 \hat{A}_{-} \pa \hat{A}_{+} +
  2\hat{A}_{+} \pa \hat{A}_{-}
- (\hat{G}_{11} \hat{G}_{22} +\hat{G}_{21} \hat{G}_{12} -
      2 \pa \hat{T})\right)
      \nonu \\
&+& c_{18} \, \pa^2 \hat{B}_3
 +P^{(3)}\Bigg](w)
\nonu \\
&+& \frac{1}{(z-w)}
\Bigg[ \frac{1}{6} \pa (\mbox{pole-2})
+  
c_{19} \, \left(\hat{T} \hat{A}_{-}\hat{A}_{+} -
  \frac{3}{2} \pa \hat{A}_{-}\pa \hat{A}_{+} -
  \frac{1}{2}i\pa \hat{T} \hat{A}_3 \right)
\nonu \\
&+& c_{20} \,
\left(\hat{T} \hat{B}_{-}\hat{B}_{+} -
  \frac{3}{2} \pa \hat{B}_{-}\pa \hat{B}_{+} -
  \frac{1}{2}i\pa \hat{T} \hat{B}_3 \right)
+ c_{21} \,
\left(\hat{T} T^{(2)} -
  \frac{3}{10} \pa^2 T^{(2)} \right)
\nonu \\
&+& c_{22} \,
\left(\hat{T}\hat{T} -
  \frac{3}{10} \pa^2 \hat{T} \right)
+ c_{23} \,
\left(\hat{T}\hat{A}_3\hat{A}_3 -
  \frac{3}{10} \pa^2 (\hat{A}_3 \hat{A}_3) \right)
 \nonu \\
&+& c_{24} \,
\left(\hat{T}\hat{A}_3\hat{B}_3 -
  \frac{3}{10} \pa^2 (\hat{A}_3 \hat{B}_3) \right)
+ c_{25} \,
\left(\hat{T}\hat{B}_3\hat{B}_3 -
  \frac{3}{10} \pa^2 (\hat{B}_3 \hat{B}_3) \right)
 \nonu \\
   &+& c_{26} \,
\left(\hat{T}\pa \hat{A}_3 -
  \frac{1}{2} \pa \hat{T} \hat{A}_3-\frac{1}{4}\pa^3 \hat{A}_3\right)
 \nonu \\
    &+& c_{27} \, 
\left(\hat{T}\pa \hat{B}_3 -
  \frac{1}{2} \pa \hat{T} \hat{B}_3-\frac{1}{4}\pa^3 \hat{B}_3\right)
+P^{(4)}+S^{(4)} \Bigg](w) +\cdots.
\label{result}
\eea
Note that the higher spin-$3$ current $P^{(3)}(w)$ 
and the higher spin-$4$ current $P^{(4)}(w)$
appeared in (\ref{g21p7half}).
The quasiprimary fields appearing in the first-order pole (\ref{result}) 
occurred in the OPE (\ref{g21p7half}).

Therefore, 
the 
six higher spin-$4$ currents and 
the higher spin-$3$ current in (\ref{next16})
are determined.

\subsection{The four higher spin-$\frac{9}{2}$ currents}



Recall that the OPE (\ref{g21p7half})
provides the relation between the third component 
$P_{-}^{(\frac{7}{2})}(w)$
and the fourth component
$P^{(4)}(w)$ which live in the first ${\cal N}=2$ multiplet
in (\ref{next16}).
Now we can apply this description to the second ${\cal N}=2$ multiplet
of (\ref{next16}).
Then we consider the OPE
$\hat{G}_{21}(z)\,
Q^{(4)}_{-}(w)$ where 
the third component of the above ${\cal N}=2$ multiplet 
is taken with the same spin-$\frac{3}{2}$ current.
It turns out that
\bea
\hat{G}_{21}(z)\,
Q^{(4)}_{-}(w)
 & = & \frac{1}{(z-w)^4}
 \Bigg[ c_1 \, \hat{G}_{11} \Bigg](w)
 \nonu \\
&+&\frac{1}{(z-w)^3} \Bigg[ 
-\frac{1}{3} \pa(\mbox{pole-4})+
c_2 \, U^{(\frac{5}{2})}
+ c_3 \, \hat{B}_{-}\hat{G}_{12}
+ c_4 \, \hat{G}_{11}\hat{A}_{3}
\nonu \\
&+& c_5 \, \hat{G}_{11}\hat{B}_{3}
+ c_6 \, \hat{G}_{21}\hat{A}_{+}
+ c_7 \, \pa \hat{G}_{11}
\Bigg](w)
\nonu
\\
&+&\frac{1}{(z-w)^2}
 \Bigg[
c_8 \, \left(\hat{T}\hat{G}_{11}-\frac{3}{8}\pa^2 \hat{G}_{11} \right)
  +\widetilde{Q}^{(\frac{7}{2})} \Bigg](w)
\nonu \\
&+& \frac{1}{(z-w)} \Bigg[ 
\frac{1}{7} \pa (\mbox{pole-2})
+ 
 c_9 \, \left(\hat{T}U^{(\frac{5}{2})}-\frac{1}{4}\pa^2 U^{(\frac{5}{2})} \right)
\nonu \\
&+&  c_{10} \,
 \left(-\frac{1}{4}\hat{B}_{-}\pa^2 \hat{G}_{12}+\pa \hat{B}_{-}\pa \hat{G}_{12}
 -\frac{1}{2} \pa^2 \hat{B}_{-} \hat{G}_{12}+\frac{i}{60}\pa^3 \hat{G}_{11} \right)
\nonu \\
&+& c_{11} \,
\left(\hat{T}\pa \hat{G}_{11}-\frac{3}{4} \pa \hat{T} \hat{G}_{11}-\frac{1}{5} \pa^3 \hat{G}_{11} \right)
\nonu \\
&+ &
 c_{12} \,
\left(-\frac{1}{2}\hat{G}_{11}\pa^2 \hat{A}_{3}+\pa \hat{G}_{11}\pa \hat{A}_{3}
 -\frac{1}{4} \pa^2 \hat{G}_{11} \hat{A}_{3}+\frac{i}{30}\pa^3 \hat{G}_{11}
\right)
\nonu \\
&+& c_{13} \,
 \left(-\frac{1}{2}\hat{G}_{21}\pa^2 \hat{A}_{+}+\pa \hat{G}_{21}\pa \hat{A}_{+}
 -\frac{1}{4} \pa^2 \hat{G}_{21} \hat{A}_{+}+\frac{1}{15}\pa^3 \hat{G}_{11}
\right)
 \nonu \\
&+ &
 c_{14} \,
\left(-\frac{1}{4}\pa^2 \hat{G}_{11}\hat{B}_{3}+\pa \hat{G}_{11}\pa\hat{B}_{3}
 -\frac{1}{2} \hat{G}_{11}\pa^2 \hat{B}_{3} -\frac{i}{30}\pa^3 \hat{G}_{11}
\right)
+Q^{(\frac{9}{2})} \Bigg](w)
\nonu \\
&+&\cdots.
\label{g21q4-}
\eea
In the third-order pole of (\ref{g21q4-}), 
the coefficient $-\frac{1}{3}$
in the descendant field of spin-$\frac{3}{2}$ current 
located at the fourth-order pole
can be obtained from the standard procedure 
for given spins of the left hand side ($h_i=\frac{3}{2}$ and 
$h_j=4$) and the spin ($h_k=\frac{3}{2}$) of 
the spin-$\frac{3}{2}$ current appearing in the fourth-order pole. 
There is no descendant field for the spin-$\frac{5}{2}$ field (appearing 
in the third-order pole) in the second-order pole 
$(h_k=\frac{5}{2})$. 
In the second-order pole of (\ref{g21q4-}), there exists 
a new higher spin-$\frac{7}{2}$ current $\widetilde{Q}^{(\frac{7}{2})}(w)$.
In the OPE $T_{+}^{(\frac{5}{2})}(z) \, U_{-}^{(3)}(w)$
in next Appendix $C$, we also observe the appearance of this 
higher spin-$\frac{7}{2}$ current $\widetilde{Q}^{(\frac{7}{2})}(w)$.
In the first-order pole, the coefficient $\frac{1}{7}$
in the descendant field of spin-$\frac{7}{2}$ current located at 
the second-order pole ($h_k=\frac{7}{2}$)
can be obtained according to previous analysis.
There are various quasiprimary fields. Two of them contain 
the stress energy tensor and the remaining four  of them 
do not contain the stress energy 
tensor. 



Now we can apply the above description to the third ${\cal N}=2$ multiplet
of (\ref{next16}).
Then we consider the OPE
$\hat{G}_{21}(z)\,
R^{(4)}_{-}(w)$ where 
the third component of the above ${\cal N}=2$ multiplet 
is taken with the same spin-$\frac{3}{2}$ current.
It turns out that we obtain
\bea
\hat{G}_{21}(z)\,
R^{(4)}_{-}(w)
 & = & \frac{1}{(z-w)^4}
 \Bigg[ c_1 \, \hat{G}_{22} \Bigg](w)
 \nonu \\
&+&\frac{1}{(z-w)^3} \Bigg[ 
-\frac{1}{3} \pa(\mbox{pole-4})+
c_2 \, V^{(\frac{5}{2})}
+ c_3 \, \hat{A}_{3}\hat{G}_{22}
+ c_4 \, \hat{A}_{-}\hat{G}_{12}
+ c_5 \, \hat{B}_{3}\hat{G}_{22}
\nonu \\
&+& c_6 \, \hat{G}_{21}\hat{B}_{+}
+  c_7 \, \pa \hat{G}_{22}
\Bigg](w)
\nonu \\
&+&\frac{1}{(z-w)^2}
 \Bigg[ 
 c_8 \left(\hat{T}\hat{G}_{22}-\frac{3}{8}\pa^2 \hat{G}_{22}\right)
 +\widetilde{R}^{(\frac{7}{2})}
\Bigg](w)
\nonu \\
&+& \frac{1}{(z-w)}
\Bigg[ \frac{1}{7} \pa (\mbox{pole-2})
+ 
c_9 \, \left(\hat{T}V^{(\frac{5}{2})}-\frac{1}{4}\pa^2 V^{(\frac{5}{2})}\right)
\nonu \\
&+&  c_{10} \,\left(\hat{T}\pa\hat{G}_{22}-\frac{3}{4}\pa \hat{T}\hat{G}_{22}
-\frac{1}{5}\pa^3 \hat{G}_{22}\right)
\nonu \\
&+ & c_{11} \, \left(
-\frac{1}{4}\hat{A}_{-}\pa^2 \hat{G}_{12}+\pa \hat{A}_{-}\pa \hat{G}_{12}
 -\frac{1}{2} \pa^2 \hat{A}_{-} \hat{G}_{12}+\frac{i}{60}\pa^3 \hat{G}_{22}
\right)
\nonu \\
&+ & c_{12} \,
 \left(-\frac{1}{2}\hat{G}_{22}\pa^2 \hat{A}_{3}+\pa \hat{G}_{22}\pa \hat{A}_{3}
 -\frac{1}{4} \pa^2 \hat{G}_{22} \hat{A}_{3}-\frac{i}{30}\pa^3 \hat{G}_{22}
\right)
\nonu \\
&+ & 
c_{13} \,
\left(-\frac{1}{2}\hat{G}_{22}\pa^2 \hat{B}_{3}+\pa \hat{G}_{22}\pa \hat{B}_{3}
 -\frac{1}{4} \pa^2 \hat{G}_{22} \hat{B}_{3}+
\frac{i}{30}\pa^3 \hat{G}_{22} \right)
\nonu \\
 &+ & 
c_{14}\,
\left(
-\frac{1}{2}\hat{G}_{21}\pa^2 \hat{B}_{+}+\pa \hat{G}_{21}\pa \hat{B}_{+}
 -\frac{1}{4} \pa^2 \hat{G}_{21} \hat{B}_{+}+\frac{i}{15}\pa^3 \hat{G}_{22}
\right)
+R^{(\frac{9}{2})} \Bigg](w)
\nonu \\
&+&\cdots.
\label{g21r4-}
\eea
In the second-order pole of (\ref{g21r4-}), there exists 
a new higher spin-$\frac{7}{2}$ current $\widetilde{R}^{(\frac{7}{2})}(w)$.
In the OPE $T_{-}^{(\frac{5}{2})}(z) \, V_{+}^{(3)}(w)$
in next Appendix $C$, we also observe the appearance of this 
higher spin-$\frac{7}{2}$ current $\widetilde{R}^{(\frac{7}{2})}(w)$.
In the first-order pole, 
there are various quasiprimary fields which can be analyzed before.


Recall that the OPE (\ref{g21q7half})
provides the relation between the first component 
$Q^{(\frac{7}{2})}(w)$
and the second component
$Q_{+}^{(4)}(w)$ which live in the second ${\cal N}=2$ multiplet
in (\ref{next16}).
Now we can apply this description to the fourth ${\cal N}=2$ multiplet
of (\ref{next16}).
Then we consider the OPE
$\hat{G}_{21}(z)\,
S^{(4)}(w)$ where 
the first component of the above ${\cal N}=2$ multiplet 
is taken with the same spin-$\frac{3}{2}$ current.
Then we obtain
\bea
\hat{G}_{21}(z)\,
S^{(4)}(w)
 & = & \frac{1}{(z-w)^4}
 \Bigg[c_1 \, \hat{G}_{21}  \Bigg](w)
 \nonu \\
&+&\frac{1}{(z-w)^3} \Bigg[ 
-\frac{1}{3} \pa(\mbox{pole-4})+
 c_2 \, T^{(\frac{5}{2})}_{+}
+ c_3 \, \hat{B}_{-}\hat{G}_{22}
+ c_4 \, \hat{G}_{11}\hat{A}_{-}
\nonu \\
&+ & c_5 \, \hat{G}_{21}\hat{A}_{3}
+ c_6 \, \hat{G}_{21}\hat{B}_{3}+
 c_7 \, \pa \hat{G}_{21}
\Bigg](w)
\nonu \\
&+&\frac{1}{(z-w)^2}
 \Bigg[ 
c_8 \, \left(\hat{T}\hat{G}_{21}-\frac{3}{8}\pa^2 \hat{G}_{21} \right)
 +\widetilde{S}^{(\frac{7}{2})}_{+}
\Bigg](w)
\nonu \\
&+& \frac{1}{(z-w)} \Bigg[ 
\frac{1}{7} \pa (\mbox{pole-2})
+ c_9 \,
\left(\hat{T}T^{(\frac{5}{2})}_{+}-\frac{1}{4}\pa^2 T^{(\frac{5}{2})}_{+}\right)
\nonu \\
&+ & c_{10} \,
\left(\hat{T}\pa\hat{G}_{21}-\frac{3}{4}\pa \hat{T}\hat{G}_{21}
-\frac{1}{5}\pa^3 \hat{G}_{21}\right)
\nonu \\
&+ & c_{11}\,
\left(-\frac{1}{4}\hat{B}_{-}\pa^2 \hat{G}_{22}+\pa \hat{B}_{-}\pa \hat{G}_{22}
 -\frac{1}{2} \pa^2 \hat{B}_{-} \hat{G}_{22}-\frac{i}{60}\pa^3 \hat{G}_{21}\right)
 \nonu \\
&+ &
c_{12}\,
\left(-\frac{1}{2}\hat{G}_{21}\pa^2 \hat{A}_{3}+\pa \hat{G}_{21}\pa \hat{A}_{3}
 -\frac{1}{4} \pa^2 \hat{G}_{21} \hat{A}_{3}-\frac{i}{30}\pa^3 \hat{G}_{21}\right)
\nonu \\
&+ & c_{13}\,
\left(-\frac{1}{2}\hat{G}_{21}\pa^2 \hat{B}_{3}+\pa \hat{G}_{21}\pa \hat{B}_{3}
 -\frac{1}{4} \pa^2 \hat{G}_{21} \hat{B}_{3}-\frac{i}{30}\pa^3 \hat{G}_{21}\right)
\nonu \\
 &+ & c_{14}\,
\left(
-\frac{1}{2}\hat{G}_{11}\pa^2 \hat{A}_{-}+\pa \hat{G}_{11}\pa \hat{A}_{-}
 -\frac{1}{4} \pa^2 \hat{G}_{11} \hat{A}_{-}+\frac{i}{15}\pa^3 \hat{G}_{21}
\right)
+S^{(\frac{9}{2})}_{+} \Bigg](w)
\nonu \\
&+&\cdots.
\label{g21s4}
\eea
In the second-order pole of (\ref{g21s4}), there exists 
a new higher spin-$\frac{7}{2}$ current $\widetilde{S}_{+}^{(\frac{7}{2})}(w)$.
In the first-order pole, 
there are various quasiprimary fields which can be analyzed before.
The first-order pole in (\ref{g21s4})
gives the higher spin-$\frac{9}{2}$ current $S^{(\frac{9}{2})}_{+}(w)$.


Recall that the OPE (\ref{g12q7half})
provides the relation between the first component 
$Q^{(\frac{7}{2})}(w)$
and the third component
$Q_{-}^{(4)}(w)$ which live in the second ${\cal N}=2$ multiplet
in (\ref{next16}).
Now we can apply this description to the fourth ${\cal N}=2$ multiplet
of (\ref{next16}).
Then we consider the OPE
$\hat{G}_{12}(z)\,
S^{(4)}(w)$ where 
the first component of the above ${\cal N}=2$ multiplet 
is taken with the same spin-$\frac{3}{2}$ current.
Then we obtain
\bea
\hat{G}_{12}(z)\,
S^{(4)}(w)
 & = & \frac{1}{(z-w)^4}
  \Bigg[ c_1 \, \hat{G}_{12} \Bigg](w)
 \nonu \\
&+&\frac{1}{(z-w)^3} \Bigg[ 
-\frac{1}{3} \pa(\mbox{pole-4})+
 c_2 \, T^{(\frac{5}{2})}_{-}
+ c_3 \, \hat{A}_{3}\hat{G}_{12}
+ c_4 \, \hat{A}_{+}\hat{G}_{22}
\nonu \\
&+ & c_5 \, \hat{B}_{3}\hat{G}_{12}
+ c_6 \, \hat{G}_{11}\hat{B}_{+}
+ c_7 \, \pa\hat{G}_{12}
\Bigg](w)
\nonu \\
&+&\frac{1}{(z-w)^2}
 \Bigg[ 
 c_8 \, \left(\hat{T}\hat{G}_{12}-\frac{3}{8}\pa^2 \hat{G}_{12}\right)
+\widetilde{S}^{(\frac{7}{2})}_{-}
\Bigg](w)
\nonu \\
&+& \frac{1}{(z-w)}
\Bigg[ \frac{1}{7} \pa (\mbox{pole-2})
+ 
c_9 \, 
\left(\hat{T}T^{(\frac{5}{2})}_{-}-\frac{1}{4}\pa^2 T^{(\frac{5}{2})}_{-}\right)
\nonu \\
&+ &
c_{10}
\, \left(-\frac{1}{4}\hat{A}_{+}\pa^2 \hat{G}_{22}+\pa \hat{A}_{+}\pa \hat{G}_{22}
 -\frac{1}{2} \pa^2 \hat{A}_{+} \hat{G}_{22}+\frac{i}{60}\pa^3 \hat{G}_{12}
\right)
 \nonu \\
&+ & c_{11} \, \left(
-\frac{1}{4}\hat{A}_{3}\pa^2 \hat{G}_{12}+\pa \hat{A}_{3}\pa \hat{G}_{12}
 -\frac{1}{2} \pa^2 \hat{A}_{3} \hat{G}_{12}-\frac{i}{120}\pa^3 \hat{G}_{12}
\right)
\nonu \\
&+ & c_{12} \, \left(
-\frac{1}{4}\hat{B}_{3}\pa^2 \hat{G}_{12}+\pa \hat{B}_{3}\pa \hat{G}_{12}
 -\frac{1}{2} \pa^2 \hat{B}_{3} \hat{G}_{12}-\frac{i}{120}\pa^3 \hat{G}_{12}
\right)
\nonu \\
&+ &  c_{13} \, \left(
-\frac{1}{4}\pa^2 \hat{G}_{11}\hat{B}_{+}+\pa \hat{G}_{11}\pa \hat{B}_{+}
 -\frac{1}{2}  \hat{G}_{11}\pa^2 \hat{B}_{+}-\frac{i}{15}\pa^3 \hat{G}_{12}
\right)
 \nonu \\
&+ &
 c_{14} \, \left(\hat{T}\pa\hat{G}_{12}-\frac{3}{4}\pa \hat{T}\hat{G}_{12}
-\frac{1}{5}\pa^3 \hat{G}_{12} \right)
+S^{(\frac{9}{2})}_{-} \Bigg](w)
+\cdots.
\label{g12s4}
\eea
In the second-order pole of (\ref{g12s4}), there exists 
a new higher spin-$\frac{7}{2}$ current $\widetilde{S}_{-}^{(\frac{7}{2})}(w)$.
The first-order pole in (\ref{g12s4})
gives the higher spin-$\frac{9}{2}$ current $S^{(\frac{9}{2})}_{-}(w)$.

Therefore, 
the 
four higher spin-$\frac{9}{2}$ currents  in (\ref{next16})
are determined.

\subsection{The 
higher spin-$5$ current}

Let us consider the following OPE
\bea
\hat{G}_{21}(z)\,
S^{(\frac{9}{2})}_{-}(w)
 & = & \frac{1}{(z-w)^5}
 \Bigg[c_1 \, \hat{A}_3 +c_2 \, \hat{B}_3 \Bigg](w)
 \nonu \\
&+&
\frac{1}{(z-w)^4}
  \Bigg[-\pa (\mbox{pole-5})
\nonu \\
&+& c_3 \,T^{(2)} +c_4 \, \hat{T} +c_5 \, \hat{A}_3 \hat{A}_3 +
c_6 \, \hat{A}_3 \hat{B}_3
+c_7 \, \hat{A}_{-} \hat{A}_{+}+c_8 \,\hat{B}_{3} \hat{B}_{3}
\nonu \\
&+&c_9 \, \hat{B}_{-} \hat{B}_{+}+c_{10}\, 
\pa \hat{A}_3 +c_{11} \, \pa \hat{B}_3 \Bigg](w)
 \nonu \\
&+&\frac{1}{(z-w)^3} \Bigg[
-\frac{1}{4}\pa (\mbox{pole-4})-
\frac{1}{12}\pa^2 (\mbox{pole-5})
\nonu \\
&+& c_{12} \,T^{(3)} +c_{13}\, W^{(3)} +c_{14} \, \hat{A}_3 T^{(2)} +
c_{15} \, \hat{A}_3 \hat{T}
+ c_{16} \, \hat{A}_3 \hat{A}_3\hat{A}_3
+ c_{17} \, \hat{A}_3 \hat{A}_3\hat{B}_3
\nonu \\
& + &  c_{18} \, \hat{A}_3 \hat{B}_3\hat{B}_3
+ c_{19} \, \hat{A}_3 \pa \hat{A}_3
+ c_{20} \, \hat{A}_3 \pa \hat{B}_3
+ c_{21} \, \hat{A}_{-} \hat{A}_3\hat{A}_{+}
+ c_{22} \, \hat{A}_{-} \hat{B}_3\hat{A}_{+}
\nonu \\
& + &  c_{23} \, \hat{A}_{-} \pa \hat{A}_{+}
+ c_{24} \, \hat{A}_{+} \pa \hat{A}_{-}
+ c_{25} \, \hat{B}_{3} T^{(2)}
+ c_{26} \, \hat{B}_{3} \hat{T}
+  c_{27} \, \hat{B}_{3} \hat{B}_3\hat{B}_{3}
\nonu \\
& + &  c_{28} \, \hat{B}_{3} \pa \hat{A}_{3}
+  c_{29} \, \hat{B}_{3} \pa \hat{B}_{3}
+  c_{30} \, \hat{B}_{-} \hat{A}_3\hat{B}_{+}
+  c_{31} \, \hat{B}_{-} \hat{B}_3\hat{B}_{+}
+  c_{32} \, \hat{B}_{-} \pa \hat{B}_{+}
\nonu \\
& + &   c_{33} \, \hat{B}_{+} \pa \hat{B}_{-}
+c_{34} \, \hat{G}_{11} \hat{G}_{22}
+  c_{35} \, \hat{G}_{21} \hat{G}_{12}
+  c_{36} \, \pa \hat{T}
+  c_{37} \, \pa^2 \hat{A}_{3}
\nonu \\
& + &  c_{38} \, \pa^2 \hat{B}_{3}
+ c_{39} \left(\hat{T} \hat{A}_3-\frac{1}{2}\pa^2 \hat{A}_3\right)
+  
c_{40} \, \left(\hat{T} \hat{B}_3-\frac{1}{2}\pa^2 \hat{B}_3 \right)
+ c_{41} \, P^{(3)} \Bigg](w)
\nonu \\
&+&\frac{1}{(z-w)^2}
 \Bigg[
c_{42} \, \left(
\hat{T}\pa \hat{A}_3-\frac{1}{2}\pa \hat{T}\hat{A}_3-\frac{1}{4}\pa^3 
\hat{A}_3 \right)
+c_{43} \, \left(
\hat{T}\pa \hat{B}_3-\frac{1}{2}\pa \hat{T}\hat{B}_3-\frac{1}{4}\pa^3 
\hat{B}_3 \right)
\nonu \\
&+&c_{44} \, \left(\hat{T} T^{(2)}-\frac{3}{10}\pa^2 T^{(2)} \right)
+c_{45} \, \left(\hat{T} \hat{T}-\frac{3}{10}\pa^2 \hat{T}\right)
\nonu \\
&+&c_{46} \left(
\hat{T}\hat{A}_3 \hat{A}_3-\frac{3}{10}\pa^2 (\hat{A}_3\hat{A}_3)\right)
+c_{47} \, \left(
\hat{T}\hat{A}_3 \hat{B}_3-\frac{3}{10}\pa^2 (\hat{A}_3\hat{B}_3)\right)
\nonu \\
&+&c_{48} \, 
\left(\hat{T}\hat{A}_{-}\hat{A}_{+}-\frac{3}{2}\pa \hat{A}_{-}\pa\hat{A}_{+}-\frac{i}{2}\pa \hat{T} \hat{A}_3 \right)
+c_{49} \, 
\left(\hat{T}\hat{B}_3 \hat{B}_3-\frac{3}{10}\pa^2 (\hat{B}_3\hat{B}_3)\right)
\nonu \\
&+&c_{50} \,
\left(\hat{T}\hat{B}_{-}\hat{B}_{+}-\frac{3}{2}\pa \hat{B}_{-}\pa\hat{B}_{+}-\frac{i}{2}\pa \hat{T} \hat{B}_3 \right)
+\widetilde{S}^{(4)} \Bigg](w)
\nonu \\
&+& \frac{1}{(z-w)} \Bigg[ \frac{1}{8} \pa (\mbox{pole-2})
\nonu \\
&+&
c_{51} \, \left(\hat{T}T^{(3)}-\frac{3}{14}\pa^2 T^{(3)} \right)
+c_{52} \, \left(\hat{T}W^{(3)}-\frac{3}{14}\pa^2 W^{(3)} \right)
\nonu \\
&+& c_{53} \,
\left(\hat{T}\hat{A}_3 T^{(2)}-\frac{1}{2}\pa^2 \hat{A}_3 T^{(2)}-\frac{3}{10}\hat{A}_3 \pa^2 T^{(2)}\right)
\nonu \\
&+&c_{54} \, \left
(\hat{T}\hat{A}_3 \hat{T}-\frac{1}{2}\pa^2 \hat{A}_3 \hat{T}-\frac{3}{10}\hat{A}_3 \pa^2 \hat{T} \right)
+c_{55} \, \left(
\hat{T}\hat{A}_3 \hat{A}_3 \hat{A}_3-\frac{9}{4}\pa \hat{A}_3 \hat{A}_3 \pa \hat{A}_3\right)
\nonu \\
&+& c_{56} \, 
\left(\hat{T}\hat{A}_3 \hat{A}_3 \hat{B}_3-\frac{3}{2} \hat{A}_3 \pa \hat{A}_3 \pa \hat{B}_3-\frac{3}{4} \pa \hat{A}_3 \pa \hat{A}_3 \hat{B}_3\right)
\nonu \\
&+& c_{57} \, \left
(\hat{T}\hat{A}_3 \hat{B}_3 \hat{B}_3-\frac{3}{2} \pa \hat{A}_3  \hat{B}_3 \pa \hat{B}_3-\frac{3}{4}  \hat{A}_3 \pa \hat{B}_3 \pa \hat{B}_3 \right)
\nonu \\
&+& c_{58} \, \left(
\hat{T}\hat{A}_3 \pa \hat{A}_3-\frac{1}{2} \pa\hat{T} \hat{A}_3  \hat{A}_3 -\frac{1}{2}\pa  \hat{A}_3 \pa^2 \hat{A}_3-\frac{1}{6} \pa^3 \hat{A}_3 \hat{A}_3
\right)
\nonu \\
&+& c_{59} \, \left(
\hat{T}\hat{A}_3 \pa \hat{B}_3-\frac{1}{2} \pa\hat{T} \hat{A}_3  \hat{B}_3 -\frac{1}{2}\pa  \hat{A}_3 \pa^2 \hat{B}_3-\frac{1}{6} \hat{A}_3 \pa^3 \hat{B}_3
\right)
\nonu \\
&+& c_{60} \, \left(
-\frac{3}{4} \hat{A}_{-} \pa \hat{A}_3 \pa \hat{A}_{+} +\hat{T}\hat{A}_{-}\hat{A}_{3}\hat{A}_{+}-\frac{3}{4}\pa \hat{A}_{-} \hat{A}_{3} \pa \hat{A}_{+}
-\frac{3}{4}\pa \hat{A}_{-} \pa \hat{A}_{3}  \hat{A}_{+}
\right.
\nonu \\
&-& \frac{i}{2} \pa \hat{T} \hat{A}_3 \hat{A}_3+\frac{i}{2} \pa \hat{T} \hat{A}_{-}\hat{A}_{+}
+\frac{i}{4}\pa^2 \hat{A}_3 \pa \hat{A}_3-\frac{i}{4}\pa^2 \hat{A}_{-} \pa \hat{A}_{+}+\frac{1}{10}\pa^2 \hat{T} \hat{A}_3
\nonu \\
&-& \left.
\frac{i}{24} \pa^3 \hat{A}_3 \hat{A}_3+\frac{i}{24} \pa^3 \hat{A}_{-} 
\hat{A}_{+} \right)
\nonu \\
&+&c_{61} \, \left(
-\frac{i}{24} \hat{A}_3 \pa^3 \hat{B}_3-\frac{3}{4} \hat{A}_{-} \pa \hat{B}_3 \pa \hat{A}_{+}+\hat{T} \hat{A}_{-} \hat{B}_3 \hat{A}_{+}
+\frac{i}{4}\pa \hat{A}_3 \pa^2 \hat{B}_3-\frac{3}{4} \pa \hat{A}_{-}\hat{B}_3 \pa \hat{A}_{+}
\right.
\nonu \\
&-& \left.\frac{3}{4} \pa \hat{A}_{-} \pa \hat{B}_3  \hat{A}_{+}-\frac{i}{2}\pa \hat{T} \hat{A}_3 \hat{B}_3 \right)
\nonu \\
&+& c_{62} \, \left(
\hat{A}_3 \pa \hat{A}_{-}\pa \hat{A}_{+}+\frac{1}{3}\pa^2 \hat{A}_{3} \hat{A}_{-}\hat{A}_{+}-\pa\hat{A}_{3}\pa \hat{A}_{-}\hat{A}_{+}
-\frac{1}{3}\hat{A}_{3}\hat{A}_{-}\pa^2 \hat{A}_{+} \right)
\nonu \\
&+& c_{63} \, \left(
\hat{T} \pa \hat{A}_{-}\hat{A}_{+}-\frac{1}{2}\pa \hat{T}\hat{A}_{-}\hat{A}_{+}-\frac{1}{2}\pa^2 \hat{A}_{-}\pa \hat{A}_{+}
+\frac{i}{10} \pa^2 \hat{T} \hat{A}_{3}-\frac{1}{6}\pa^3 \hat{A}_{-}\hat{A}_{+}
\right)
\nonu \\
&+&c_{64} \, \left(
\hat{T}\hat{B}_{3} T^{(2)}-\frac{1}{2}\pa^2 \hat{B}_3 T^{(2)}-\frac{3}{10} \hat{B}_3 \pa^2 T^{(2)} \right)
\nonu \\
&+& c_{65} \, \left(
\hat{T}\hat{B}_{3}\hat{T}-\frac{1}{2}\pa^2 \hat{B}_3 \hat{T}-\frac{3}{10} \hat{B}_3 \pa^2\hat{T} \right)
+ c_{66} \, \left(
\hat{T}\hat{B}_{3}\hat{B}_3 \hat{B}_3-\frac{9}{4}\pa \hat{B}_3 \hat{B}_3 \pa \hat{B}_3 \right)
\nonu \\
&+& c_{67} \, \left(
\hat{T}\hat{B}_{3}\pa \hat{A}_3-\frac{1}{2}\pa \hat{T}\hat{A}_3 \hat{B}_3-\frac{1}{2} \pa^2 \hat{A}_3 \pa \hat{B}_3 -\frac{1}{6} \pa^3 \hat{A}_3 \hat{B}_3
\right)
\nonu \\
&+& c_{68} \, \left(
\hat{T}\hat{B}_3 \pa \hat{B}_3-\frac{1}{2}\pa \hat{T} \hat{B}_3 \hat{B}_3-\frac{1}{2} \pa \hat{B}_3 \pa^2 \hat{B}_3-\frac{1}{6} \pa^3 \hat{B}_3 \hat{B}_3
\right)
\nonu \\
&+&c_{69} \, \left(
-\frac{i}{24} \hat{A}_3 \pa^3 \hat{B}_3-\frac{3}{4} \hat{B}_{-} \pa \hat{A}_3 \pa \hat{B}_{+}+\hat{T} \hat{B}_{-} \hat{A}_3 \hat{B}_{+}
+\frac{i}{4}\pa \hat{A}_3 \pa^2 \hat{B}_3-\frac{3}{4} \pa \hat{B}_{-}\hat{A}_3 \pa \hat{B}_{+} \right.
\nonu \\
&-& \left. \frac{3}{4} \pa \hat{B}_{-} \pa \hat{A}_3  \hat{B}_{+}-\frac{i}{2}\pa \hat{T} \hat{A}_3 \hat{B}_3 \right)
\nonu \\
&+& c_{70} \, \left(
-\frac{3}{4} \hat{B}_{-} \pa \hat{B}_3 \pa \hat{B}_{+} -i\hat{T}\hat{B}_{3} \pa \hat{B}_{3}+\hat{T} \hat{B}_{-} \hat{B}_3 \hat{B}_{+}
-\frac{3}{4}\pa \hat{B}_{-} \hat{B}_3 \pa \hat{B}_{+}
\right.
\nonu \\
&-&\frac{3}{4}\pa \hat{B}_{-} \pa \hat{B}_3 \hat{B}_{+}+\frac{i}{2}\pa \hat{T} \hat{B}_{-}\hat{B}_{+}+\frac{3i}{4}\pa^2 \hat{B}_3 \pa \hat{B}_3
\nonu \\
&-& \left.
\frac{i}{4} \pa^2 \hat{B}_{-} \pa \hat{B}_{+}+\frac{1}{10}\pa^2 \hat{T} \hat{B}_3 +\frac{i}{8}\pa^3 \hat{B}_3 \hat{B}_3+\frac{i}{24}\pa^3 \hat{B}_{-}\hat{B}_{+}
\right)
\nonu \\
&+& c_{71} \, \left(
\hat{B}_3 \pa \hat{B}_{-} \pa \hat{B}_{+}+\frac{1}{3} \pa^2 \hat{B}_3\hat{B}_{-} \hat{B}_{+}-\pa \hat{B}_3 \pa \hat{B}_{-} \hat{B}_{+}-\frac{1}{3} \hat{B}_3 \hat{B}_{-} \pa^2 \hat{B}_{+}\right)
\nonu \\
&+& c_{72} \, \left(
\hat{T} \pa \hat{B}_{-} \hat{B}_{+}-\frac{1}{2} \pa \hat{T} \hat{B}_{-}\hat{B}_{+}-\frac{1}{2}\pa^2 \hat{B}_{-} \pa \hat{B}_{+}
+\frac{i}{10}\pa^2 \hat{T} \hat{B}_3 -\frac{1}{6} \pa^3 \hat{B}_{-} \hat{B}_{+}
\right)
\nonu \\
&+&c_{73} \, \left(
\hat{T} \hat{G}_{11} \hat{G}_{22}-\hat{T}\pa \hat{T}-
\frac{2 i N}{3(N+2+k)}
\hat{T} \pa^2 \hat{A}_3+\frac{2ik}{3(N+2+k)}\hat{T} \pa^2 \hat{B}_3-\pa \hat{G}_{11} \pa \hat{G}_{22} \right.
\nonu \\
&-& \frac{1}{(N+2+k)}\pa \hat{T} \hat{A}_3 \hat{A}_3-
\frac{2}{(N+2+k)}\pa \hat{T} \hat{A}_3 \hat{B}_3-\frac{1}{(N+2+k)}
\pa \hat{T} \hat{A}_{-} \hat{A}_{+}
\nonu \\
& - & \frac{1}{(N+2+k)}
\pa \hat{T} \hat{B}_{3} \hat{B}_{3}
-\frac{1}{(N+2+k)}\pa \hat{T} \hat{B}_{-} \hat{B}_{+}+
\frac{i}{(N+2+k)}\pa \hat{T} \pa \hat{A}_3 
\nonu \\
& + & 
\left.
\frac{i}{(N+2+k)}\pa \hat{T}\pa  \hat{B}_3 +
\frac{1}{6}\pa^3 \hat{T}+\frac{i N}{10(N+2+k)} \pa^4 \hat{A}_3
- \frac{i k}{10(N+2+k)} \pa^4 \hat{B}_3 \right)
\nonu \\
&+&c_{74} \, \left(
\hat{T} \hat{G}_{21} \hat{G}_{12}-\hat{T}\pa \hat{T}+
\frac{2i N}{3(N+2+k)}\hat{T} \pa^2 \hat{A}_3+
\frac{2ik}{3(N+2+k)}\hat{T} \pa^2 \hat{B}_3-\pa \hat{G}_{21} \pa \hat{G}_{12}
\right.
\nonu \\
&-& \frac{1}{(N+2+k)}\pa \hat{T} \hat{A}_3 \hat{A}_3+
\frac{2}{(N+2+k)}\pa \hat{T} \hat{A}_3 \hat{B}_3-
\frac{1}{(N+2+k)}\pa \hat{T} \hat{A}_{-} \hat{A}_{+}
\nonu \\
& - & 
\frac{1}{(N+2+k)}\pa \hat{T} \hat{B}_{3} \hat{B}_{3}
-\frac{1}{(N+2+k)}\pa \hat{T} \hat{B}_{-} \hat{B}_{+}+
\frac{i}{(N+2+k)}\pa \hat{T} \pa \hat{A}_3 
\nonu \\
& + & \left.
\frac{i}{(N+2+k)}\pa \hat{T} \pa \hat{B}_3 +
\frac{1}{6}\pa^3 \hat{T}-\frac{i N}{10(N+2+k)} \pa^4 \hat{A}_3
- \frac{i k}{10(N+2+k)} \pa^4 \hat{B}_3 \right)
\nonu \\
& + &  c_{75} \, \left(
\hat{T}\pa T^{(2)}-\pa \hat{T} T^{(2)}-\frac{1}{6} \pa^3 T^{(2)} \right)
\nonu \\
&+& c_{76} \, \left(
\hat{T}\pa^2 \hat{A}_3-\frac{3}{2} \pa \hat{T} \pa \hat{A}_3+\frac{3}{10} \pa^2 \hat{T} \hat{A}_3 -\frac{3}{20} \pa^4 \hat{A}_3 \right)
\nonu \\
&+& c_{77} \, 
\left(
\hat{T}\pa^2 \hat{B}_3-\frac{3}{2} \pa \hat{T} \pa \hat{B}_3+\frac{3}{10} \pa^2 \hat{T} \hat{B}_3 -\frac{3}{20} \pa^4 \hat{B}_3 \right)
\nonu \\
&+& 
c_{78} \, \left(\hat{T}P^{(3)}-\frac{3}{14}\pa^2 P^{(3)} \right)+S^{(5)}
\Bigg](w)
+\cdots.
\label{hg21sm9half}
\eea
In the fourth-order pole, the coefficient $-1$
in the descendant field of spin-$1$ current located at the fifth-order pole
can be obtained from the standard procedure 
for given spins of the left hand side ($h_i=\frac{3}{2}$ and 
$h_j=\frac{9}{2}$) and the spin ($h_k=1$) of 
the spin-$1$ current appearing in the fifth-order pole. 
There is no descendant field for the spin-$3$ field (appearing 
in the third-order pole) in the second-order pole $(h_k=3)$. 
Furthermore, there exists a new higher spin-$4$ current 
$\widetilde{S}^{(4)}(w)$.
In the OPE $T^{(2)}(z) \, W^{(4)}(w)$
in next Appendix $C$, we also observe the appearance of this 
higher spin-$4$ current $\widetilde{S}^{(4)}(w)$.
In the first-order pole, the coefficient $\frac{1}{8}$
in the descendant field of spin-$4$ current located at 
the second-order pole ($h_k=4$)
can be obtained according to previous analysis.
There are also various quasiprimary fields. Two of them 
have their $N$-dependence on the coefficient functions.
The first-order pole in (\ref{hg21sm9half})
gives the higher spin-$5$ current $S^{(5)}(w)$.

Therefore, 
the 
higher spin-$5$ current  in (\ref{next16})
is determined.

\section{The next higher spin currents appearing in the OPEs between 
the lowest higher spin currents  }

As described before, in section $2$,
the four next higher spin-$\frac{7}{2}$ currents in (\ref{next16})
were obtained and 
in Appendix $B$, the remaining $12$ next higher spin currents  in
(\ref{next16})
were obtained.
In this Appendix, we would like to see them in the OPEs 
between the $16$ lowest higher spin currents.  
All the structure constants appearing in the OPEs for $N=4$ are known.
We do not present them (which are rather complicated fractional functions 
of $k$) in this paper.

In subsection $2.4$, 
we have seen the four higher spin-$\frac{7}{2}$ currents.
We will see how the remaining higher spin currents appear in the right hand 
side of 
OPEs between the $16$ higher spin currents. 
The lowest higher spin-$3$ current will appear at the end of this Appendix.
Then we can start with the higher spin-$4$ currents. 

$\bullet$
 The higher spin-$4$ current 
in the OPE
$T^{(2)}(z) \, U^{(3)}_{-}(w)$

Let us consider the following OPE
\bea
T^{(2)}(z)U^{(3)}_{-}(w)
 & = & \frac{1}{(z-w)^4}
  \Bigg[ c_1 \, \hat{A}_{+}\Bigg](w)
\nonu
\\
&+&\frac{1}{(z-w)^2}
 \Bigg[ c_2 \, U^{(3)}_{-}
+
c_3 \, \hat{A}_{3}\pa \hat{A}_{+}
+ c_4 \, \hat{A}_{+}\hat{T}
+ c_5 \,
\hat{A}_{+}\hat{A}_{3}\hat{A}_{3}
+ c_6 \,
\hat{A}_{+}\hat{A}_{+} \hat{A}_{-}
\nonu \\
& + &  c_7 \, \hat{A}_{+} \hat{B}_3 \hat{B}_3
+ c_8 \, \hat{A}_{+}\hat{B}_{+} \hat{B}_{-}
+ c_9 \, 
\hat{A}_{+} \pa  \hat{A}_3
+ c_{10} \,
\hat{A}_{+} \pa \hat{B}_{3}
+ c_{11} \,
\hat{B}_{3} \pa \hat{A}_{+}
\nonu \\
&+ & c_{12} \, 
\hat{G}_{11}  \hat{G}_{12}
+ c_{13} \, 
\pa^2\hat{A}_{+} -\widetilde{Q}^{(3)}_{-} \Bigg](w)
\nonu \\
&+& \frac{1}{(z-w)} \Bigg[ \frac{1}{3} \pa (\mbox{pole-2})+
c_{14} \left( \hat{T} \pa \hat{A}_{+}-\frac{1}{2}\pa \hat{T}  \hat{A}_{+}
-\frac{1}{4} \pa ^3 \hat{A}_{+}\right)
+\widetilde{Q}^{(4)}_{-}+Q^{(4)}_{-} \Bigg](w)
\nonu \\
& + & \cdots.
\label{t2u3-}
\eea
In the first-order pole of (\ref{t2u3-}), the coefficient $\frac{1}{3}$
in the descendant field of spin-$3$ current 
located at the second-order pole
can be obtained from the standard procedure 
for given spins of the left hand side ($h_i=2$ and 
$h_j=3$) and the spin ($h_k=3$) of 
the spin-$3$ current appearing in the second-order pole.
There exists a new higher spin-$4$ current $\widetilde{Q}^{(4)}_{-}(w)$.

$\bullet$
 The higher spin-$4$ current 
in the OPE
$T^{(2)}(z) \, V^{(3)}_{+}(w)$

We calculate the following OPE
\bea
T^{(2)}(z)V^{(3)}_{+}(w)
 & = & \frac{1}{(z-w)^4}
  \Bigg[ c_1 \, \hat{A}_{-} \Bigg](w)
\nonu
\\
&+&\frac{1}{(z-w)^2}
 \Bigg[ 
c_2 \, V^{(3)}_{+}
+c_3 \, \hat{A}_{3}\pa \hat{A}_{-}
+ c_4 \, \hat{A}_{-}\hat{T}
+ c_5 \, \hat{A}_{-}\hat{A}_{3}\hat{A}_{3}
+ c_6 \, \hat{A}_{-}\hat{B}_{3} \hat{B}_{3}
\nonu \\
& + &  c_7 \, \hat{A}_{-} \hat{B}_{+} \hat{B}_{-}
+ c_8 \, \hat{A}_{-} \pa \hat{A}_{3}
+ c_9 \,
\hat{A}_{-} \pa  \hat{B}_3
+ c_{10} \, \hat{A}_{+} \hat{A}_{-} \hat{A}_{-}
+ c_{11} \, \hat{B}_{3} \pa \hat{A}_{-}
\nonu \\
&+ & c_{12} \,
\hat{G}_{22}  \hat{G}_{21}
+ c_{13} \, \pa^2\hat{A}_{-}
-\widetilde{R}^{(3)}_{+} \Bigg](w)
\nonu \\
&+& 
 \frac{1}{(z-w)}
\Bigg[ \frac{1}{3} \pa (\mbox{pole-2}) 
+ c_{14} \, \left(\hat{T} \pa \hat{A}_{-}-\frac{1}{2}\pa \hat{T}  \hat{A}_{-}
-\frac{1}{4} \pa ^3 \hat{A}_{-} \right)
+\widetilde{R}^{(4)}_{+}+R^{(4)}_{+} \Bigg](w)
\nonu \\
& + & \cdots.
\label{t2v3+}
\eea
Again the first-order pole of (\ref{t2v3+})
contains the new higher spin-$4$ current $\widetilde{R}^{(4)}_{+}(w)$.

$\bullet$
 The higher spin-$4$ current
in the OPE
$T^{(\frac{5}{2})}_{+}(z) \, U^{(\frac{5}{2})}(w)$

Let us calculate the following OPE
\bea
T^{(\frac{5}{2})}_{+}(z)U^{(\frac{5}{2})}(w)
 & = & \frac{1}{(z-w)^4}
  \Bigg[ c_1 \, \hat{B}_{-}\Bigg](w)
 \nonu \\
&+&\frac{1}{(z-w)^3} \Bigg[ \frac{1}{2}\pa (\mbox{pole-4})+
c_2 \, \hat{A}_3\hat{B}_{-}
 \Bigg](w)
\nonu \\
&+&\frac{1}{(z-w)^2}
 \Bigg[ 
- 
\frac{1}{12}\pa^2 (\mbox{pole-4})+
\frac{1}{2} \pa (\mbox{pole-3})
 +
c_3 \, U^{(3)}_{+}
+ c_4 \, 
\hat{A}_{3}\hat{A}_{3}\hat{B}_{-}
\nonu \\
&+ & c_5 \, \hat{A}_{3}\pa \hat{B}_{-}
+c_6 \, 
\hat{A}_{+}\hat{A}_{-}\hat{B}_{-}
+ c_7 \, 
\hat{B}_{3}\pa \hat{B}_{-}
+ c_8 \, \hat{B}_{-} T^{(2)}
\nonu \\
&+& c_9 \, \hat{B}_{-} \hat{T}
+ c_{10} \,
\hat{B}_{-} \hat{B}_3 \hat{B}_3
+ c_{11} \,
\hat{B}_{-} \pa \hat{A}_3
+ c_{12} \,
\hat{B}_{-} \pa \hat{B}_3
\nonu \\
&+ & c_{13} \,
\hat{B}_{+} \hat{B}_{-} \hat{B}_{-}
+ c_{14} \, \hat{G}_{11} \hat{G}_{21}
+ c_{15} \,  \pa^2\hat{B}_{-}
\Bigg](w)
\nonu \\
&+& \frac{1}{(z-w)} \Bigg[ 
\frac{1}{120} \pa^3 (\mbox{pole-4})-
\frac{1}{10} \pa^2 (\mbox{pole-3})
 +\frac{1}{2} \pa (\mbox{pole-2})
\nonu \\
&+& 
c_{16} \, \hat{A}_{3}\hat{A}_{3}\hat{B}_{-}\hat{B}_3
+ c_{17} \,
\hat{A}_{3}\hat{A}_{3} \pa \hat{B}_{-}
+ c_{18} \, \hat{A}_{3}\hat{B}_{3} \pa \hat{B}_{-}
+ c_{19} \, \hat{A}_{3}\hat{B}_{-}\hat{T}
\nonu \\
&+ & c_{20} \, \hat{A}_{3}\hat{B}_{-}\pa \hat{A}_{3}
+ c_{21} \, \hat{A}_{3}\hat{B}_{-} \pa \hat{B}_{3}
+ c_{22} \, \hat{A}_{3}\hat{B}_{+}\hat{B}_{-} \hat{B}_{-}
+ c_{23} \, \hat{A}_{3}\pa^2 \hat{B}_{-}
\nonu \\
&+ &  c_{24} \, \hat{A}_{-}\hat{B}_{-} \pa \hat{A}_{+}
+ c_{25} \,
\hat{A}_{+}\hat{A}_{-}\hat{A}_3 \hat{B}_{-}
+ c_{26} \, \hat{A}_{+}\hat{A}_{-}\hat{B}_{-} \hat{B}_{3}
+ c_{27} \,  \hat{A}_{+}\hat{A}_{-}\pa \hat{B}_{-}
\nonu \\
& + & c_{28} \, \hat{A}_{+}\hat{B}_{-}\pa \hat{A}_{-}
+ c_{29} \,  \hat{B}_3 U^{(3)}_{+}
+ c_{30} \,
\hat{B}_{3}\hat{B}_{3}\pa \hat{B}_{-}
+ c_{31} \, \hat{B}_3 \hat{G}_{11}\hat{G}_{21}
\nonu \\
&+ &
 c_{32} \, \hat{B}_{3}\pa^2 \hat{B}_{-}
+ c_{33} \, \hat{B}_{-}T^{(3)}
+ c_{34} \, \hat{B}_{-}W^{(3)}
+ c_{35} \, \hat{B}_{-}\hat{B}_{3}\pa \hat{A}_{3}
\nonu \\
&+ &
c_{36} \, \hat{B}_{-}\hat{B}_{3}\pa \hat{B}_{3}
+ c_{37} \,
\hat{B}_{-}\hat{B}_{-}\pa \hat{B}_{+}
+ c_{38} \,  \hat{B}_{-}\hat{G}_{11}\hat{G}_{22}
+ c_{39} \, \hat{B}_{-}\hat{G}_{12} \hat{G}_{21}
\nonu \\
& + & c_{40} \, \hat{B}_{-}\pa T^{(2)}
+ c_{41} \,
\hat{B}_{-} \pa \hat{T}
+ c_{42} \, \hat{B}_{-}\pa^2 \hat{A}_{3}
+ c_{43} \,
\hat{B}_{-}\pa^2 \hat{B}_{3}
\nonu \\
& + & c_{44} \, 
\hat{B}_{+}\hat{B}_{-}\pa \hat{B}_{-}
+ c_{45} \,\hat{G}_{11}T^{(\frac{5}{2})}_{+}
+ c_{46} \,
\hat{G}_{11}\pa \hat{G}_{21}
+ c_{47} \, \hat{G}_{21} U^{(\frac{5}{2})}
\nonu \\
& + & c_{48} \, 
\hat{G}_{21}\pa \hat{G}_{11}
+ c_{49} \,
\pa \hat{A}_{3}\pa \hat{B}_{-}
+ c_{50} \, \pa \hat{B}_{-}T^{(2)}
+ c_{51} \, 
\pa \hat{B}_{-} \hat{T}
\nonu \\
& + & c_{52} \, \pa U^{(3)}_{+}
+  c_{53} \, \pa^3 \hat{B}_{-}
+Q^{(4)}_{+} \Bigg](w)
+\cdots.
\label{t5half+u5half}
\eea
In the third-order pole of (\ref{t5half+u5half}), 
the coefficient $\frac{1}{2}$
in the descendant field of spin-$1$ current 
located at the fourth-order pole
can be obtained from the standard procedure 
for given spins of the left hand side ($h_i=\frac{5}{2}$ and 
$h_j=\frac{5}{2}$) and the spin ($h_k=1$) of 
the spin-$1$ current appearing in the fourth-order pole.

$\bullet$
 The higher spin-$4$ current 
in the OPE
$T^{(\frac{5}{2})}_{-}(z) \, V^{(\frac{5}{2})}(w)$

Similarly we have the following OPE 
\bea
T^{(\frac{5}{2})}_{-}(z)V^{(\frac{5}{2})}(w)
 & = & \frac{1}{(z-w)^4}
  \Bigg[ c_1 \, \hat{B}_{+} \Bigg](w)
 \nonu \\
&+&\frac{1}{(z-w)^3} \Bigg[\frac{1}{2}\pa (\mbox{pole-4})+ 
c_2 \, \hat{A}_3\hat{B}_{+} \Bigg](w)
\nonu \\
&+&\frac{1}{(z-w)^2}
 \Bigg[- 
\frac{1}{12}\pa^2 (\mbox{pole-4})+
\frac{1}{2} \pa (\mbox{pole-3})
 +
c_3 \, V^{(3)}_{-}
+c_4 \, 
\hat{A}_{3}\hat{A}_{3}\hat{B}_{+}
\nonu \\
& + & c_5 \, \hat{A}_{3}\pa \hat{B}_{+}
+c_6 \,
\hat{A}_{+}\hat{A}_{-}\hat{B}_{+}
+c_7 \, \hat{B}_{3}\pa \hat{B}_{+}
+ c_8 \, \hat{B}_{+} T^{(2)}
+c_9 \, \hat{B}_{+} \hat{T}
\nonu \\
&+ & c_{10} \, \hat{B}_{+} \hat{B}_3 \hat{B}_3
+c_{11} \,
\hat{B}_{+} \hat{B}_{+} \hat{B}_{-}
+ c_{12} \,
\hat{B}_{+} \pa \hat{A}_3
+ c_{13} \,
\hat{B}_{+} \pa \hat{B}_{3}
\nonu \\
&+ & c_{14} \, \hat{G}_{22} \hat{G}_{12}
+ c_{15} \,  \pa^2\hat{B}_{+}
\Bigg](w)
\nonu \\
&+& \frac{1}{(z-w)} \Bigg[ 
\frac{1}{120} \pa^3 (\mbox{pole-4})-
\frac{1}{10} \pa^2 (\mbox{pole-3})
 +\frac{1}{2} \pa (\mbox{pole-2})
\nonu \\
&+& 
c_{16} \, \hat{A}_{3}\hat{A}_{3} \pa \hat{B}_{+}
+ c_{17} \, \hat{A}_{3}\hat{B}_{3} \pa \hat{B}_{+}
+c_{18} \,
\hat{A}_{3}\hat{B}_{+}  \hat{T}
+ c_{19} \,
\hat{A}_{3}\hat{B}_{+} \pa \hat{A}_3
\nonu \\
& + &  c_{20} \,
\hat{A}_{3}\hat{B}_{+}\pa \hat{B}_{3}
+  c_{21} \,  \hat{A}_{3} \pa^2 \hat{B}_{+}
+ c_{22}  \, \hat{A}_{-}\hat{B}_{+}\pa \hat{A}_{+}
+ c_{23} \, \hat{A}_{-} \hat{G}_{12}\hat{G}_{12}
\nonu \\
&+ &
c_{24} \, \hat{A}_{+}\hat{A}_{-} \hat{A}_3 \hat{B}_{+}
+
c_{25} \, \hat{A}_{+}\hat{A}_{-}\pa \hat{B}_{+}
+ c_{26} \,  \hat{A}_{+}\hat{B}_{+}\pa \hat{A}_{-}
+ c_{27} \hat{B}_{3}V^{(3)}_{-}
\nonu \\
& + &
c_{28} \,  \hat{B}_{3}\hat{B}_{3}\pa \hat{B}_{+}
+ c_{29} \,  \hat{B}_3 \hat{G}_{22} \hat{G}_{12}
+
c_{30} \,  \hat{B}_{3}\pa^2 \hat{B}_{+}
+ c_{31} \,  \hat{B}_{+} T^{(3)}
\nonu \\
& + &  c_{32} \,  \hat{B}_{+} W^{(3)}
+  c_{33} \, \hat{B}_{+} \hat{B}_3 \pa \hat{A}_3
+ c_{34} \,  \hat{B}_{+} \hat{B}_3 \pa \hat{B}_3
+ c_{35} \,
\hat{B}_{+}\hat{B}_{-}\pa \hat{B}_{+}
\nonu \\
& + & 
c_{36} \, \hat{B}_{+}\hat{B}_{+}\pa \hat{B}_{-}
+ c_{37} \, \hat{B}_{+}\hat{G}_{11}\hat{G}_{22}
+ c_{38} \, \hat{B}_{+}\hat{G}_{12}\hat{G}_{21}
+ c_{39} \, \hat{B}_{+}\pa T^{(2)}
\nonu \\
&+ & c_{40} \,
\hat{B}_{+} \pa \hat{T}
+ c_{41} \,
\hat{B}_{+}\pa^2 \hat{A}_{3}
+ c_{42}\,
\hat{B}_{+}\pa^2 \hat{B}_{3}
+ c_{43} \, \hat{G}_{12}V^{(\frac{5}{2})}
\nonu \\
&+ & c_{44} \,
\hat{G}_{21} \pa \hat{G}_{22}
+ c_{45} \, \hat{G}_{22}T^{(\frac{5}{2})}_{-}
+ c_{46} \,
\hat{G}_{22}\pa \hat{G}_{12}
+ c_{47}\, \pa \hat{B}_{+} T^{(2)}
\nonu \\
&+ & c_{48} \, \pa \hat{B}_{+} \hat{T}
+ c_{49} \, \pa \hat{B}_{+}\pa \hat{B}_{3}
+ c_{50} \, \pa V^{(3)}_{-}
+ c_{51}\, \pa^3 \hat{B}_{+} +R^{(4)}_{-} \Bigg](w)
\nonu \\
& + & \cdots.
\label{t5half-v5half}
\eea
We can also describe the numerical factors in the derivative terms 
in (\ref{t5half-v5half}) as before.

Therefore, we have seen the four higher spin-$4$ currents 
$Q_{\pm}^{(4)}(w)$ and $R_{\pm}^{(4)}(w)$ in (\ref{next16}). The remaining 
two higher spin-$4$ currents will appear at the end of this Appendix. 

$\bullet$ 
The higher spin-$\frac{9}{2}$ current 
in the OPE
 $T^{(2)}(z) \, W^{(\frac{7}{2})}_{+}(w)$

Let us consider the following OPE
\bea
T^{(2)}(z)\, W^{(\frac{7}{2})}_{+}(w)
 & = & \frac{1}{(z-w)^4}
 \Bigg[ c_1 \, \hat{G}_{21} \Bigg](w)
 \nonu \\
&+&\frac{1}{(z-w)^3} \Bigg[ 
c_2 \, T^{(\frac{5}{2})}_{+}
+ c_3 \, \hat{A}_{3}\hat{G}_{21}
+ c_4 \, \hat{A}_{-}\hat{G}_{11}
+ 
c_5 \, \hat{B}_{3}\hat{G}_{21}
\nonu \\
&
+ & c_6 \, 
\hat{B}_{-}\hat{G}_{22}
+c_7 \, \pa \hat{G}_{21}
 \Bigg](w)
\nonu \\
&+&\frac{1}{(z-w)^2}
 \Bigg[ \frac{1}{5} \pa(\mbox{pole-3})
+ c_8 \, W^{(\frac{7}{2})}_{+}
+ c_9 \, \hat{A}_{3}T^{(\frac{5}{2})}_{+}
+  c_{10} \,
\hat{A}_{3}\hat{A}_{3}\hat{G}_{21}
\nonu \\
& + &  c_{11} \, \hat{A}_{3}\hat{B}_{3}\hat{G}_{21}
+
c_{12} \, \hat{A}_{3}\hat{B}_{-}\hat{G}_{22}
+ c_{13} \, \hat{A}_3 \pa \hat{G}_{21}
+ c_{14} \, \hat{A}_{-} U^{(\frac{5}{2})}
\nonu \\
& + & c_{15} \, \hat{A}_{-}\hat{A}_{3}\hat{G}_{11}
+ c_{16} \, \hat{A}_{-}\hat{B}_{3}\hat{G}_{11}
+ c_{17} \, \hat{A}_{-}\hat{B}_{-}\hat{G}_{12}
+  c_{18} \, \hat{A}_{-} \pa \hat{G}_{11}
\nonu \\
& + & c_{19} \, \hat{A}_{+}\hat{A}_{-}\hat{G}_{21}
+ c_{20} \, \hat{B}_{3}T^{(\frac{5}{2})}_{+}
+ c_{21} \,
\hat{B}_{3}\hat{B}_{3}\hat{G}_{21}
+
c_{22} \, \hat{B}_{3} \pa \hat{G}_{21}
\nonu \\
& + & c_{23} \, \hat{B}_{-} V^{(\frac{5}{2})}
+ c_{24} \, \hat{B}_{-}\hat{B}_{3}\hat{G}_{22}
+ c_{25} \,
\hat{B}_{-} \pa \hat{G}_{22}
+ c_{26} \,
\hat{B}_{+}\hat{B}_{-}\hat{G}_{21}
\nonu \\
& + &
c_{27} \, \hat{G}_{11} \pa \hat{A}_{-}
+ c_{28} \, \hat{G}_{21}T^{(2)}
+ c_{29} \, 
\hat{G}_{21} \hat{T}
+ c_{30} \,
\hat{G}_{21} \pa \hat{A}_3
\nonu \\
& + &
c_{31} \, \hat{G}_{21} \pa \hat{B}_3
+
c_{32} \, \hat{G}_{22} \pa \hat{B}_{-}
+ c_{33} \, \pa T^{(\frac{5}{2})}_{+}
+ c_{34} \, \pa^2 \hat{G}_{21}
+  c_{35} \, P^{(\frac{7}{2})}_{+} \Bigg](w)
\nonu \\
&+& \frac{1}{(z-w)} \Bigg[ 
 -\frac{1}{42} \pa^2 
(\mbox{pole-3})+\frac{2}{7}\pa (\mbox{pole-2})
\nonu \\
&+& 
c_{36} \,
\left(\hat{T}T^{(\frac{5}{2})}_{+}-\frac{1}{4}\pa^2 T^{(\frac{5}{2})}_{+}\right)
\nonu \\
&+ & c_{37} \,
\left(\pa \hat{B}_{-}\pa \hat{G}_{22} -\frac{1}{2} \pa^2 \hat{B}_{-} \hat{G}_{22}-\frac{1}{4}\hat{B}_{-}\pa^2 \hat{G}_{22}-\frac{i}{60}\pa^3 \hat{G}_{21}\right)
\nonu \\
&+ & c_{38} \,
\left(\pa \hat{G}_{21}\pa \hat{A}_{3} -\frac{1}{4} \pa^2 \hat{G}_{21} \hat{A}_{3}-\frac{1}{2}\hat{G}_{21}\pa^2 \hat{A}_{3}-\frac{i}{30}\pa^3 \hat{G}_{21}\right)
\nonu \\
& + & c_{39} \, \left(\pa \hat{G}_{21}\pa \hat{B}_{3}
 -\frac{1}{4} \pa^2 \hat{G}_{21} \hat{B}_{3}-\frac{1}{2}\hat{G}_{21}\pa^2 \hat{B}_{3}-\frac{i}{30}\pa^3 \hat{G}_{21}\right)
\nonu \\
&+ & c_{40}\, \left(\pa \hat{G}_{11}\pa \hat{A}_{-}
 -\frac{1}{4} \pa^2 \hat{G}_{11} \hat{A}_{-}-\frac{1}{2}\hat{G}_{11}\pa^2 \hat{A}_{-}+\frac{i}{15}\pa^3 \hat{G}_{21}\right)
\nonu \\
&+ & c_{41} \,
\left(\hat{T}\pa\hat{G}_{21}-\frac{3}{4}\pa \hat{T}\hat{G}_{21}-\frac{1}{5}\pa^3 \hat{G}_{21}\right)
+\widetilde{S}^{(\frac{9}{2})}_{+}+S^{(\frac{9}{2})}_{+} \Bigg](w)
+\cdots.
\label{t2w7half+}
\eea
There is no descendant field for the spin-$\frac{3}{2}$ field (appearing 
in the fourth-order pole) in the third-order pole  
$(h_k=\frac{3}{2})$.
In the second-order pole of (\ref{t2w7half+}), 
the coefficient $\frac{1}{5}$
in the descendant field of spin-$\frac{5}{2}$ current 
located at the third-order pole
can be obtained from the standard procedure 
for given spins of the left hand side ($h_i=2$ and 
$h_j=\frac{7}{2}$) and the spin ($h_k=\frac{5}{2}$) of 
the spin-$\frac{5}{2}$ current appearing in the third-order pole. 
In the first-order pole, the coefficient $\frac{2}{7}$
in the descendant field of spin-$\frac{7}{2}$ current located at 
the second-order pole ($h_k=\frac{7}{2}$)
can be obtained according to previous analysis.
Note that there exists a new higher spin-$\frac{9}{2}$ current
$\widetilde{S}^{(\frac{9}{2})}_{+}(w)$ which belongs to other 
${\cal N}=4$ multiplet.
We have seen the various quasiprimary fields appearing in the 
first-order pole before.

$\bullet$
 The higher spin-$\frac{9}{2}$ current 
in the OPE
$T^{(2)}(z) \, W^{(\frac{7}{2})}_{-}(w)$

Similarly we consider the following OPE
\bea
T^{(2)}(z)W^{(\frac{7}{2})}_{-}(w)
 & = & \frac{1}{(z-w)^4}
  \Bigg[ c_1 \, \hat{G}_{12}  \Bigg](w)
 \nonu \\
&+&\frac{1}{(z-w)^3} \Bigg[
c_2 \, T_{-}^{\frac{5}{2}}+
c_3\,  \hat{A}_{3}\hat{G}_{12}
+c_4 \, \hat{A}_{+}\hat{G}_{22}
+ c_5 \, \hat{B}_{3}\hat{G}_{12}
+ c_6 \, \hat{B}_{+}\hat{G}_{11}
+  c_7 \, \pa \hat{G}_{12}
 \Bigg](w)
\nonu \\
&+&\frac{1}{(z-w)^2}
 \Bigg[ \frac{1}{5} \pa(\mbox{pole-3})+
c_8 \,  W^{(\frac{7}{2})}_{-}
+ c_9 \, \hat{A}_{3}T^{(\frac{5}{2})}_{-}
+
c_{10} \, \hat{A}_{3}\hat{A}_{3}\hat{G}_{12}
+ c_{11} \, \hat{A}_{3}\hat{B}_{3}\hat{G}_{12}
\nonu \\
&+ & c_{12} \, \hat{A}_{3}\hat{B}_{+}\hat{G}_{11}
+ c_{13} \, \hat{A}_3 \pa \hat{G}_{12}
+ c_{14} \, \hat{A}_{+} V^{(\frac{5}{2})}
+ c_{15} \, \hat{A}_{+}\hat{A}_{3}\hat{G}_{22}
\nonu \\
&+ & c_{16} \,
\hat{A}_{+}\hat{A}_{-}\hat{G}_{12}
+ c_{17} \, \hat{A}_{+}\hat{B}_{3}\hat{G}_{22}
+ c_{18} \, \hat{A}_{+}\hat{B}_{+} \hat{G}_{21}
+ c_{19} \, \hat{A}_{+}\pa \hat{G}_{22}
\nonu \\
&+ & c_{20} \, \hat{B}_{3}T^{(\frac{5}{2})}_{-}
+ c_{21} \,
\hat{B}_{3}\hat{B}_{3}\hat{G}_{12}
+ c_{22} \,
\hat{B}_{3} \pa \hat{G}_{12}
+ c_{23} \, \hat{B}_{+} U^{(\frac{5}{2})}
\nonu \\
&+ & c_{24} \, \hat{B}_{+}\hat{B}_{3}\hat{G}_{11}
+ c_{25} \,
\hat{B}_{+}\hat{B}_{-}\hat{G}_{12}
+ c_{26} \,
\hat{B}_{+} \pa \hat{G}_{11}
+ c_{27}\,
\hat{G}_{11} \pa \hat{B}_{+}
\nonu \\
&+ & c_{28} \, \hat{G}_{12}T^{(2)}
+ c_{29} \,
\hat{G}_{12} \hat{T}
+ c_{30}\,
\hat{G}_{12} \pa \hat{A}_3
+ c_{31} \,
\hat{G}_{12} \pa \hat{B}_3
\nonu \\
&+ & c_{32} \,
\hat{G}_{22} \pa \hat{A}_{+}
+ c_{33} \,
\pa T^{(\frac{5}{2})}_{-}
+ c_{34} \, \pa^2 \hat{G}_{12}
+ c_{35} \, P^{(\frac{7}{2})}_{-} \Bigg](w)
\nonu \\
&+& \frac{1}{(z-w)} \Bigg[ -\frac{1}{42} \pa^2 
(\mbox{pole-3})+\frac{2}{7}\pa (\mbox{pole-2})
\nonu \\
&+& 
c_{36} \, 
\left(\hat{T}T^{(\frac{5}{2})}_{-}-\frac{1}{4}\pa^2 T^{(\frac{5}{2})}_{-}\right)
\nonu \\
&+ &
c_{37} \,
\left(\pa \hat{A}_{+}\pa \hat{G}_{22} -\frac{1}{2} \pa^2 \hat{A}_{+} \hat{G}_{22}-\frac{1}{4}\hat{A}_{+}\pa^2 \hat{G}_{22}+\frac{i}{60}\pa^3 \hat{G}_{12}\right)
\nonu \\
&+ &
c_{38} \, 
\left(\pa \hat{A}_{3}\pa \hat{G}_{12} -\frac{1}{4} \hat{A}_{3}\pa^2 \hat{G}_{12} -\frac{1}{2}\pa^2 \hat{A}_{3}\hat{G}_{12}-\frac{i}{120}\pa^3 \hat{G}_{12}\right)
\nonu \\
&+ &
c_{39} \,
\left(\pa \hat{B}_{3}\pa \hat{G}_{12}
 -\frac{1}{4} \hat{B}_{3} \pa^2 \hat{G}_{12}-\frac{1}{2}\pa^2 \hat{B}_{3}\hat{G}_{12}-\frac{i}{120}\pa^3 \hat{G}_{12}\right)
\nonu \\
&+ &
c_{40} \,
\left(\pa \hat{G}_{11}\pa \hat{B}_{+}
 -\frac{1}{4} \pa^2 \hat{G}_{11} \hat{B}_{+}-\frac{1}{2}\hat{G}_{11}\pa^2 \hat{B}_{+}-\frac{i}{15}\pa^3 \hat{G}_{12}\right)
\nonu \\
&+&
c_{41} \,
\left(\hat{T}\pa\hat{G}_{12}-\frac{3}{4}\pa \hat{T}\hat{G}_{12}-\frac{1}{5}\pa^3 \hat{G}_{12}\right)
+  \widetilde{S}^{(\frac{9}{2})}_{-}+S^{(\frac{9}{2})}_{-} \Bigg](w)
+\cdots.
\label{t2w7half-}
\eea
Note that there exists a new higher spin-$\frac{9}{2}$ current
$\widetilde{S}^{(\frac{9}{2})}_{-}(w)$ in the first-order pole of
(\ref{t2w7half-}) which belongs to other 
${\cal N}=4$ multiplet.
There are various quasiprimary fields.

$\bullet$ The higher spin-$\frac{9}{2}$ current
in the OPE $T^{(\frac{5}{2})}_{+}(z) \, U^{(3)}_{-}(w)$

Let us consider the OPE
\bea
T^{(\frac{5}{2})}_{+}(z)U^{(3)}_{-}(w)
 & = & \frac{1}{(z-w)^4}
\Bigg[  c_1 \,   \hat{G}_{11} \Bigg](w)
 \nonu \\
&+&\frac{1}{(z-w)^3}
\Bigg [\frac{1}{3}\pa (\mbox{pole-4})+
c_2 \, U^{(\frac{5}{2})}
+ c_3 \, \hat{A}_{3}\hat{G}_{11}
+ c_4 \,
\hat{A}_{+}\hat{G}_{21}
+ c_5 \,
\hat{B}_{3}\hat{G}_{11}
\nonu \\
& + &  c_6 \,
\hat{B}_{-}\hat{G}_{12}
+ c_7 \,
\pa \hat{G}_{11}
\Bigg](w)
\nonu \\
&+&\frac{1}{(z-w)^2}
 \Bigg[
-\frac{1}{20} \pa^2 (\mbox{pole-4})+
\frac{2}{5} \pa (\mbox{pole-3})
+ 
c_8 \, U^{(\frac{7}{2})}
+ c_9 \, \hat{A}_{3}U^{(\frac{5}{2})}
\nonu \\
& + & 
c_{10} \, \hat{A}_{3}\hat{A}_{3}\hat{G}_{11}
+ c_{11} \,
\hat{A}_{3}\hat{B}_{3}\hat{G}_{11}
+ c_{12} \,
\hat{A}_{3}\hat{B}_{-}\hat{G}_{12}
+ c_{13} \,
\hat{A}_3 \pa \hat{G}_{11}
\nonu \\
&+ &
 c_{14} \, \hat{A}_{+} T^{(\frac{5}{2})}_{+}
+ c_{15} \, \hat{A}_{+}\hat{A}_{3}\hat{G}_{21}
+
c_{16} \, \hat{A}_{+}\hat{A}_{-}\hat{G}_{11}
+
c_{17} \, \hat{A}_{+}\hat{B}_{3}\hat{G}_{21}
\nonu \\
&+ & c_{18} \, \hat{A}_{+}\hat{B}_{+} \hat{G}_{22}
+ c_{19} \,
\hat{A}_{+}\pa \hat{G}_{21}
+ c_{20} \,
\hat{B}_{3}U^{(\frac{5}{2})}
+ c_{21} \,
\hat{B}_{3}\hat{B}_{3}\hat{G}_{11}
\nonu \\
&+ & 
c_{22} \, \hat{B}_{3} \pa \hat{G}_{11}
+
c_{23} \, \hat{B}_{-} T^{(\frac{5}{2})}_{-}
+ c_{24} \,
\hat{B}_{-}\hat{B}_{3}\hat{G}_{12}
+ c_{25} \,
\hat{B}_{-}\pa \hat{G}_{12}
\nonu \\
&+ & c_{26} \,
\hat{B}_{+}\hat{B}_{-} \pa \hat{G}_{11}
+ c_{27} \, \hat{G}_{11}T^{(2)}
+ c_{28} \,
\hat{G}_{11}\hat{T}
+ c_{29} \,
\hat{G}_{11} \pa \hat{A}_3
\nonu \\
&+ &
c_{30} \, \hat{G}_{11} \pa \hat{B}_3
+ c_{31} \,
\hat{G}_{12} \pa \hat{B}_{-}
+ c_{32} \,
\hat{G}_{21} \pa \hat{A}_{+}
+ c_{33} \,
\pa U^{(\frac{5}{2})}
\nonu \\
&+ &
c_{34} \,  \pa^2 \hat{G}_{11}
+ c_{35}\, \widetilde{Q}^{(\frac{7}{2})}
+ c_{36} \, Q^{(\frac{7}{2})} \Bigg](w)
\nonu \\
&+& \frac{1}{(z-w)} \Bigg[
\frac{1}{210} \pa^3 (\mbox{pole-4})-\frac{1}{14} 
\pa^2 (\mbox{pole-3})
+ \frac{3}{7}\pa (\mbox{pole-2})
\nonu \\
&+& c_{37} \,
 \left(\hat{T}U^{(\frac{5}{2})}-\frac{1}{4}\pa^2 U^{(\frac{5}{2})} \right)
\nonu \\
&+ & c_{38} \,
 \left(\pa \hat{B}_{-}\pa \hat{G}_{12}
 -\frac{1}{2} \pa^2 \hat{B}_{-} \hat{G}_{12}-\frac{1}{4}\hat{B}_{-}\pa^2 \hat{G}_{12}+\frac{i}{60}\pa^3 \hat{G}_{11} \right)
\nonu \\
&+ & c_{39} \,
 \left(\pa \hat{G}_{11}\pa \hat{A}_{3}
 -\frac{1}{4} \pa^2 \hat{G}_{11} \hat{A}_{3}-\frac{1}{2}\hat{G}_{11}\pa^2 \hat{A}_{3}+\frac{i}{30}\pa^3 \hat{G}_{11} \right)
 \nonu \\
&+ & c_{40} \,
\left(\pa \hat{G}_{11}\pa\hat{B}_{3}
 -\frac{1}{2} \hat{G}_{11}\pa^2 \hat{B}_{3}-\frac{1}{4}\pa^2 \hat{G}_{11}\hat{B}_{3} -\frac{i}{30}\pa^3 \hat{G}_{11} \right)
\nonu \\
&+ & c_{41} \,
\left(\pa \hat{G}_{21}\pa \hat{A}_{+}
 -\frac{1}{4} \pa^2 \hat{G}_{21} \hat{A}_{+}-\frac{1}{2}\hat{G}_{21}\pa^2 \hat{A}_{+}+\frac{i}{15}\pa^3 \hat{G}_{11} \right)
 \nonu \\
&+& c_{42} \,
 \left(\hat{T}\pa \hat{G}_{11}-\frac{3}{4} \pa \hat{T} \hat{G}_{11}-\frac{1}{5} \pa^3 \hat{G}_{11} \right)
+  \widetilde{Q}^{(\frac{9}{2})}+Q^{(\frac{9}{2})} \Bigg](w)
+\cdots.
\label{t5half+u3-}
\eea
In the third-order pole of (\ref{t5half+u3-}), the coefficient $\frac{1}{3}$
in the descendant field of spin-$\frac{3}{2}$ current 
located at the fourth-order pole
can be obtained from the standard procedure 
for given spins of the left hand side ($h_i=\frac{5}{2}$ and 
$h_j=3$) and the spin ($h_k=\frac{3}{2}$) of 
the spin-$\frac{3}{2}$ current appearing in the fourth-order pole. 
In the second-order pole, the coefficient $\frac{2}{5}$
in the descendant field of spin-$\frac{5}{2}$ current 
located at the third-order pole
can be obtained.
In the first-order pole, the coefficient $\frac{3}{7}$
in the descendant field of spin-$\frac{7}{2}$ current located at 
the second-order pole ($h_k=\frac{7}{2}$)
can be obtained according to previous analysis.
Note that there exists a new higher spin-$\frac{9}{2}$ current
$\widetilde{Q}^{(\frac{9}{2})}(w)$ in the first-order pole of
(\ref{t5half+u3-}) which belongs to other 
${\cal N}=4$ multiplet.
There are various quasiprimary fields.

$\bullet$
 The higher spin-$\frac{9}{2}$ current 
in the OPE
$T^{(\frac{5}{2})}_{-}(z) \, V^{(3)}_{+}(w)$

Furthermore, we have the following OPE 
\bea
T^{(\frac{5}{2})}_{-}(z)V^{(3)}_{+}(w)
 & = & \frac{1}{(z-w)^4}
  \Bigg[ c_1 \, \hat{G}_{22}  \Bigg](w)
 \nonu \\
&+&\frac{1}{(z-w)^3} \Bigg[
\frac{1}{3}\pa (\mbox{pole-4})+
 c_2 \, V^{(\frac{5}{2})}
+ c_3 \, \hat{A}_{3}\hat{G}_{22}
+ c_4 \, \hat{A}_{-}\hat{G}_{12}
+ c_5 \, \hat{B}_{3}\hat{G}_{22}
\nonu \\
& + &  c_6 \, \hat{B}_{+}\hat{G}_{21}
+ c_7 \, \pa \hat{G}_{22}
\Bigg](w)
\nonu \\
&+&\frac{1}{(z-w)^2}
 \Bigg[ 
-\frac{1}{20} \pa^2 (\mbox{pole-4})+
\frac{2}{5} \pa (\mbox{pole-3})
\nonu \\
&+& c_8 \, 
V^{(\frac{7}{2})}
+ c_9 \, \hat{A}_{3}V^{(\frac{5}{2})}
+ c_{10} \,
\hat{A}_{3}\hat{A}_{3}\hat{G}_{22}
+ c_{11} \,
\hat{A}_{3}\hat{B}_{3}\hat{G}_{22}
+ c_{12} \,
\hat{A}_{3}\hat{B}_{+}\hat{G}_{21}
\nonu \\
&+ & c_{13} \, \hat{A}_3 \pa \hat{G}_{22}
+ c_{14} \, \hat{A}_{-} T^{(\frac{5}{2})}_{-}
+ c_{15} \, \hat{A}_{-}\hat{A}_{3}\hat{G}_{12}
+ c_{16}\, \hat{A}_{-}\hat{B}_{3}\hat{G}_{12}
\nonu \\
&+ & c_{17} \, \hat{A}_{-}\hat{B}_{+}\hat{G}_{11}
+  c_{18} \, \hat{A}_{-}\pa \hat{G}_{12}
+  c_{19} \, \hat{A}_{+}\hat{A}_{-} \hat{G}_{22}
+ c_{20} \, \hat{B}_{3}\hat{B}_{3}\hat{G}_{22}
\nonu \\
& + &  c_{21} \, \hat{B}_{3} \pa \hat{G}_{22}
+ c_{22} \, \hat{B}_{+} T^{(\frac{5}{2})}_{+}
+ c_{23} \,
\hat{B}_{+}\hat{B}_{3}\hat{G}_{21}
+ c_{24} \, \hat{B}_{+}\hat{B}_{-} \pa \hat{G}_{22}
\nonu \\
& + & c_{25} \, \hat{B}_{+} \pa \hat{G}_{21}
+ c_{26} \, \hat{G}_{12}\pa \hat{A}_{-}
+  c_{27} \, \hat{G}_{21}\pa \hat{B}_{+}
+ c_{28} \, \hat{G}_{22} T^{(2)}
\nonu \\
&+ & c_{29} \, \hat{G}_{22}\pa  \hat{T}
+ c_{30} \, \hat{G}_{22} \pa \hat{A}_{3}
+ c_{31} \, \hat{G}_{22} \pa \hat{B}_{3}
+  c_{32} \, \pa V^{(\frac{5}{2})}
\nonu \\
&+ & c_{33} \, \pa^2 \hat{G}_{22}
+ c_{34} \, \hat{B}_3 V^{(\frac{5}{2})}
+\widetilde{R}^{(\frac{7}{2})}-4R^{(\frac{7}{2})} \Bigg](w)
\nonu \\
&+& \frac{1}{(z-w)} \Bigg[ 
\frac{1}{210} \pa^3 (\mbox{pole-4})-\frac{1}{14} 
\pa^2 (\mbox{pole-3})
+ \frac{3}{7}\pa (\mbox{pole-2})
\nonu \\
&+& 
c_{35} \, \left( \hat{T}V^{(\frac{5}{2})}-\frac{1}{4}\pa^2 V^{(\frac{5}{2})}\right)
\nonu \\
&+ & c_{36} \, 
 \left(\pa \hat{A}_{-}\pa \hat{G}_{12}
 -\frac{1}{2} 
\pa^2 \hat{A}_{-} \hat{G}_{12}-\frac{1}{4}\hat{A}_{-}\pa^2 \hat{G}_{12}+
\frac{i}{60}\pa^3 \hat{G}_{22} \right)
\nonu \\
&+ & c_{37} \,
 \left(\pa \hat{G}_{22}\pa \hat{A}_{3}
 -\frac{1}{4} \pa^2 \hat{G}_{22} \hat{A}_{3}-\frac{1}{2}\hat{G}_{22}\pa^2 \hat{A}_{3}-\frac{i}{30}\pa^3 \hat{G}_{22} \right)
\nonu \\
&+ & 
 c_{38} \left(\pa \hat{G}_{22}\pa \hat{B}_{3}
 -\frac{1}{4} \pa^2 \hat{G}_{22} \hat{B}_{3}-\frac{1}{2}\hat{G}_{22}\pa^2 \hat{B}_{3}+\frac{i}{30}\pa^3 \hat{G}_{22} \right)
\nonu \\
&+ & 
 c_{39} \left(\pa \hat{G}_{21}\pa \hat{B}_{+}
 -\frac{1}{4} \pa^2 \hat{G}_{21} \hat{B}_{+}-\frac{1}{2}\hat{G}_{21}\pa^2 \hat{B}_{+}+\frac{i}{15}\pa^3 \hat{G}_{22} \right)
\nonu \\
 &+ &
 c_{40} \left(\hat{T}\pa\hat{G}_{22}-\frac{3}{4}\pa \hat{T}\hat{G}_{22}
-\frac{1}{5}\pa^3 \hat{G}_{22} \right)
+\widetilde{R}^{(\frac{9}{2})}+R^{(\frac{9}{2})} \Bigg](w)
+\cdots.
\label{tm5halfvp3}
\eea
Note that there exists a new higher spin-$\frac{9}{2}$ current
$\widetilde{R}^{(\frac{9}{2})}(w)$ in the first-order pole of
(\ref{tm5halfvp3}) which belongs to other 
${\cal N}=4$ multiplet.
There are various quasiprimary fields.

$\bullet$  The higher spin currents of spins $s=3,4,5$ in the OPE
 $T^{(2)}(z) \, W^{(4)}(w)$

Let us consider the final OPE
\bea
T^{(2)}(z)\,W^{(4)}(w)
 & = & \frac{1}{(z-w)^5}
\Bigg[ c_1 \, \hat{A}_3 +c_2 \, \hat{B}_3  \Bigg](w)
 \nonu \\
&+& \frac{1}{(z-w)^4}
 \Bigg[-\frac{1}{2} \pa  (\mbox{pole-5})+
 c_3 \, T^{(2)} +c_4  \, \hat{T} +c_5  \, \hat{A}_3 \hat{A}_3 +
c_6  \, \hat{A}_3 \hat{B}_3
+c_7 \, \hat{A}_{-} \hat{A}_{+}\nonu \\
& + & c_8 \,  \hat{B}_{3} \hat{B}_{3}
+c_9 \, \hat{B}_{-} \hat{B}_{+}+c_{10} \, \pa \hat{A}_3 +
c_{11} \,  \pa \hat{B}_3
 \Bigg](w)
\nonu \\
&+&\frac{1}{(z-w)^3} \Bigg[
c_{12} \, T^{(3)}
 +c_{13}\,  W^{(3)} +
c_{14} \, \hat{A}_3 T^{(2)} +
c_{15} \,  \hat{A}_3 \hat{T}
+c_{16} \, \hat{A}_3 \hat{A}_3\hat{A}_3
\nonu \\
& + & c_{17} \, \hat{A}_3 \hat{A}_3\hat{B}_3
+c_{18} \, \hat{A}_3 \hat{B}_3\hat{B}_3
+ c_{19} \, \hat{A}_3 \hat{B}_{+}\hat{B}_{-}
+c_{20} \, \hat{A}_3 \pa \hat{A}_3
+c_{21} \, \hat{A}_3 \pa \hat{B}_3
\nonu \\
& + & c_{22} \, \hat{A}_{-} \pa \hat{A}_{+}
+c_{23} \, \hat{A}_{+} \hat{A}_{-} \hat{A}_3
+
c_{24} \, \hat{A}_{+} \hat{A}_{-} \hat{B}_3
+c_{25} \, \hat{A}_{+}\pa \hat{A}_{-}
+c_{26} \,  \hat{B}_3 T^{(2)}
\nonu \\
& + & c_{27} \, \hat{B}_{3} \hat{T}
+c_{28} \, \hat{B}_{3}\hat{B}_{3}\hat{B}_{3}
+c_{29} \, \hat{B}_{3} \pa \hat{A}_{3}
+c_{30} \, \hat{B}_3 \pa \hat{B}_{3}
+c_{31} \, \hat{B}_{-} \pa \hat{B}_{+}
\nonu \\
& + & c_{32} \, \hat{B}_{+}  \hat{B}_{-} \hat{B}_3
+c_{33} \, \hat{B}_{+} \pa \hat{B}_{-}
+c_{34}\, \hat{G}_{11} \hat{G}_{22}
+c_{35} \, \hat{G}_{12} \hat{G}_{21}
+c_{36} \, \pa \hat{T}
\nonu \\
& + & c_{37} \, \pa^2 \hat{A}_{3}
+c_{38} \,  \pa^2 \hat{B}_{3}
+ c_{39} \, P^{(3)} \Bigg](w)
\nonu \\
&+& \frac{1}{(z-w)^2} \Bigg[\frac{1}{6} \pa (\mbox{pole-3})
\nonu \\
& + &
c_{40} \, 
\left(\hat{T}\pa \hat{A}_3-\frac{1}{2}\pa \hat{T}\hat{A}_3-\frac{1}{4}\pa^3 \hat{A}_3 \right)
+  c_{41} \, 
\left(\hat{T}\pa \hat{B}_3-\frac{1}{2}\pa \hat{T}\hat{B}_3-\frac{1}{4}\pa^3 \hat{B}_3 \right)
\nonu \\
&+&c_{42} \, \left(\hat{T} T^{(2)}-\frac{3}{10}\pa^2 T^{(2)} \right)
+c_{43} \, \left(\hat{T} \hat{T}-\frac{3}{10}\pa^2 \hat{T}\right)
\nonu \\
&+& c_{44} \, \left(\hat{T}\hat{A}_3 \hat{A}_3-\frac{3}{10}\pa^2 (\hat{A}_3
\hat{A}_3) \right)
+c_{45} \, 
\left(\hat{T}\hat{A}_3 \hat{B}_3-\frac{3}{10}\pa^2 (\hat{A}_3\hat{B}_3)
\right)
\nonu \\
&+&c_{46} \, 
\left(\hat{T}\hat{A}_{-}\hat{A}_{+}-\frac{3}{2}\pa \hat{A}_{-}\pa\hat{A}_{+}-\frac{i}{2}\pa \hat{T} \hat{A}_3\right)
+c_{47} \,
\left(\hat{T}\hat{B}_3 \hat{B}_3-\frac{3}{10}\pa^2 (\hat{B}_3\hat{B}_3)
\right)
\nonu \\
&+& c_{48} \,
 \left(\hat{T}\hat{B}_{-}\hat{B}_{+}-\frac{3}{2}\pa \hat{B}_{-}\pa\hat{B}_{+}-\frac{i}{2}\pa \hat{T} \hat{B}_3 \right)
\nonu \\
&+& P^{(4)}+\widetilde{P}^{(4)}+S^{(4)}+\widetilde{S}^{(4)} \Bigg](w)
\nonu \\
&+&\frac{1}{(z-w)} \Bigg[-\frac{1}{56} \pa^2  (\mbox{pole-3})
+ \frac{1}{4} \pa (\mbox{pole-2})
\nonu \\
&+& 
c_{49} \, \left(\hat{T}T^{(3)}-\frac{3}{14}\pa^2 T^{(3)} \right)
+c_{50}\, \left(\hat{T}W^{(3)}-\frac{3}{14}\pa^2 W^{(3)} \right)
\nonu \\
&+&c_{51} \, \left(\hat{T}\hat{A}_3 T^{(2)}-\frac{1}{2}\pa^2 \hat{A}_3 T^{(2)}-\frac{3}{10}\hat{A}_3 \pa^2 T^{(2)} \right)
\nonu \\
&+& c_{52} \, \left(
\hat{T}\hat{A}_3 \hat{T}-\frac{1}{2}\pa^2 \hat{A}_3 \hat{T}-\frac{3}{10}\hat{A}_3 \pa^2 \hat{T} \right)
+c_{53} \, 
\left(\hat{T}\hat{A}_3 \hat{A}_3 \hat{A}_3-\frac{9}{4}\pa \hat{A}_3 \hat{A}_3 \pa \hat{A}_3 \right)
\nonu \\
&+&c_{54} \,
\left(\hat{T}\hat{A}_3 \hat{A}_3 \hat{B}_3-\frac{3}{2} \hat{A}_3 \pa \hat{A}_3 \pa \hat{B}_3-\frac{3}{4} \pa \hat{A}_3 \pa \hat{A}_3 \hat{B}_3\right)
\nonu \\
&+&c_{55} \,
\left(\hat{T}\hat{A}_3 \hat{B}_3 \hat{B}_3-\frac{3}{2} \pa \hat{A}_3  \hat{B}_3 \pa \hat{B}_3-\frac{3}{4}  \hat{A}_3 \pa \hat{B}_3 \pa \hat{B}_3\right)
\nonu \\
&+& c_{56} \, \left(
\hat{T}\hat{A}_3 \pa \hat{A}_3-\frac{1}{2} \pa\hat{T} \hat{A}_3  \hat{A}_3 -\frac{1}{2}\pa  \hat{A}_3 \pa^2 \hat{A}_3-\frac{1}{6} \pa^3 \hat{A}_3 \hat{A}_3
\right)
\nonu \\
&+& c_{57} \, \left(
\hat{T}\hat{A}_3 \pa \hat{B}_3-\frac{1}{2} \pa\hat{T} \hat{A}_3  \hat{B}_3 -\frac{1}{2}\pa  \hat{A}_3 \pa^2 \hat{B}_3-\frac{1}{6} \hat{A}_3 \pa^3 \hat{B}_3
\right)
\nonu \\
&+&c_{58} \, \left(
-\frac{3}{4} \hat{A}_{-} \pa \hat{A}_3 \pa \hat{A}_{+} +\hat{T}\hat{A}_{-}\hat{A}_{3}\hat{A}_{+}-\frac{3}{4}\pa \hat{A}_{-} \hat{A}_{3} \pa \hat{A}_{+}
-\frac{3}{4}\pa \hat{A}_{-} \pa \hat{A}_{3}  \hat{A}_{+}
\right.
\nonu \\
&-& \frac{i}{2} \pa \hat{T} \hat{A}_3 \hat{A}_3+\frac{i}{2} \pa \hat{T} \hat{A}_{-}\hat{A}_{+}
+\frac{i}{4}\pa^2 \hat{A}_3 \pa \hat{A}_3-\frac{i}{4}\pa^2 \hat{A}_{-} \pa \hat{A}_{+}+\frac{1}{10}\pa^2 \hat{T} \hat{A}_3
\nonu \\
&-& \left.
\frac{i}{24} \pa^3 \hat{A}_3 \hat{A}_3+\frac{i}{24} \pa^3 \hat{A}_{-} \hat{A}_{+}\right)
\nonu \\
&+&c_{59} \, \left(
-\frac{i}{24} \hat{A}_3 \pa^3 \hat{B}_3-\frac{3}{4} \hat{A}_{-} \pa \hat{B}_3 \pa \hat{A}_{+}+\hat{T} \hat{A}_{-} \hat{B}_3 \hat{A}_{+}
+\frac{i}{4}\pa \hat{A}_3 \pa^2 \hat{B}_3-\frac{3}{4} \pa \hat{A}_{-}\hat{B}_3 \pa \hat{A}_{+} \right.
\nonu \\
&-& \left.
\frac{3}{4} \pa \hat{A}_{-} \pa \hat{B}_3  \hat{A}_{+}-\frac{i}{2}\pa \hat{T} \hat{A}_3 \hat{B}_3 \right)
\nonu \\
&+& c_{60} \, \left(
\hat{A}_3 \pa \hat{A}_{-}\pa \hat{A}_{+}+\frac{1}{3}\pa^2 \hat{A}_{3} \hat{A}_{-}\hat{A}_{+}-\pa\hat{A}_{3}\pa \hat{A}_{-}\hat{A}_{+}
-\frac{1}{3}\hat{A}_{3}\hat{A}_{-}\pa^2 \hat{A}_{+}\right)
\nonu \\
&+& c_{61} \, \left(
\hat{T} \pa \hat{A}_{-}\hat{A}_{+}-\frac{1}{2}\pa \hat{T}\hat{A}_{-}\hat{A}_{+}-\frac{1}{2}\pa^2 \hat{A}_{-}\pa \hat{A}_{+}
+\frac{i}{10} \pa^2 \hat{T} \hat{A}_{3}-\frac{1}{6}\pa^3 \hat{A}_{-}
\hat{A}_{+}\right)
\nonu \\
&+&c_{62}\, \left(
\hat{T}\hat{B}_{3} T^{(2)}-\frac{1}{2}\pa^2 \hat{B}_3 T^{(2)}-\frac{3}{10} \hat{B}_3 \pa^2 T^{(2)} \right)
\nonu \\
&+& c_{63} \, \left(
\hat{T}\hat{B}_{3}\hat{T}-\frac{1}{2}\pa^2 \hat{B}_3 \hat{T}-\frac{3}{10} \hat{B}_3 \pa^2\hat{T}\right)
+c_{64} \, \left(
\hat{T}\hat{B}_{3}\hat{B}_3 \hat{B}_3-\frac{9}{4}\pa \hat{B}_3 \hat{B}_3 \pa \hat{B}_3\right)
\nonu \\
&+& c_{65} \,  \left(
\hat{T}\hat{B}_{3}\pa \hat{A}_3-\frac{1}{2}\pa \hat{T}\hat{A}_3 \hat{B}_3-\frac{1}{2} \pa^2 \hat{A}_3 \pa \hat{B}_3 -\frac{1}{6} \pa^3 \hat{A}_3 \hat{B}_3
\right)
\nonu \\
&+& c_{66} \,
\left(
\hat{T}\hat{B}_3 \pa \hat{B}_3-\frac{1}{2}\pa \hat{T} \hat{B}_3 \hat{B}_3-\frac{1}{2} \pa \hat{B}_3 \pa^2 \hat{B}_3-\frac{1}{6} \pa^3 \hat{B}_3 \hat{B}_3
\right)
\nonu \\
&+&c_{67}\, \left(
-\frac{i}{24} \hat{A}_3 \pa^3 \hat{B}_3-\frac{3}{4} \hat{B}_{-} \pa \hat{A}_3 \pa \hat{B}_{+}+\hat{T} \hat{B}_{-} \hat{A}_3 \hat{B}_{+}
+\frac{i}{4}\pa \hat{A}_3 \pa^2 \hat{B}_3-\frac{3}{4} \pa \hat{B}_{-}\hat{A}_3 \pa \hat{B}_{+}
\right.
\nonu \\
&-& \left.
\frac{3}{4} \pa \hat{B}_{-} \pa \hat{A}_3  \hat{B}_{+}-\frac{i}{2}\pa \hat{T} \hat{A}_3 \hat{B}_3 \right)(w)
\nonu \\
&+&c_{68} \, \left(
-\frac{3}{4} \hat{B}_{-} \pa \hat{B}_3 \pa \hat{B}_{+} -i\hat{T}\hat{B}_{3} \pa \hat{B}_{3}+\hat{T} \hat{B}_{-} \hat{B}_3 \hat{B}_{+}
-\frac{3}{4}\pa \hat{B}_{-} \hat{B}_3 \pa \hat{B}_{+}
\right. \nonu \\
&-&\frac{3}{4}\pa \hat{B}_{-} \pa \hat{B}_3 \hat{B}_{+}+\frac{i}{2}\pa \hat{T} \hat{B}_{-}\hat{B}_{+}+\frac{3i}{4}\pa^2 \hat{B}_3 \pa \hat{B}_3
\nonu \\
&-& \left.
\frac{i}{4} \pa^2 \hat{B}_{-} \pa \hat{B}_{+}+\frac{1}{10}\pa^2 \hat{T} \hat{B}_3 +\frac{i}{8}\pa^3 \hat{B}_3 \hat{B}_3+\frac{i}{24}\pa^3 \hat{B}_{-}
\hat{B}_{+}\right)
\nonu \\
&+& c_{69} \, \left(
\hat{B}_3 \pa \hat{B}_{-} \pa \hat{B}_{+}+\frac{1}{3} \pa^2 \hat{B}_3\hat{B}_{-} \hat{B}_{+}-\pa \hat{B}_3 \pa \hat{B}_{-} \hat{B}_{+}-\frac{1}{3} \hat{B}_3 \hat{B}_{-} \pa^2 \hat{B}_{+}\right)
\nonu \\
&+& c_{70} \, \left(
\hat{T} \pa \hat{B}_{-} \hat{B}_{+}-\frac{1}{2} \pa \hat{T} \hat{B}_{-}\hat{B}_{+}-\frac{1}{2}\pa^2 \hat{B}_{-} \pa \hat{B}_{+}
+\frac{i}{10}\pa^2 \hat{T} \hat{B}_3 -\frac{1}{6} \pa^3 \hat{B}_{-} 
\hat{B}_{+}\right)
\nonu \\
&+&c_{71} \, \left(
\hat{T} \hat{G}_{11} \hat{G}_{22}-\hat{T}\pa \hat{T}-
\frac{2 i N}{3(N+2+k)}
\hat{T} \pa^2 \hat{A}_3+\frac{2ik}{3(N+2+k)}\hat{T} \pa^2 \hat{B}_3-\pa \hat{G}_{11} \pa \hat{G}_{22} \right.
\nonu \\
&-& \frac{1}{(N+2+k)}\pa \hat{T} \hat{A}_3 \hat{A}_3-
\frac{2}{(N+2+k)}\pa \hat{T} \hat{A}_3 \hat{B}_3-\frac{1}{(N+2+k)}
\pa \hat{T} \hat{A}_{-} \hat{A}_{+}
\nonu \\
& - & \frac{1}{(N+2+k)}
\pa \hat{T} \hat{B}_{3} \hat{B}_{3}
-\frac{1}{(N+2+k)}\pa \hat{T} \hat{B}_{-} \hat{B}_{+}+
\frac{i}{(N+2+k)}\pa \hat{T} \pa \hat{A}_3 
\nonu \\
& + & 
\left.
\frac{i}{(N+2+k)}\pa \hat{T}\pa  \hat{B}_3 +
\frac{1}{6}\pa^3 \hat{T}+\frac{i N}{10(N+2+k)} \pa^4 \hat{A}_3
- \frac{i k}{10(N+2+k)} \pa^4 \hat{B}_3 \right)
\nonu \\
&+&c_{72} \, \left(
\hat{T} \hat{G}_{21} \hat{G}_{12}-\hat{T}\pa \hat{T}+
\frac{2i N}{3(N+2+k)}\hat{T} \pa^2 \hat{A}_3+
\frac{2ik}{3(N+2+k)}\hat{T} \pa^2 \hat{B}_3
\right. \nonu \\
& - & \pa \hat{G}_{21} \pa \hat{G}_{12}
\nonu \\
&-& \frac{1}{(N+2+k)}\pa \hat{T} \hat{A}_3 \hat{A}_3+
\frac{2}{(N+2+k)}\pa \hat{T} \hat{A}_3 \hat{B}_3-
\frac{1}{(N+2+k)}\pa \hat{T} \hat{A}_{-} \hat{A}_{+}
\nonu \\
& - & 
\frac{1}{(N+2+k)}\pa \hat{T} \hat{B}_{3} \hat{B}_{3}
-\frac{1}{(N+2+k)}\pa \hat{T} \hat{B}_{-} \hat{B}_{+}+
\frac{i}{(N+2+k)}\pa \hat{T} \pa \hat{A}_3 
\nonu \\
& + & \left.
\frac{i}{(N+2+k)}\pa \hat{T} \pa \hat{B}_3 +
\frac{1}{6}\pa^3 \hat{T}-\frac{i N}{10(N+2+k)} \pa^4 \hat{A}_3
- \frac{i k}{10(N+2+k)} \pa^4 \hat{B}_3 \right)
\nonu \\
&+&c_{73}\, \left(
\hat{T}\pa T^{(2)}-\pa \hat{T} T^{(2)}-\frac{1}{6} \pa^3 T^{(2)}\right)
\nonu \\
&+& c_{74} \, \left(
\hat{T}\pa^2 \hat{A}_3-\frac{3}{2} \pa \hat{T} \pa \hat{A}_3+\frac{3}{10} \pa^2 \hat{T} \hat{A}_3 -\frac{3}{20} \pa^4 \hat{A}_3\right)
\nonu \\
&+& c_{75} \,
\left(\hat{T}\pa^2 \hat{B}_3-\frac{3}{2} \pa \hat{T} \pa \hat{B}_3+
\frac{3}{10} \pa^2 \hat{T} \hat{B}_3 -\frac{3}{20} 
\pa^4 \hat{B}_3 \right)
\nonu \\
& + & 
c_{76} \,
\left(
\hat{T}P^{(3)}-\frac{3}{14}\pa^2 P^{(3)}\right)+
S^{(5)}+\widetilde{S}^{(5)} \Bigg](w)
+ \cdots.
\label{t2w4}
\eea
In the fourth-order pole of (\ref{t2w4}), the coefficient $-\frac{1}{2}$
in the descendant field of spin-$1$ current 
located at the fifth-order pole
can be obtained from the standard procedure 
for given spins of the left hand side ($h_i=2$ and 
$h_j=4$) and the spin ($h_k=1$) of 
the spin-$1$ current appearing in the fifth-order pole. 
There is no descendant field for the spin-$2$ field (appearing 
in the fourth-order pole) in the third-order pole 
$(h_k=2)$. 
In the second-order pole, the coefficient $\frac{1}{6}$
in the descendant field of spin-$3$ current located at 
the third-order pole ($h_k=3$)
can be obtained according to previous analysis.
There are new higher spin-$4$ currents $\widetilde{P}^{(4)}(w)$
and $\widetilde{S}^{(4)}(w)$ (appeared in Appendix $B$).
In the first-order pole, the coefficient $\frac{1}{4}$
in the descendant field of spin-$4$ current located at 
the second-order pole ($h_k=4$)
can be obtained similarly.
Note that there exists a new higher spin-$5$ current
$\widetilde{S}^{(5)}(w)$ in the first-order pole of
(\ref{t2w4}) which belongs to other 
${\cal N}=4$ multiplet.
In particular, the correct presence of various quasiprimary fields 
is very important to obtain the final higher spin-$5$ current which is the 
highest higher spin current in (\ref{next16}).
Two of the quasiprimary fields have the explicit $N$-dependence in their 
expressions.  

Therefore, we have observed the presence of the next $16$ lowest
higher spin currents in the right hand side of the OPEs between the 
$16$ lowest higher spin currents.

\section{ The complete
OPEs between the $16$ currents  and the $16$ lowest
higher spin currents for generic $N$ }

In this Appendix, we describe
the complete
OPEs between the $16$ currents  (of large ${\cal N}=4$
 linear superconformal algebra) and the $16$ lowest
higher spin currents for generic $N$ from the results of
$N=4,5,8,9$.
Except the few cases, these are linear.

\subsection{ The OPEs between the spin-$\frac{1}{2}$ currents and
the $16$ lowest higher spin currents }

We perform the various OPEs 
between the four spin-$\frac{1}{2}$ currents, $F_{11}(z) 
\equiv {\bf F_{11}}(z)$, $F_{22}(z) \equiv {\bf F_{22}}(z)$, 
$F_{12}(z) \equiv {\bf F_{12}}(z)$, and $F_{21}(z) \equiv {\bf F_{21}}(z)$,
and the $16$ higher spin currents obtained previously
as follows:
\bea
\left(
\begin{array}{c}
F_{11} \nonu \\
F_{22} \end{array}
\right)(z) \,  {\bf T^{(3)}}(w) & = & \frac{1}{(z-w)}
\left(
\begin{array}{c}
{\bf U^{(\frac{5}{2})}} \nonu \\
-{\bf V^{(\frac{5}{2})}}
\end{array}
\right)
(w) +\cdots,
\nonu \\
\left(
\begin{array}{c}
F_{11} \nonu \\
F_{22}
\end{array}
\right)(z) \,
\left(
\begin{array}{c}
{\bf V_{+}^{(3)}}
\nonu \\
{\bf U_{-}^{(3)}}
\end{array}
\right)(w) & = &
\frac{1}{(z-w)}   {\bf T_{\pm}^{(\frac{5}{2})}}
(w) +\cdots,
\nonu \\
\left(
\begin{array}{c}
F_{11} \nonu \\
F_{22}
\end{array}
\right)(z) \,
\left(
\begin{array}{c}
{\bf V_{-}^{(3)}}
\nonu \\
{\bf U_{+}^{(3)}}
\end{array}
\right)(w) & = &
\frac{1}{(z-w)}   {\bf T_{\mp}^{(\frac{5}{2})}}
(w) +\cdots,
\nonu \\
\left(
\begin{array}{c}
F_{11} \nonu \\
F_{22}
\end{array}
\right)(z) \,
\left(
\begin{array}{c}
{\bf V^{(\frac{7}{2})}}
\nonu \\
{\bf U^{(\frac{7}{2})}}
\end{array}
\right)(w) & = &
\pm \frac{1}{(z-w)^2} 4{\bf T^{(2)}}(w)
+\frac{1}{(z-w)} \Bigg[ \mp  \pa {\bf T^{(2)}}- {\bf W^{(3)}}
\Bigg](w) + \cdots,
\nonu \\
\left(
\begin{array}{c}
F_{11} \nonu \\
F_{22}
\end{array}
\right)(z) \,
{\bf W_{\pm}^{(\frac{7}{2})}}(w) & = &  \mp
\frac{1}{(z-w)} \left(
\begin{array}{c}
{\bf U_{+}^{(3)}} \nonu \\
{\bf V_{-}^{(3)}}
\end{array}
\right)(w) + \cdots,
\nonu \\
\left(
\begin{array}{c}
F_{11} \nonu \\
F_{22}
\end{array}
\right)(z) \,
{\bf W_{\mp}^{(\frac{7}{2})}}(w) & = &
\pm \frac{1}{(z-w)} \left(
\begin{array}{c}
{\bf U_{-}^{(3)}} \nonu \\
{\bf V_{+}^{(3)}}
\end{array}
\right)(w) + \cdots,
\nonu \\
\left(
\begin{array}{c}
F_{11} \nonu \\
F_{22}
\end{array}
\right)(z)
{\bf W^{(4)}}(w) & = &
\frac{1}{(z-w)^2} \Bigg[ 5
\left(
\begin{array}{c}
{\bf U^{(\frac{5}{2})}}
\nonu \\
{\bf V^{(\frac{5}{2})}}
\end{array}
\right)
\nonu \\
&+&\frac{36 (-N+k)}{((37N+59)+(15N+37)k)} \left(
\begin{array}{c}
F_{11} \nonu \\
F_{22} \end{array}
\right) {\bf T^{(2)}}
 \Bigg](w)
\nonu \\
&- & \frac{1}{(z-w)} \Bigg[
\pa\left(
\begin{array}{c}
{\bf U^{(\frac{5}{2})}}
\nonu \\
{\bf V^{(\frac{5}{2})}}
\end{array}
\right)
\nonu \\
&+&\frac{36 (-N+k)}{((37N+59)+(15N+37)k)} \pa \left(
\begin{array}{c}
F_{11} \nonu \\
F_{22} \end{array}
\right) {\bf T^{(2)}}
\Bigg](w) +\cdots,
\nonu \\
\left(
\begin{array}{c}
F_{12} \nonu \\
F_{21}
\end{array}
\right)(z) \,
\left(
\begin{array}{c}
{\bf U_{+}^{(3)}}
\nonu \\
{\bf V_{-}^{(3)}}
\end{array}
\right)(w) & = &
-\frac{1}{(z-w)}  \left(
\begin{array}{c}
{\bf U^{(\frac{5}{2})}}
\nonu \\
{\bf V^{(\frac{5}{2})}}
\end{array}
\right)
(w) +\cdots,
\nonu \\
\left(
\begin{array}{c}
F_{12} \nonu \\
F_{21}
\end{array}
\right)(z) \,
\left(
\begin{array}{c}
{\bf V_{+}^{(3)}}
\nonu \\
{\bf U_{-}^{(3)}}
\end{array}
\right)(w) & = &
\frac{1}{(z-w)}   \left(
\begin{array}{c}
{\bf V^{(\frac{5}{2})}}
\nonu \\
{\bf U^{(\frac{5}{2})}}
\end{array}
\right)
(w) +\cdots,
\nonu \\
\left(
\begin{array}{c}
F_{12} \nonu \\
F_{21}
\end{array}
\right)(z) \,
\left(
\begin{array}{c}
{\bf U^{(\frac{7}{2})}}
\nonu \\
{\bf V^{(\frac{7}{2})}}
\end{array}
\right)(w) & = &
\mp \frac{1}{(z-w)}
\left(
\begin{array}{c}
{\bf U_{-}^{(3)}}
\nonu \\
{\bf V_{+}^{(3)}}
\end{array}
\right)
(w) + \cdots,
\nonu \\
\left(
\begin{array}{c}
F_{12} \nonu \\
F_{21}
\end{array}
\right)(z) \,
\left(
\begin{array}{c}
{\bf V^{(\frac{7}{2})}}
\nonu \\
{\bf U^{(\frac{7}{2})}}
\end{array}
\right)(w) & = &
\pm \frac{1}{(z-w)}
\left(
\begin{array}{c}
{\bf V_{-}^{(3)}}
\nonu \\
{\bf U_{+}^{(3)}}
\end{array}
\right)
(w) + \cdots,
\nonu \\
\left(
\begin{array}{c}
F_{12} \nonu \\
F_{21} \end{array}
\right)(z) \,  {\bf W^{(3)}}(w) & = & \pm \frac{1}{(z-w)}
{\bf T_{\mp}^{(\frac{5}{2})}}(w) +\cdots,
\nonu \\
\left(
\begin{array}{c}
F_{12} \nonu \\
F_{21}
\end{array}
\right)(z) \,
{\bf W_{\pm}^{(\frac{7}{2})}}(w) & = &
\pm \frac{4}{(z-w)^2} {\bf T^{(2)}}(w)
+  \frac{1}{(z-w)} \Bigg[ \mp  \pa {\bf T^{(2)}} -
{\bf T^{(3)}}  \Bigg](w) + \cdots,
\nonu \\
\nonu \\
\left(
\begin{array}{c}
F_{12} \nonu \\
F_{21}
\end{array}
\right)(z)
{\bf W^{(4)}}(w) & = &
\frac{1}{(z-w)^2} \Bigg[ 5
\left(
\begin{array}{c}
{\bf T^{(\frac{5}{2})}_{-}}
\nonu \\
{\bf T^{(\frac{5}{2})}_{+}}
\end{array}
\right)
\nonu \\
&+&\frac{36 (-N+k)}{((37N+59)+(15N+37)k)} \left(
\begin{array}{c}
F_{12} \nonu \\
F_{21} \end{array}
\right) {\bf T^{(2)}}
\Bigg](w)
\nonu \\
&- & \frac{1}{(z-w)} \Bigg[
\pa\left(
\begin{array}{c}
{\bf T^{(\frac{5}{2})}_{-}}
\nonu \\
{\bf T^{(\frac{5}{2})}_{+}}
\end{array}
\right)
\nonu \\
&+&\frac{36 (-N+k)}{((37N+59)+(15N+37)k)} \pa \left(
\begin{array}{c}
F_{12}  \\
F_{21} \end{array}
\right) {\bf T^{(2)}}
\Bigg](w) +\cdots.
\label{opehalf-16}
\eea
The nonlinear terms appear in the OPEs 
containing the higher spin-$4$ current.
As done in the unitary coset theory \cite{Ahn1504}, 
by adding the extra quasiprimary field
of spin-$4$ 
containing the higher spin-$2$ current to the above
higher spin-$4$ current, the nonlinear terms disappear.
See also the subsection $4.4$.
We also have checked that the above OPEs  
(\ref{opehalf-16}) are equivalent to those OPEs in \cite{BCG1404}.
The $N$-dependence on the structure constants can be obtained easily
because the fractional $k$-dependent terms for $N=4,5,8,9$ are simple 
and the numerators and the denominators are linear in $k$. 
  
\subsection{ The OPEs between the spin-$1$ currents and
the $16$ lowest higher spin currents  }

Let us perform the various OPEs between the spin-$1$ current, $U(z) \equiv
{\bf U}(z)$,
and the $16$ higher spin currents
as follows:
\bea
U(z) \left(
\begin{array}{c}
{\bf U^{(\frac{7}{2})}} \nonu \\
{\bf V^{(\frac{7}{2})}} \end{array}
\right)(w) & = &
\frac{1}{(z-w)^2} \left(
\begin{array}{c}
{\bf U^{(\frac{5}{2})}} \nonu \\
-{\bf V^{(\frac{5}{2})}}
\end{array}
\right)
(w) +\cdots,
\nonu \\
U(z) \left(
\begin{array}{c}
{\bf W_{+}^{(\frac{7}{2})}} \nonu \\
{\bf W_{-}^{(\frac{7}{2})}} \end{array}
\right)(w) & = &
\mp \frac{1}{(z-w)^2}
{\bf T_{\pm}^{(\frac{5}{2})}}
(w) +\cdots,
\nonu \\
U(z) \, {\bf W^{(4)}}(w) & = &
-\frac{1}{(z-w)^3} 8 {\bf T^{(2)}}(w)
\label{opeu-16}
\\
& + & \frac{1}{(z-w)^2}  \Bigg[ 2 \pa {\bf T^{(2)}}
 + \frac{72 (-N+k)}{((37N+59)+(15N+37)k)}
U \, {\bf T^{(2)}} \Bigg](w)
+\cdots.
\nonu
\eea
Again, by introducing the quasiprimary field of spin $4$, the above 
nonlinear terms disappear.


The OPEs between the three spin-$1$ currents, $A_{\pm}(z) \equiv
{\bf A_{\pm}}(z)$ and $A_{3}(z)\equiv {\bf A_3}(z)$,
and the $16$ higher spin currents
are
\bea
A_{\pm}(z) \, {\bf T_{\pm}^{(\frac{5}{2})}}(w) & = &
 \mp  \frac{1}{(z-w)} i
\left(
\begin{array}{c}
{\bf U^{(\frac{5}{2})}} \nonu \\
 {\bf V^{(\frac{5}{2})}}
\end{array}
\right)(w)+\cdots,
\nonu \\
A_{\pm}(z) \, {\bf T^{(3)}}(w) & = &
-\frac{1}{(z-w)} i
\left(
\begin{array}{c}
{\bf U_{-}^{(3)}} \nonu \\
 {\bf V_{+}^{(3)}}
\end{array}
\right)(w)+\cdots,
\nonu \\
A_{\pm}(z)
\left(
\begin{array}{c}
{\bf V^{(\frac{5}{2})}} \nonu \\
 {\bf U^{(\frac{5}{2})}}
\end{array}
\right)(w) & = &  \pm
\frac{1}{(z-w)} i  {\bf T_{\mp}^{(\frac{5}{2})}}(w)
+ \cdots,
\nonu \\
A_{\pm}(z) \left(
\begin{array}{c}
{\bf V_{+}^{(3)}} \nonu \\
 {\bf U_{-}^{(3)}}
\end{array}
\right)(w) & = &
\mp \frac{1}{(z-w)^2} 4i {\bf T^{(2)}}(w)
+  \frac{1}{(z-w)} i \Bigg[ {\bf T^{(3)}} + {\bf W^{(3)}} \Bigg](w) +
\cdots,
\nonu \\
A_{\pm}(z) \left(
\begin{array}{c}
{\bf V^{(\frac{7}{2})}} \nonu \\
{\bf U^{(\frac{7}{2})}}
\end{array}
\right)(w) & = &
\frac{1}{(z-w)^2} \frac{2i(12N+25+13k)}{5(N+2+k)}
{\bf T_{\mp}^{(\frac{5}{2})}}
\mp \frac{1}{(z-w)} i {\bf W_{\mp}^{(\frac{7}{2})}}(w) +\cdots,
\nonu \\
A_{\pm}(z) \, {\bf W^{(3)}}(w) & = &
\frac{1}{(z-w)} i
 \left(
\begin{array}{c}
{\bf U_{-}^{(3)}} \nonu \\
 {\bf V_{+}^{(3)}}
\end{array}
\right)(w) + \cdots,
\nonu \\
A_{\pm}(z) \,
{\bf W_{\pm}^{(\frac{7}{2})}}(w) & = &
-\frac{1}{(z-w)^2} \frac{2i(12N+25+13k)}{5(N+2+k)}
\left(
\begin{array}{c}
{\bf U^{(\frac{5}{2})}} \nonu \\
 {\bf V^{(\frac{5}{2})}}
\end{array}
\right)
\pm \frac{1}{(z-w)} i  \left(
\begin{array}{c}
{\bf U^{(\frac{7}{2})}} \nonu \\
 {\bf V^{(\frac{7}{2})}}
\end{array}
\right)(w) \nonu \\
& + & \cdots,
\nonu \\
A_{\pm}(z) \, {\bf W^{(4)}}(w) & = &
\frac{1}{(z-w)^2} \Bigg[  \pm 6 i  \left(
\begin{array}{c}
{\bf U_{-}^{(3)}} \nonu \\
 {\bf V_{+}^{(3)}}
\end{array}
\right) +\frac{72 (-N+k)}{((37N+59)+(15N+37)k)} A_{\pm } {\bf T^{(2)}}
\Bigg](w) \nonu \\
& + & \cdots,
\nonu \\
A_3 (z) \, {\bf T_{\pm}^{(\frac{5}{2})}}(w) & = &
\pm \frac{1}{(z-w)} \frac{i}{2}   {\bf T_{\pm}^{(\frac{5}{2})}}(w)+
\cdots,
\nonu \\
A_3(z) \,
\left(
\begin{array}{c}
{\bf T^{(3)}} \nonu \\
{\bf W^{(3)}}
\end{array}
\right)(w) & = &
\frac{1}{(z-w)^2} 2i {\bf T^{(2)}}(w) +\cdots,
\nonu \\
A_3(z) \left(
\begin{array}{c}
{\bf U^{(\frac{5}{2})}} \nonu \\
 {\bf V^{(\frac{5}{2})}}
\end{array}
\right)(w) & = &
 \mp   \frac{1}{(z-w)} \frac{i}{2}
 \left(
\begin{array}{c}
{\bf U^{(\frac{5}{2})}} \nonu \\
{\bf V^{(\frac{5}{2})}}
\end{array}
\right)(w)
 +\cdots,
\nonu \\
A_3(z)   \left(
\begin{array}{c}
{\bf U_{-}^{(3)}} \nonu \\
 {\bf V_{+}^{(3)}}
\end{array}
\right)(w) & = & \mp \frac{1}{(z-w)} i
\left(
\begin{array}{c}
{\bf U_{-}^{(3)}} \nonu \\
 {\bf V_{+}^{(3)}}
\end{array}
\right)(w) + \cdots,
\nonu \\
A_3(z) \left(
\begin{array}{c}
{\bf U^{(\frac{7}{2})}} \nonu \\
 {\bf V^{(\frac{7}{2})}}
\end{array}
\right)(w) & = &
\frac{1}{(z-w)^2} \frac{i(12N+25+13k)}{5(N+2+k)}
 \left(
\begin{array}{c}
{\bf U^{(\frac{5}{2})}} \nonu \\
 {\bf V^{(\frac{5}{2})}}
\end{array}
\right) (w)
\mp \frac{1}{(z-w)} \frac{i}{2}  \left(
\begin{array}{c}
{\bf U^{(\frac{7}{2})}} \nonu \\
 {\bf V^{(\frac{7}{2})}}
\end{array}
\right) (w) \nonu \\
& + & \cdots,
\nonu \\
A_3(z) \, {\bf W_{\pm}^{(\frac{7}{2})}}(w)
& = &
\frac{1}{(z-w)^2} \frac{i(12N+25+13k)}{5(N+2+k)}
 {\bf T_{\pm}^{(\frac{5}{2})}}(w)
 \pm
\frac{1}{(z-w)} \frac{i}{2}  {\bf W_{\pm}^{(\frac{7}{2})}}(w) 
+ 
\cdots,
\nonu \\
A_3(z) \, {\bf W^{(4)}}(w) & = &
\frac{1}{(z-w)^2} \Bigg[  3 i {\bf T^{(3)}}
+ 3i {\bf W^{(3)}}
+\frac{72 (-N+k)}{((37N+59)+(15N+37)k)} A_3 {\bf T^{(2)}} \Bigg](w) 
\nonu \\
& + & \cdots.
\label{a-16}
\eea


The OPEs between the other three spin-$1$ currents,  $B_{\pm}(z) \equiv 
{\bf B_{\pm}}(z)$ and 
$B_{3}(z) \equiv {\bf B_3}(z)$,
and  the $16$ higher spin currents obtained
are
\bea
B_{\pm}(z) \, {\bf T_{\pm}^{(\frac{5}{2})}}(w) & = &
 \mp  \frac{1}{(z-w)} i
\left(
\begin{array}{c}
{\bf V^{(\frac{5}{2})}} \nonu \\
 {\bf U^{(\frac{5}{2})}}
\end{array}
\right)(w)+\cdots,
\nonu \\
B_{\pm}(z) \, {\bf T^{(3)}}(w) & = &
\frac{1}{(z-w)} i
\left(
\begin{array}{c}
{\bf V_{-}^{(3)}} \nonu \\
 {\bf U_{+}^{(3)}}
\end{array}
\right)(w)+\cdots,
\nonu \\
B_{\pm}(z)
\left(
\begin{array}{c}
{\bf U^{(\frac{5}{2})}} \nonu \\
 {\bf V^{(\frac{5}{2})}}
\end{array}
\right)(w) & = &  \pm
\frac{1}{(z-w)} i  {\bf T_{\mp}^{(\frac{5}{2})}}(w)
+ \cdots,
\nonu \\
B_{\pm}(z) \left(
\begin{array}{c}
{\bf U_{+}^{(3)}} \nonu \\
 {\bf V_{-}^{(3)}}
\end{array}
\right)(w) & = &
\mp \frac{1}{(z-w)^2} 4i {\bf T^{(2)}}(w)
+  \frac{1}{(z-w)} i \Bigg[ {\bf T^{(3)}}- {\bf W^{(3)}} \Bigg](w) +
\cdots,
\nonu \\
B_{\pm}(z) \left(
\begin{array}{c}
{\bf U^{(\frac{7}{2})}} \nonu \\
{\bf V^{(\frac{7}{2})}}
\end{array}
\right)(w) & = &
\frac{1}{(z-w)^2} \frac{2i(13N+25+12k)}{5(N+2+k)}
{\bf T_{\mp}^{(\frac{5}{2})}}
\pm \frac{1}{(z-w)} i {\bf W_{\mp}^{(\frac{7}{2})}}(w) +\cdots,
\nonu \\
B_{\pm}(z) \, {\bf W^{(3)}}(w) & = &
-\frac{1}{(z-w)} i
 \left(
\begin{array}{c}
{\bf V_{-}^{(3)}} \nonu \\
 {\bf U_{+}^{(3)}}
\end{array}
\right)(w) + \cdots,
\nonu \\
B_{\pm}(z) \,
{\bf W_{\pm}^{(\frac{7}{2})}}(w) & = &
\frac{1}{(z-w)^2} \frac{2i(13N+25+12k)}{5(N+2+k)}
\left(
\begin{array}{c}
{\bf V^{(\frac{5}{2})}} \nonu \\
 {\bf U^{(\frac{5}{2})}}
\end{array}
\right)
\mp \frac{1}{(z-w)}\left(
\begin{array}{c}
{\bf V^{(\frac{7}{2})}} \nonu \\
 {\bf U^{(\frac{7}{2})}}
\end{array}
\right) (w) +\cdots,
\nonu \\
B_{\pm}(z) \, {\bf W^{(4)}}(w) & = &
\frac{1}{(z-w)^2} \Bigg[  \mp 6 i  \left(
\begin{array}{c}
{\bf V_{-}^{(3)}} \nonu \\
 {\bf U_{+}^{(3)}}
\end{array}
\right) +\frac{72 (-N+k)}{((37N+59)+(15N+37)k)} B_{\pm } {\bf T^{(2)}}
\Bigg](w) \nonu \\
& + & \cdots,
\nonu \\
B_3 (z) \, {\bf T_{\pm}^{(\frac{5}{2})}}(w) & = &
\pm \frac{1}{(z-w)} \frac{i}{2}   {\bf T_{\pm}^{(\frac{5}{2})}}(w)+
\cdots,
\nonu \\
B_3(z) \,
\left(
\begin{array}{c}
{\bf T^{(3)}} \nonu \\
{\bf W^{(3)}}
\end{array}
\right)(w) & = &
\pm\frac{1}{(z-w)^2} 2i {\bf T^{(2)}}(w) +\cdots,
\nonu \\
B_3(z) \left(
\begin{array}{c}
{\bf U^{(\frac{5}{2})}} \nonu \\
 {\bf V^{(\frac{5}{2})}}
\end{array}
\right)(w) & = &
 \pm   \frac{1}{(z-w)} \frac{i}{2}
 \left(
\begin{array}{c}
{\bf U^{(\frac{5}{2})}} \nonu \\
{\bf V^{(\frac{5}{2})}}
\end{array}
\right)(w)
 +\cdots,
\nonu \\
B_3(z)   \left(
\begin{array}{c}
{\bf U_{+}^{(3)}} \nonu \\
 {\bf V_{-}^{(3)}}
\end{array}
\right)(w) & = & \pm \frac{1}{(z-w)} i
\left(
\begin{array}{c}
{\bf U_{+}^{(3)}} \nonu \\
 {\bf V_{-}^{(3)}}
\end{array}
\right)(w) + \cdots,
\nonu \\
B_3(z) \left(
\begin{array}{c}
{\bf U^{(\frac{7}{2})}} \nonu \\
 {\bf V^{(\frac{7}{2})}}
\end{array}
\right)(w) & = &
\frac{1}{(z-w)^2} \frac{i(13N+25+12k)}{5(N+2+k)}
 \left(
\begin{array}{c}
{\bf U^{(\frac{5}{2})}} \nonu \\
 {\bf V^{(\frac{5}{2})}}
\end{array}
\right) (w)
\pm \frac{1}{(z-w)} \frac{i}{2}  \left(
\begin{array}{c}
{\bf U^{(\frac{7}{2})}} \nonu \\
 {\bf V^{(\frac{7}{2})}}
\end{array}
\right) (w) \nonu \\
& + & \cdots,
\nonu \\
B_3(z) \, {\bf W_{\pm}^{(\frac{7}{2})}}(w)
& = &
-\frac{1}{(z-w)^2} \frac{i(13N+25+12k)}{5(N+2+k)}
 {\bf T_{\pm}^{(\frac{5}{2})}}(w)
 \pm
\frac{1}{(z-w)} \frac{i}{2}  {\bf W_{\pm}^{(\frac{7}{2})}}(w) +
\cdots,
\nonu \\
B_3(z) \, {\bf W^{(4)}}(w) & = &
\frac{1}{(z-w)^2} \Bigg[ - 3 i {\bf T^{(3)}}
+ 3i {\bf W^{(3)}} +\frac{72 (-N+k)}{((37N+59)+(15N+37)k)} B_3 {\bf T^{(2)}} 
\Bigg](w)
\nonu \\
&+&\cdots.
\label{b-16}
\eea

As done before, the nonlinear terms 
appearing in (\ref{opeu-16}), (\ref{a-16}) or (\ref{b-16})
can be removed by introducing the 
extra quasiprimary field of spin-$4$ in the expression of 
the higher spin-$4$ current.
See also the subsection $4.4$.
Via explicit field identifications between the fields in this 
paper and those in \cite{BCG1404}, we have checked that the above 
OPEs (\ref{opeu-16}), (\ref{a-16}) and (\ref{b-16}) 
are the same as the ones in \cite{BCG1404}.

\subsection{ The OPEs between the spin-$\frac{3}{2}$ currents and
the $16$ lowest higher spin currents }

The OPEs between the four spin-$\frac{3}{2}$ currents currents, 
$G_{11}(z) \equiv {\bf G_{11}}(z)$, 
$G_{22}(z) \equiv {\bf G_{22}}(z)$, 
$G_{12}(z) \equiv {\bf G_{12}}(z)$ and 
$ G_{21}(z) \equiv {\bf G_{21}}(z)$,
and the $16$ higher spin currents obtained
are
\bea
\left(
\begin{array}{c}
G_{11} \nonu \\
G_{22}  \end{array}
\right)(z) \, {\bf T^{(2)}}(w)
& = & \frac{1}{(z-w)}
\left(
\begin{array}{c}
{\bf U^{(\frac{5}{2})}} \nonu \\
{\bf V^{(\frac{5}{2})}}  \end{array}
\right)
(w) +\cdots,
\nonu \\
 \left(
\begin{array}{c}
G_{11} \nonu \\
G_{22}
 \end{array}
\right)(z) \, {\bf T_{\pm}^{(\frac{5}{2})}}(w)
& = &
-\frac{1}{(z-w)} \left(
\begin{array}{c}
{\bf U_{+}^{(3)}} \nonu \\
{\bf V_{-}^{(3)}}
 \end{array}
\right)(w) +
\cdots,
\nonu \\
 \left(
\begin{array}{c}
G_{11} \nonu \\
G_{22}
 \end{array}
\right)(z) \, {\bf T_{\mp}^{(\frac{5}{2})}}(w)
& = &
-\frac{1}{(z-w)}  \left(
\begin{array}{c}
{\bf U_{-}^{(3)}} \nonu \\
{\bf V_{+}^{(3)}}
 \end{array}
\right) (w) +
\cdots,
\nonu \\
 \left(
\begin{array}{c}
G_{11} \nonu \\
G_{22}
 \end{array}
\right)(z) \, {\bf T^{(3)}}(w) & = &
\pm \frac{1}{(z-w)^2} \frac{(-N+k)}{(N+2+k)}
 \left(
\begin{array}{c}
{\bf U^{(\frac{5}{2})}} \nonu \\
{\bf V^{(\frac{5}{2})}}  \end{array}
\right) (w)
\nonu \\
&+&  \frac{1}{(z-w)} \Bigg[
\frac{1}{5} \pa (\mbox{pole-2}) +
\left(
\begin{array}{c}
{\bf U^{(\frac{7}{2})}} \nonu \\
{\bf V^{(\frac{7}{2})}}  \end{array}
\right)
\Bigg](w) +\cdots,
\nonu \\
\left(
\begin{array}{c}
G_{11} \nonu \\
G_{22}
 \end{array}
\right)(z)  \left(
\begin{array}{c}
{\bf V^{(\frac{5}{2})}} \nonu \\
{\bf U^{(\frac{5}{2})}}
\end{array}
\right)(w)
& = &
\frac{1}{(z-w)^2}4 {\bf T^{(2)}}(w)
\nonu \\
&+& \frac{1}{(z-w)} \Bigg[
\pa {\bf T^{(2)}}
\mp {\bf W^{(3)}}  \Bigg](w) +\cdots,
\nonu \\
\left(
 \begin{array}{c}
G_{11} \nonu \\
G_{22}
 \end{array}
\right)(z) \,
\left(
\begin{array}{c}
{\bf V_{+}^{(3)}} \nonu \\
{\bf U_{-}^{(3)}}
 \end{array}
\right)(w) & = &
-\frac{1}{(z-w)^2} \frac{2(2N+5+3k)}{(N+2+k)}  {\bf T_{\pm}^{(\frac{5}{2})}}
(w)
\nonu \\
& + & \frac{1}{(z-w)} \Bigg[
\frac{1}{5} \pa \, \mbox{(pole-2)}
\pm  {\bf W_{\pm}^{(\frac{7}{2})}} \Bigg](w) + \cdots,
\nonu \\
\left(
 \begin{array}{c}
G_{11} \nonu \\
G_{22}
 \end{array}
\right)(z) \,
\left(
\begin{array}{c}
{\bf V_{-}^{(3)}} \nonu \\
{\bf U_{+}^{(3)}}
 \end{array}
\right)(w) & = &
-\frac{1}{(z-w)^2} \frac{2(3N+5+2k)}{(N+2+k)}   {\bf T_{\mp}^{(\frac{5}{2})}}
(w)
\nonu \\
& + & \frac{1}{(z-w)} \Bigg[
\frac{1}{5} \pa \, \mbox{(pole-2)}
\pm  {\bf W_{\mp}^{(\frac{7}{2})}} \Bigg](w) + \cdots,
\nonu \\
\left(
 \begin{array}{c}
G_{11} \nonu \\
G_{22}
 \end{array}
\right)(z) \,
 \left(
\begin{array}{c}
 {\bf V^{(\frac{7}{2})}} \nonu \\
{\bf U^{(\frac{7}{2})}}
\end{array}
\right)(w)
& = &
\pm \frac{1}{(z-w)^3} \frac{48(-N+k)}{5(N+2+k)} {\bf T^{(2)}}(w)
\nonu \\
& + & \frac{1}{(z-w)^2} \Bigg[ 6 {\bf T^{(3)}} -
\frac{6(-N+k)}{5(N+2+k)} {\bf W^{(3)}} \Bigg](w)
\nonu \\
&+& \frac{1}{(z-w)} \Bigg[
\frac{1}{6} \pa \, \mbox{(pole-2)}
\mp{\bf W^{(4)}}
 \nonu \\
& \pm & 
\frac{72 (-N+k)}{((37N+59)+(15N+37)k)}
\left( T {\bf T^{(2)}} -\frac{3}{10} \pa^2 {\bf T^{(2)}}
\right)
\Bigg](w) +\cdots,
\nonu \\
 \left(
\begin{array}{c}
G_{11} \nonu \\
G_{22}
 \end{array}
\right)(z) \, {\bf W^{(3)}}(w) & = &
\pm \frac{1}{(z-w)^2}  5 \left(
\begin{array}{c}
{\bf U^{(\frac{5}{2})}} \nonu \\
{\bf V^{(\frac{5}{2})}}  \end{array}
\right) (w)
+  \frac{1}{(z-w)}
\frac{1}{5} \pa (\mbox{pole-2})(w)
+\cdots,
\nonu \\
 \left(
\begin{array}{c}
G_{11} \nonu \\
G_{22}
 \end{array}
\right)(z) \, {\bf W_{\pm}^{(\frac{7}{2})}}(w) & = & \mp
\frac{1}{(z-w)^2} \frac{12(2N+5+3k)}{5(N+2+k)}
\left(
\begin{array}{c}
{\bf U_{+}^{(3)}} \nonu \\
{\bf V_{-}^{(3)}}
 \end{array}
\right)(w) \nonu \\
& + & \frac{1}{(z-w)}
\frac{1}{6} \pa \, \mbox{(pole-2)}(w)  +\cdots,
\nonu \\
  \left(
\begin{array}{c}
G_{11} \nonu \\
G_{22}
 \end{array}
\right)(z) \, {\bf W_{\mp}^{(\frac{7}{2})}}(w) & = & \mp
\frac{1}{(z-w)^2} \frac{12(3N+5+2k)}{5(N+2+k)}
\left(
\begin{array}{c}
{\bf U_{-}^{(3)}} \nonu \\
{\bf V_{+}^{(3)}}
 \end{array}
\right)(w) \nonu \\
& + & \frac{1}{(z-w)}
\frac{1}{6} \pa \, \mbox{(pole-2)}(w)  +\cdots,
\nonu \\
\left(
 \begin{array}{c}
G_{11} \nonu \\
G_{22}
 \end{array}
\right)(z) \, {\bf W^{(4)}}(w) & = &
\frac{1}{(z-w)^3} \frac{12(-N+k)(41N+70+(12N+41)k)}{(N+2+k)((37N+59)+(15N+37)k)}
 \left(
\begin{array}{c}
{\bf U^{(\frac{5}{2})}} \nonu \\
{\bf V^{(\frac{5}{2})}}  \end{array}
\right) (w)
\nonu \\
&+& \frac{1}{(z-w)^2} \Bigg[ \pm 7  \left(
\begin{array}{c}
{\bf U^{(\frac{7}{2})}} \nonu \\
{\bf V^{(\frac{7}{2})}}  \end{array}
\right)
 \nonu \\
& - & \frac{216(-N+k)}{5((37N+59)+(15N+37)k)}
 \pa \left(
\begin{array}{c}
{\bf U^{(\frac{5}{2})}} \nonu \\
{\bf V^{(\frac{5}{2})}}  \end{array}
\right)
 \nonu \\
& + &  \frac{108(-N+k)}{((37N+59)+(15N+37)k)} \left(
\begin{array}{c} G_{11}
\nonu \\
G_{22}
\end{array}
\right) {\bf T^{(2)}}
\Bigg](w)
\nonu \\
&+&\frac{1}{(z-w)} \Bigg[ \pm \pa \left(
\begin{array}{c}
{\bf U^{(\frac{7}{2})}} \nonu \\
{\bf V^{(\frac{7}{2})}}  \end{array}
\right)
 \nonu \\
& -& \frac{18(-N+k)}{5((37N+59)+(15N+37)k)}
 \pa^2 \left(
\begin{array}{c}
{\bf U^{(\frac{5}{2})}} \nonu \\
{\bf V^{(\frac{5}{2})}}  \end{array}
\right)+
 \nonu \\
& + &  \frac{36(-N+k)}{((37N+59)+(15N+37)k))}
 {\bf T^{(2)}} \pa \left(
\begin{array}{c} G_{11}
\nonu \\
G_{22}
\end{array}
\right)
 \nonu \\
& + & 
\frac{72(-N+k)}{((37N+59)+(15N+37)k)}
 T \left(
\begin{array}{c}
{\bf U^{(\frac{5}{2})}} \nonu \\
{\bf V^{(\frac{5}{2})}}  \end{array}
\right)
\Bigg](w) +\cdots,
\nonu \\
\left(
\begin{array}{c}
G_{12} \nonu \\
G_{21}  \end{array}
\right)(z) \, {\bf T^{(2)}}(w)
& = &  \frac{1}{(z-w)}
{\bf T_{\mp}^{(\frac{5}{2})}}
(w) +\cdots,
\nonu \\
 \left(
\begin{array}{c}
G_{12} \nonu \\
G_{21}
 \end{array}
\right)(z) \, {\bf T_{\pm}^{(\frac{5}{2})}}(w)
& = &
\frac{1}{(z-w)^2} 4 {\bf T^{(2)}}
(w) + \frac{1}{(z-w)}
\Bigg[\pa \, {\bf T^{(2)}}
\mp {\bf T^{(3)}} \Bigg](w) +
\cdots,
\nonu \\
 \left(
\begin{array}{c}
G_{12} \nonu \\
G_{21}
 \end{array}
\right)(z) \, {\bf T^{(3)}}(w) & = &
\pm \frac{1}{(z-w)^2} 5 {\bf T_{\mp}^{(\frac{5}{2})}} (w)
+  \frac{1}{(z-w)}
\frac{1}{5} \pa \, \mbox{(pole-2)}(w) +\cdots,
\nonu \\
\left(
\begin{array}{c}
G_{12} \nonu \\
G_{21}
 \end{array}
\right)(z)  \left(
\begin{array}{c}
{\bf U^{(\frac{5}{2})}} \nonu \\
{\bf V^{(\frac{5}{2})}}
\end{array}
\right)(w)
& = &
 \frac{1}{(z-w)}  \left(
\begin{array}{c}
{\bf U_{-}^{(3)}} \nonu \\
{\bf V_{+}^{(3)}}
 \end{array}
\right)
(w) +\cdots,
\nonu \\
\left(
\begin{array}{c}
G_{12} \nonu \\
G_{21}
 \end{array}
\right)(z)  \left(
\begin{array}{c}
{\bf V^{(\frac{5}{2})}} \nonu \\
{\bf U^{(\frac{5}{2})}}
\end{array}
\right)(w)
& = &
 \frac{1}{(z-w)}  \left(
\begin{array}{c}
{\bf V_{-}^{(3)}} \nonu \\
{\bf U_{+}^{(3)}}
 \end{array}
\right)
(w) +\cdots,
\nonu \\
\left(
 \begin{array}{c}
G_{12} \nonu \\
G_{21}
 \end{array}
\right)(z) \,
\left(
\begin{array}{c}
{\bf U_{+}^{(3)}} \nonu \\
{\bf V_{-}^{(3)}}
 \end{array}
\right)(w) & = &
\frac{1}{(z-w)^2} \frac{2(3N+5+2k)}{(N+2+k)}   \left(
\begin{array}{c}
{\bf U^{(\frac{5}{2})}} \nonu \\
{\bf V^{(\frac{5}{2})}}
\end{array}
\right)(w)
\nonu \\
& + & \frac{1}{(z-w)} \Bigg[
\frac{1}{5} \pa \, \mbox{(pole-2)}
+   \left(
\begin{array}{c}
-{\bf U^{(\frac{7}{2})}} \nonu \\
{\bf V^{(\frac{7}{2})}}
\end{array}
\right) \Bigg](w) + \cdots,
\nonu \\
\left(
 \begin{array}{c}
G_{12} \nonu \\
G_{21}
 \end{array}
\right)(z) \,
\left(
\begin{array}{c}
{\bf V_{+}^{(3)}} \nonu \\
{\bf U_{-}^{(3)}}
 \end{array}
\right)(w) & = &
\frac{1}{(z-w)^2} \frac{2(2N+5+3k)}{(N+2+k)}   \left(
\begin{array}{c}
{\bf V^{(\frac{5}{2})}} \nonu \\
{\bf U^{(\frac{5}{2})}}
\end{array}
\right)(w)
\nonu \\
& + & \frac{1}{(z-w)} \Bigg[
\frac{1}{5} \pa \, \mbox{(pole-2)}
+   \left(
\begin{array}{c}
-{\bf V^{(\frac{7}{2})}} \nonu \\
{\bf U^{(\frac{7}{2})}}
\end{array}
\right) \Bigg](w) + \cdots,
\nonu \\
\left(
 \begin{array}{c}
G_{12} \nonu \\
G_{21}
 \end{array}
\right)(z) \,
 \left(
\begin{array}{c}
{\bf U^{(\frac{7}{2})}} \nonu \\
{\bf V^{(\frac{7}{2})}}
\end{array}
\right)(w)
& = & \pm
\frac{1}{(z-w)^2} \frac{12(3N+5+2k)}{5(N+2+k)} \left(
\begin{array}{c}
{\bf U_{-}^{(3)}} \nonu \\
{ \bf V_{+}^{(3)}}
 \end{array}
\right) (w)
\nonu \\
&+& \frac{1}{(z-w)}
\frac{1}{6} \pa \, \mbox{(pole-2)}(w) +\cdots,
\nonu \\
\left(
 \begin{array}{c}
G_{12} \nonu \\
G_{21}
 \end{array}
\right)(z) \,
 \left(
\begin{array}{c}
{\bf V^{(\frac{7}{2})}} \nonu \\
{\bf U^{(\frac{7}{2})}}
\end{array}
\right)(w)
& = &
\pm
\frac{1}{(z-w)^2} \frac{12(2N+5+3k)}{5(N+2+k)} \left(
\begin{array}{c}
{\bf V_{-}^{(3)}} \nonu \\
{\bf U_{+}^{(3)}}
 \end{array}
\right) (w)
\nonu \\
&+&\frac{1}{(z-w)}\frac{1}{6} \pa \, \mbox{(pole-2)}(w) +\cdots,
\nonu \\
\left(
\begin{array}{c}
G_{12} \nonu \\
G_{21}
 \end{array}
\right)(z) \, {\bf W^{(3)}}(w) & = &
\mp \frac{1}{(z-w)^2} \frac{(N-k)}{(N+2+k)} { \bf T_{\mp}^{(\frac{5}{2})}}(w)
\nonu \\
&+ &  \frac{1}{(z-w)} \Bigg[
\frac{1}{5} \pa \, \mbox{(pole-2)}
  +
{\bf W_{\mp}^{(\frac{7}{2})}} \Bigg](w) +\cdots,
\nonu \\
 \left(
\begin{array}{c}
G_{12} \nonu \\
G_{21}
 \end{array}
\right)(z) \, {\bf W_{\pm}^{(\frac{7}{2})}}(w) & = &
\pm
\frac{1}{(z-w)^3} \frac{48(-N+k)}{5(N+26+k)} {\bf T^{(2)}}(w)
\nonu \\
& + &
\frac{1}{(z-w)^2} \Bigg[ -\frac{6(-N+k)}{5(N+2+k)} {\bf T^{(3)}} +
6 {\bf W^{(3)}} \Bigg](w) \nonu \\
& + & \frac{1}{(z-w)} \Bigg[
\frac{1}{6} \pa \, \mbox{(pole-2)}
\mp {\bf W^{(4)}}
 \nonu \\
& \pm & \frac{72 (-N+k)}{(37N+59+(15N+37)k)} \left(
T \, {\bf T^{(2)}} -\frac{3}{10} \pa^2 {\bf T^{(2)}} \right) \Bigg](w)  +\cdots,
\nonu \\
\left(
 \begin{array}{c}
G_{12} \nonu \\
G_{21}
 \end{array}
\right)(z) \, {\bf W^{(4)}}(w) & = &
\frac{1}{(z-w)^3} \frac{12(-N+k)(41N+70+(12N+41)k)}{
(N+2+k)(37N+59+(15N+37)k)}  {\bf T_{\mp}^{(\frac{5}{2})}}(w)
\nonu \\
&+& \frac{1}{(z-w)^2} \Bigg[ 
\pm 7  {\bf W_{\mp}^{(\frac{7}{2})}}- \frac{216(-N+k)}{5(37N+59+(15N+37)k)}
 \pa
 {\bf T_{\mp}^{(\frac{5}{2})}}
 \nonu \\
& + &    \frac{108(-N+k)}{(37N+59+(15N+37)k)} \left(
\begin{array}{c}
G_{12}
\nonu \\
G_{21}
\end{array}
\right) {\bf T^{(2)}}
\Bigg](w)
\nonu \\
& + & \frac{1}{(z-w)} \Bigg[
\pm \pa  {\bf W_{\mp}^{(\frac{7}{2})}}+\frac{72 (-N+k)}{(37N+59+(15N+37)k))}  T
{\bf T_{\mp}^{(\frac{5}{2})}}
\nonu \\
 & +&
\frac{36 (-N+k)}{(37N+59+(15N+37)k)}{\bf T^{(2)}} \pa   \left(
 \begin{array}{c}
G_{12} \nonu \\
G_{21}
 \end{array}
\right)
\nonu \\
&-& \frac{18(-N+k)}{5(37N+59+(15N+37)k)} \pa^2
 {\bf T_{\mp}^{(\frac{5}{2})}}
\Bigg](w) +\cdots.
\label{opeg-16}
\eea
The nonlinear terms 
appearing in (\ref{opeg-16})
can be removed by adding the 
extra quasiprimary field of spin-$4$ to 
the higher spin-$4$ current.
See also the subsection $4.4$.
Via the explicit field identifications between the fields in this 
paper and those in \cite{BCG1404}, the above 
OPEs (\ref{opeg-16}) 
are the same as the ones in \cite{BCG1404}.

\section{The OPEs between the $16$ currents and the 
$16$ higher spin currents  in 
component approach with different basis}

Let us present the description of \cite{BCG1404} as follows: 
\bea
G^a(z) \, V_0^{(s)}(w)  & = & \frac{1}{(z-w)} \, V_{\frac{1}{2}}^{(s), a}(w) 
+\cdots,  
\nonu \\
A^{\pm, i}(z) \, V_{\frac{1}{2}}^{(s), a}(w)  & = & \frac{1}{(z-w)} \,
\alpha_{ab}^{\pm, i} \, V_{\frac{1}{2}}^{(s), b}(w)+\cdots,
\nonu \\
G^a(z) \, V_{\frac{1}{2}}^{(s), b}(w)  & = & \frac{1}{(z-w)^2} \, 2 s 
\, \delta^{ab} \, V_0^{(s)}(w) \nonu \\
& + &
\frac{1}{(z-w)} \, \Bigg[ \alpha_{ab}^{+, i} \, V_{1}^{(s),+, i} +
\alpha_{ab}^{-, i} \, V_{1}^{(s),-, i} +\delta^{ab} \,
\pa V_{0}^{(s)} \Bigg](w) 
+\cdots,  
\nonu \\
Q^a(z) \,  V_{1}^{(s),\pm, i}(w)  & = & \pm \frac{1}{(z-w)} \,
 2 \, \alpha_{ab}^{\pm, i} \, V_{\frac{1}{2}}^{(s),b}(w) +\cdots,
\nonu \\
A^{\pm, i}(z) \, V_{1}^{(s), \pm, j}(w)  & = & \frac{1}{(z-w)^2} \,
2s \, \delta^{ij} \, V_0^{(s)}(w) + \frac{1}{(z-w)} \,
\ep^{ijk} \, V_{1}^{(s), \pm, k}(w)+\cdots,
\nonu \\
G^a(z) \, V_1^{(s), \pm, i}(w)  & = & \frac{1}{(z-w)^2} \, 
4 \, (s + \gamma_{\mp})  \, \alpha_{ab}^{\pm, i} \, V_{\frac{1}{2}}^{(s), b}(w)  
\nonu \\
& + &
\frac{1}{(z-w)} \, \Bigg[ \frac{1}{(2s+1)} \pa (\mbox{pole-2}) \mp
\alpha_{ab}^{\pm, i} \, V_{\frac{3}{2}}^{(s), b} \Bigg](w) 
+\cdots,  
\nonu \\
U (z) \, V_{\frac{3}{2}}^{(s), a}(w)  & = & -\frac{1}{(z-w)^2} \, 
2 \, V_{\frac{1}{2}}^{(s), a}(w) +\cdots,
\nonu \\
Q^a(z) \,  V_{\frac{3}{2}}^{(s),b}(w)  & = & 
\frac{1}{(z-w)^2} \,
4 \, s \, \delta^{ab} \, V_0^{(s)}(w)
\nonu \\
& + &   \frac{1}{(z-w)} \,
2 \Bigg[ \alpha_{ab}^{+, i} \, V_{1}^{(s),+, i} 
+ \alpha_{ab}^{-, i} \, V_{1}^{(s),-, i} -\delta^{ab} \, 
\pa V_{0}^{(s)} \Bigg](w)  +\cdots,
\nonu \\
A^{\pm, i}(z) \, V_{\frac{3}{2}}^{(s), a}(w)  & = & \pm \frac{1}{(z-w)^2} \,
\left[ \frac{8s(s+1) + 4 \gamma_{\mp}}{(2s+1)} \right]  \,
\alpha_{ab}^{\pm, i} \,
V_{\frac{1}{2}}^{(s), b}(w) \nonu \\
& + &  \frac{1}{(z-w)} \,
\alpha_{ab}^{\pm, i} \, 
V_{\frac{3}{2}}^{(s), b}(w)+\cdots,
\nonu \\
G^a(z) \, V_{\frac{3}{2}}^{(s), b}(w)  & = & 
- \frac{1}{(z-w)^3} \, \Bigg[ \frac{16s(s+1)(2\gamma-1)}
{(2s+1)}\Bigg] \delta^{ab} \, V_0^{(s)}(w)  
\nonu \\
& - &  \frac{1}{(z-w)^2} \, \frac{8(s+1)}{(2s+1)} \,
\Bigg[ (s + \gamma_{+}) \,  \alpha_{ab}^{+, i} \, V_{1}^{(s), +, i}  
- (s + \gamma_{-})  \, \alpha_{ab}^{-, i} \, V_{1}^{(s), -, i}  
\Bigg](w)
\nonu \\
& + &
\frac{1}{(z-w)} \, \Bigg[ \frac{1}{2(s+1)}\, \pa (\mbox{pole-2}) +
\delta^{ab} \, V_{2}^{(s)} \Bigg](w) 
+\cdots,  
\nonu \\
U (z) \, V_{2}^{(s)}(w)  & = & \frac{1}{(z-w)^3} \,
8 \, s \, V_{0}^{(s)}(w) -\frac{1}{(z-w)^2} \, 
4 \, \pa V_{0}^{(s)}(w) +\cdots,
\nonu \\
Q^a(z) \,  V_{2}^{(s)}(w)  & = & 
- \frac{1}{(z-w)^2} \,
2  \, (2s+1)  \, V_{\frac{1}{2}}^{(s), a}(w)
+    \frac{1}{(z-w)} \,
2 \,
\pa V_{\frac{1}{2}}^{(s), a}(w)  +\cdots,
\nonu \\
A^{\pm, i}(z) \, V_{2}^{(s)}(w)  & = & \pm \frac{1}{(z-w)^2} \,
2(s+1) \, V_{1}^{(s), \pm, i}(w) +\cdots,
\nonu \\
G^a(z) \, V_{2}^{(s)}(w)  & = & 
\frac{1}{(z-w)^3} \, \left[ \frac{16s(s+1)(2\gamma-1)}
{(2s+1)}\right]  \, V_{\frac{1}{2}}^{(s), a}(w)  
\nonu \\
& + &  \frac{1}{(z-w)^2} \, (2s+3) \,
V_{\frac{3}{2}}^{(s), a}(w)
+ 
\frac{1}{(z-w)} \, \pa  V_{\frac{3}{2}}^{(s), a}(w) 
+\cdots,  
\nonu \\
T(z) \, V_{2}^{(s)}(w)  & = & 
-\frac{1}{(z-w)^4} \, \left[ \frac{24s(s+1)(2\gamma-1)}
{(2s+1)}\right]  \, V_{0}^{(s)}(w)  
\nonu \\
& + &  \frac{1}{(z-w)^2} \, (s+2) \,
V_{2}^{(s)}(w)
+ 
\frac{1}{(z-w)} \, \pa  V_{2}^{(s)}(w) 
+\cdots,  
\nonu \\
T(z) \, V_{0}^{(s)}(w)  & = & 
  \frac{1}{(z-w)^2} \, s \,
V_{0}^{(s)}(w)
+ 
\frac{1}{(z-w)} \, \pa  V_{0}^{(s)}(w) 
+\cdots,  
\nonu \\
T(z) \, V_{\frac{1}{2}}^{(s), a}(w)  & = & 
  \frac{1}{(z-w)^2} \, (s+\frac{1}{2}) \,
V_{\frac{1}{2}}^{(s), a}(w)
+ 
\frac{1}{(z-w)} \, \pa  V_{\frac{1}{2}}^{(s), a}(w) 
+\cdots,  
\nonu \\
T(z) \, V_{1}^{(s), \pm, i}(w)  & = & 
  \frac{1}{(z-w)^2} \, (s+1) \,
V_{1}^{(s), \pm, i}(w)
+ 
\frac{1}{(z-w)} \, \pa  V_{1}^{(s), \pm, i}(w) 
+\cdots,  
\nonu \\
T(z) \, V_{\frac{3}{2}}^{(s), a}(w)  & = & 
  \frac{1}{(z-w)^2} \, (s+\frac{3}{2}) \,
V_{\frac{3}{2}}^{(s), a}(w)
+ 
\frac{1}{(z-w)} \, \pa  V_{\frac{3}{2}}^{(s), a}(w) 
+\cdots.  
\label{bcgprimary}
\eea
Here the two parameters are introduced as follows:
$\gamma_{+} =\gamma= \frac{k^{-}}{(k^{+}+k^{-})}$
and $\gamma_{-} = 1-\gamma = \frac{k^{+}}{(k^{+}+k^{-})}$
where $k^{+} =k+1$ and $k^{-}=N+1$.
From the OPEs in (\ref{bcgprimary}), the higher spin-$s, (s+\frac{1}{2}),
(s+1), (s+\frac{3}{2})$ currents are primary fields under the stress 
energy tensor $T(z)$. 
Note that the higher spin-$(s+2)$ current $V_2^{(s)}(w)$ is not a primary
current because there is a fourth-order pole term.
We can consider the extra composite field 
$T^{(s)} T(w)$ in order to make the above 
higher spin-$(s+2)$ current transforming as a primary field. 
See also the subsection $4.4$.
We can analyze what has been done in \cite{Ahn1504} in order to 
see the explicit relations between the higher spin currents in Appendix $D$
and those in Appendix $E$.
The final expressions are given in the subsection $4.4$.



\begin{thebibliography}{99}

\bibitem{AK1506} 
  C.~Ahn and H.~Kim,
  ``Three Point Functions in the Large N=4 Holography,''
  arXiv:1506.00357 [hep-th].

\bibitem{GG1305} 
  M.~R.~Gaberdiel and R.~Gopakumar,
  ``Large $\mathcal{N}=4$ Holography,''  JHEP {\bf 1309}, 036 (2013)  [arXiv:1305.4181 [hep-th]].  

\bibitem{CH1506} 
  T.~Creutzig and Y.~Hikida,
  ``Higgs phenomenon for higher spin fields on $AdS_3$,''
  arXiv:1506.04465 [hep-th].

\bibitem{HR1503} 
  Y.~Hikida and P.~B.~Ronne,
  ``Marginal deformations and the Higgs phenomenon in higher spin AdS$_{3}$ holography,''
  JHEP {\bf 1507}, 125 (2015)
  [arXiv:1503.03870 [hep-th]].

\bibitem{AP1410} 
  C.~Ahn and J.~Paeng,
  ``Higher Spin Currents in Orthogonal Wolf Space,''
  Class.\ Quant.\ Grav.\  {\bf 32}, no. 4, 045011 (2015)
  [arXiv:1410.0080 [hep-th]].

\bibitem{AK1308} 
  C.~Ahn and H.~Kim,
  ``Spin-5 Casimir operator its three-point functions with two scalars,''
  JHEP {\bf 1401}, 012 (2014)
  [arXiv:1308.1726 [hep-th]].

\bibitem{GG1011} 
  M.~R.~Gaberdiel and R.~Gopakumar,
  ``An $AdS_3$ Dual for Minimal Model CFTs,''  
Phys.\ Rev.\ D {\bf 83}, 066007 (2011)  
[arXiv:1011.2986 [hep-th]].  

\bibitem{GG1205} 
  M.~R.~Gaberdiel and R.~Gopakumar,
  ``Triality in Minimal Model Holography,''  JHEP {\bf 1207}, 127
  (2012)  
[arXiv:1205.2472 [hep-th]].  

\bibitem{GG1207} 
  M.~R.~Gaberdiel and R.~Gopakumar,
  ``Minimal Model Holography,''
  J.\ Phys.\ A {\bf 46}, 214002 (2013)
  [arXiv:1207.6697 [hep-th]].

\bibitem{FG} 
  K.~Ferreira and M.~R.~Gaberdiel,
  ``The $so$-Kazama-Suzuki Models at Large Level,''
  JHEP {\bf 1504}, 017 (2015)
  [arXiv:1412.7213 [hep-th]].

\bibitem{Ahn1208} 
  C.~Ahn,
  ``The Operator Product Expansion of the Lowest Higher Spin Current at Finite N,''
  JHEP {\bf 1301}, 041 (2013)
  [arXiv:1208.0058 [hep-th]].

\bibitem{Ahn1206} 
  C.~Ahn,
  ``The Large N 't Hooft Limit of Kazama-Suzuki Model,''
  JHEP {\bf 1208}, 047 (2012)
  [arXiv:1206.0054 [hep-th]].

\bibitem{ST} 
  A.~Sevrin and G.~Theodoridis,
  ``N=4 Superconformal Coset Theories,''
  Nucl.\ Phys.\ B {\bf 332}, 380 (1990).

\bibitem{AK1411} 
  C.~Ahn and H.~Kim,
  ``Higher Spin Currents in Wolf Space for Generic $N$,''
  JHEP {\bf 1412}, 109 (2014)
  [arXiv:1411.0356 [hep-th]].

\bibitem{Ahn1106} 
  C.~Ahn,
  ``The Large N 't Hooft Limit of Coset Minimal Models,''  JHEP {\bf 1110}, 125 (2011)  [arXiv:1106.0351 [hep-th]].  

\bibitem{GV1106} 
  M.~R.~Gaberdiel and C.~Vollenweider,
  ``Minimal Model Holography for SO(2N),''
  JHEP {\bf 1108}, 104 (2011)
  [arXiv:1106.2634 [hep-th]].

\bibitem{Ahn1202} 
  C.~Ahn,
  ``The Primary Spin-4 Casimir Operators in the Holographic SO(N) Coset Minimal Models,''  JHEP {\bf 1205}, 040 (2012)  [arXiv:1202.0074 [hep-th]].  

\bibitem{AP1310} 
  C.~Ahn and J.~Paeng,
  ``Higher Spin Currents in the Holographic $\mathcal{N} = 1$ Coset Minimal Model,''
  JHEP {\bf 1401}, 007 (2014)
  [arXiv:1310.6185 [hep-th]].

\bibitem{AP1301} 
  C.~Ahn and J.~Paeng,
  ``The OPEs of Spin-4 Casimir Currents in the Holographic $SO(N)$ Coset Minimal Models,''
  Class.\ Quant.\ Grav.\  {\bf 30}, 175004 (2013)
  [arXiv:1301.0208 [hep-th]].

\bibitem{Slansky} 
  R.~Slansky,
  ``Group Theory for Unified Model Building,''
  Phys.\ Rept.\  {\bf 79}, 1 (1981).

\bibitem{npb1989} 
  M.~Gunaydin, J.~L.~Petersen, A.~Taormina and A.~Van Proeyen,
  ``On The Unitary Representations Of A Class Of N=4 Superconformal Algebras,''
  Nucl.\ Phys.\ B {\bf 322}, 402 (1989).

\bibitem{Thielemans} 
  K.~Thielemans,
  ``A Mathematica package for computing operator product expansions,''  Int.\ J.\ Mod.\ Phys.\ C {\bf 2}, 787 (1991).  

\bibitem{KT1985} 
  V.~G.~Kac and I.~T.~Todorov,
  ``Superconformal Current Algebras And Their Unitary Representations,''
  Commun.\ Math.\ Phys.\  {\bf 102}, 337 (1985).

\bibitem{Wolf}
J.~A.~Wolf,
``Complex Homogeneous Contact Manifolds and Quaternionic Symmetric Spaces,''
J. \ Math. \ Mech. {\bf 14}, 1033 (1965).

\bibitem{Alek}
D.~V.~Alekseevskii,
``Classification of Quarternionic Spaces with a Transitive Solvable Group of
Motions,''
\ Math. \ USSR \ Izv. {\bf 9}, 297 (1975).

\bibitem{Salamon}
S.~Salamon,
``Quaternionic Kahler Manifolds,''
\ Invent. \ Math. {\bf 67}, 143 (1982).

\bibitem{Saulina} 
  N.~Saulina,
  ``Geometric interpretation of the large N=4 index,''
  Nucl.\ Phys.\ B {\bf 706}, 491 (2005)
  [hep-th/0409175].

\bibitem{GS} 
  P.~Goddard and A.~Schwimmer,
  ``Factoring Out Free Fermions And Superconformal Algebras,''
  Phys.\ Lett.\ B {\bf 214}, 209 (1988).

\bibitem{cqg1989} 
  A.~Van Proeyen,
  ``Realizations Of N=4 Superconformal Algebras On Wolf Spaces,''
  Class.\ Quant.\ Grav.\  {\bf 6}, 1501 (1989).

\bibitem{GK} 
  S.~J.~Gates, Jr. and S.~V.~Ketov,
  ``No N=4 strings on wolf spaces,''
  Phys.\ Rev.\ D {\bf 52}, 2278 (1995)
  [hep-th/9501140].

\bibitem{BCG1404} 
  M.~Beccaria, C.~Candu and M.~R.~Gaberdiel,
  ``The large N = 4 superconformal $W_{\infty}$ algebra,''
  JHEP {\bf 1406}, 117 (2014)
  [arXiv:1404.1694 [hep-th]].

\bibitem{BBSS1} 
  F.~A.~Bais, P.~Bouwknegt, M.~Surridge and K.~Schoutens,
  ``Extensions of the Virasoro Algebra 
Constructed from Kac-Moody Algebras Using Higher Order Casimir
  Invariants,''  
Nucl.\ Phys.\ B {\bf 304}, 348 (1988).  

\bibitem{BBSS2} 
  F.~A.~Bais, P.~Bouwknegt, M.~Surridge and K.~Schoutens,
  ``Coset Construction for Extended Virasoro Algebras,''  
Nucl.\ Phys.\ B {\bf 304}, 371 (1988).  

\bibitem{BS} 
  P.~Bouwknegt and K.~Schoutens,
  ``W symmetry in conformal field theory,''  
Phys.\ Rept.\  {\bf 223}, 183 (1993)  [hep-th/9210010].  

\bibitem{Ahn1408} 
  C.~Ahn,
  ``Higher Spin Currents in Wolf Space: Part II,''
  Class.\ Quant.\ Grav.\  {\bf 32}, no. 1, 015023 (2015)
  [arXiv:1408.0655 [hep-th]].

\bibitem{Ahn1211} 
  C.~Ahn,
  ``The Higher Spin Currents in the N=1 Stringy Coset Minimal Model,''
  JHEP {\bf 1304}, 033 (2013)
  [arXiv:1211.2589 [hep-th]].

\bibitem{Ahn1305} 
  C.~Ahn,
  ``Higher Spin Currents with Arbitrary N in the ${\cal N} = 1$ Stringy Coset Minimal Model,''
  JHEP {\bf 1307}, 141 (2013)
  [arXiv:1305.5892 [hep-th]].

\bibitem{GH1101} 
  M.~R.~Gaberdiel and T.~Hartman,
  ``Symmetries of Holographic Minimal Models,''
  JHEP {\bf 1105}, 031 (2011)
  [arXiv:1101.2910 [hep-th]].

\bibitem{STVplb} 
  A.~Sevrin, W.~Troost and A.~Van Proeyen,
  ``Superconformal Algebras in Two-Dimensions with N=4,''
  Phys.\ Lett.\ B {\bf 208}, 447 (1988).

\bibitem{AKP} 
  M.~Ammon, P.~Kraus and E.~Perlmutter,
  ``Scalar fields and three-point functions in D=3 higher spin gravity,''
  JHEP {\bf 1207}, 113 (2012)
  [arXiv:1111.3926 [hep-th]].

\bibitem{MZ1211} 
  H.~Moradi and K.~Zoubos,
  ``Three-Point Functions in N=2 Higher-Spin Holography,''
  JHEP {\bf 1304}, 018 (2013)
  [arXiv:1211.2239 [hep-th]].

\bibitem{CY1106} 
  C.~M.~Chang and X.~Yin,
  ``Higher Spin Gravity with Matter in $AdS_3$ and Its CFT Dual,''
  JHEP {\bf 1210}, 024 (2012)
  [arXiv:1106.2580 [hep-th]].

\bibitem{Ahn1111} 
  C.~Ahn,
  ``The Coset Spin-4 Casimir Operator and Its Three-Point Functions with Scalars,''  JHEP {\bf 1202}, 027 (2012)  [arXiv:1111.0091 [hep-th]].  

\bibitem{npb1988} 
  A.~Sevrin, W.~Troost, A.~Van Proeyen and P.~Spindel,
  ``EXTENDED SUPERSYMMETRIC sigma MODELS ON GROUP MANIFOLDS. 2. CURRENT ALGEBRAS,''
  Nucl.\ Phys.\ B {\bf 311}, 465 (1988).

\bibitem{Schoutensnpb} 
  K.~Schoutens,
  ``O(n) Extended Superconformal Field Theory in Superspace,''  Nucl.\ Phys.\ B {\bf 295}, 634 (1988).  

\bibitem{AK1509} 
  C.~Ahn and M.~H.~Kim,
  ``The Operator Product Expansion between the 16 Lowest Higher Spin Currents in the N=4 Superspace,''
  arXiv:1509.01908 [hep-th].

\bibitem{Ahn1311} 
  C.~Ahn,
  ``Higher Spin Currents in Wolf Space. Part I,''
  JHEP {\bf 1403}, 091 (2014)
  [arXiv:1311.6205 [hep-th]].

\bibitem{Ahn1504} 
  C.~Ahn,
  ``Higher spin currents in Wolf space: III,''
  Class.\ Quant.\ Grav.\  {\bf 32}, no. 18, 185001 (2015)
  [arXiv:1504.00070 [hep-th]].

\end{thebibliography}
\end{document}